\documentclass{aa}  

\usepackage{graphicx}
\usepackage{amsmath}
\usepackage{comment}
\usepackage{txfonts}
\usepackage{subfigure}

\def\sec\ond{{\rm s}}

\def\be{\begin{equation}}\def\bea{\begin{eqnarray}}\def\beaa{\begin{eqnarray*}}
\def\ee{\end{equation}}  \def\eea{\end{eqnarray}}  \def\eeaa{\end{eqnarray*}}

\begin{document} 



\title{Suite of hydrodynamical simulations for the Lyman-alpha forest with massive neutrinos}
\author{Graziano Rossi$^{1,2\thanks{\email{graziano@sejong.ac.kr}}}$, Nathalie Palanque-Delabrouille$^{1,3}$, Arnaud Borde$^{1}$,
Matteo Viel$^{4,5}$, Christophe Y{\`e}che$^{1}$, James S. Bolton$^{6}$, James Rich$^{1}$, Jean-Marc Le Goff$^{1}$}
\institute{$^{1}$ CEA, Centre de Saclay, Irfu/SPP, F-91191 Gif-sur-Yvette, France \\
$^{2}$ Department of Astronomy and Space Science, Sejong University, Seoul, 143-747, Korea \\
$^{3}$ Lawrence Berkeley National Laboratory, Berkeley, CA 94720, USA \\
$^{4}$ INAF, Osservatorio Astronomico di Trieste, via G. B. Tiepolo 11, 34131 Trieste, Italy \\
$^{5}$ INFN/National Institute for Nuclear Physics, via Valerio 2, 34127 Trieste, Italy \\
$^{6}$ School of Physics and Astronomy, University of Nottingham, University Park, Nottingham NG7 2RD}

\authorrunning{G. Rossi  et al. 2014}
\titlerunning{Lyman-$\alpha$ forest and massive neutrinos}
   


 \abstract{The signature left in quasar spectra by neutral hydrogen in the Universe allows constraining the sum of the 
neutrino masses with a better sensitivity than laboratory experiments and may shed new light on the neutrino mass hierarchy 
and the absolute mass-scale of neutrinos. 
Constraints on cosmological parameters and on the dark energy equation of state can also be derived from a joint parameter estimation procedure. 
However, this requires a detailed modeling of the line-of-sight power spectrum of the transmitted flux in the Lyman-$\alpha$ (Ly$\alpha$) forest on scales 
ranging from a few to hundreds of megaparsecs, which in turn demands the inclusion and careful treatment of cosmological neutrinos.
To this end, we present here a suite of state-of-the-art 
hydrodynamical simulations with cold dark matter (CDM), baryons and massive neutrinos, specifically 
targeted for modeling the low-density regions of the intergalactic medium (IGM) as probed by the Ly$\alpha$ forest at high-redshift.
The simulations span volumes ranging from $(25~h^{-1} {\rm Mpc})^3$ to $(100~h^{-1} {\rm Mpc} )^3$, 
and were made using either $3 \times 192^3 \simeq 21$ million or $3 \times 768^3 \simeq 1.4$ billion particles. 
The resolution of the various runs was further enhanced, 
so that we reached  the equivalent of $3 \times 3072^3 \simeq 87$ billion particles in a $(100~h^{-1} {\rm Mpc} )^3$ box size.
The chosen cosmological parameters are compatible with the latest Planck (2013) results, although we also
explored the effect of slight variations in the main cosmological and astrophysical parameters.
We adopted a particle-type implementation of massive neutrinos, and consider three degenerate 
 species with masses $\sum m_{\rm \nu} =0.1, 0.2, 0.3, 0.4$, and $0.8$ eV, respectively.
We improved on previous studies in several ways, in particular with 
updated routines for IGM radiative cooling and heating processes, and
initial conditions based on second-order Lagrangian perturbation theory
(2LPT) rather than the {Zel'dovich} approximation.
This allowed us to safely  start our runs at  relatively low redshift ($z=30$), which
reduced the shot-noise contamination in the neutrino component and the CPU consumption. 
In addition to providing technical details on the simulations, we present the first
analysis of the nonlinear three- and one-dimensional matter and flux power spectra from these models, 
and characterize the statistics of the transmitted flux in the Ly$\alpha$ forest including the effect of massive neutrinos. 
In synergy with recent data from the Baryon Acoustic Spectroscopic Survey (BOSS) and the Planck satellite, and with a grid of corresponding neutrino-less simulations, our realizations will allow 
us to constrain cosmological parameters
and neutrino masses directly from the Ly$\alpha$ forest with improved sensitivity.
In addition, our simulations can be useful for a broader variety of cosmological and astrophysical applications, 
ranging from the three-dimensional modeling 
of the Ly$\alpha$ forest to cross-correlations between different probes, 
studying the expansion history of the Universe 
including massive neutrinos, and particle-physics related topics.  
Moreover, while our simulations have been specifically designed to meet the requirements of the BOSS survey, 
they can also be used for upcoming or future experiments -- such as eBOSS and DESI.}
  





 \maketitle



\section{INTRODUCTION}


Neutrino science has received a boost of attention recently, because the breakthrough discovery in particle physics over the last decade that 
neutrinos are indeed massive. 
However,  at the present time we only know their mass differences, because solar, atmospheric, reactor, and accelerator observations of neutrino oscillations 
 are sensitive only to differences in the squares of neutrino masses, 
 requiring that there be at least one species with mass $m \ge 0.06$ eV.
On the other hand, cosmology offers a unique `laboratory' with the best sensitivity to the neutrino mass (see for example Lesgourgues \& Pastor 2012, and references therein), as 
primordial massive neutrinos comprise a small portion of the dark matter (DM) and therefore must
significantly alter structure formation. 
Potentially, combining cosmological and particle physics results, it is expected that we will be able to determine the absolute mass scale of  
neutrinos  in the very near future, and solve one of the key questions in neutrino physics today -- namely, the nature of their mass hierarchy and perhaps
the origin of mass.

Neutrino physics also provides one of the best examples of the interplay between particle physics and cosmology/astrophysics.
For instance, the measurement of neutrino masses
could point to a new fundamental theory, of which the
standard model (SM) is the low-energy limit (Lesgourgues \& Pastor 2006) -- hence calling for new physics beyond the SM.
In addition, astrophysical neutrino fluxes can be exploited to test the SM, with experiments of neutrino decays, oscillations,
and searches for nonzero neutrino electromagnetic moments.


In a cosmological context, the effect of massive neutrinos is essentially twofold.
Firstly, neutrinos contribute to the expansion rate during the radiation epoch as one of $N_{\rm eff}$ 
neutrinos (with $N_{\rm eff}$ the effective number of neutrino species; a recent constraint from the Planck data is $N_{\rm eff}=3.36 \pm 0.34$ -- see Ade et al. 2013) 
and later as a nonrelativistic component of matter. 
Compared with massless models, this modifies the timing of matter-radiation equality and the distance-redshift relation. 
Secondly, after they become non-relativistic, neutrinos participate in structure formation, but only on scales greater than the 
free-streaming scale.
Because of these two effects, models with neutrino masses greater than $0.1$ eV give predictions different from 
standard cold dark matter (CDM) scenarios with a cosmological constant (i.e.,  LCDM models),
which generally incorporate a minimal neutrino mass of $0.06$ eV.

Hence, while the most recent results from the cosmic microwave background (CMB), such as data from the Planck satellite (Ade et al. 2013), the Atacama 
Cosmology Telescope (ACT; Sievers et al. 2013) 
or the South Pole Telescope (SPT; Hou et al. 2012),
and from the  large-scale structure (LSS) as in the Sloan Digital Sky Survey (SDSS; York et al. 2000, Eisenstein et al. 2011) or  in the WiggleZ survey (Drinkwater et al. 2010; Blake et al. 2012)
are consistent with  the $\Lambda$CDM  model dominated by a dark energy (DE) component, with baryons constituting only $4.5\%$ of the total matter-energy content, 
a pure CDM scenario is still unsatisfactory and incomplete -- since 
 even a small amount of neutrinos can significantly impact structure formation. 
  Improving our knowledge of cosmological neutrinos is essential for an accurate and consistent minimal cosmological model, and the present study is an effort in this direction.


In cosmology, neutrinos have been studied with a large number of probes and complementary techniques.
The most direct way is through the analysis of the CMB radiation, because for the current mass limits their primordial signature does not
vanish although neutrinos are still relativistic at the time of recombination (Lesgourgues \& Pastor 2006).
While the overall sensitivity of massive neutrinos impacts the CMB temperature power spectrum very marginally, there
are non-negligible consequences in the polarization maps through the early integrated Sachs Wolfe (ISW) effect (Hinshaw et al. 2013), and distinct
signatures  from the gravitational lensing of the CMB by LSS -- both in temperature and polarization (see for instance Santos et al. 2013
or Battye \& Moss 2013). 
Other methods for quantifying the impact of massive neutrinos involve  baryonic tracers of the LSS clustering of matter, and 
high-redshift surveys. Examples include the measurement of the three-dimensional matter power spectrum obtained from galaxy surveys,
 Lyman-$\alpha$ (Ly$\alpha$), or $21$ cm probes where the underlying tracer is neutral hydrogen (HI), the study of galaxy clusters via the  Sunyaev-Zel'dovich (SZ) effect, and
the characterization of the cosmic shear through weak lensing 
(Kaiser 1992; Jain \& Seljak 1997; Zaldarriaga \& Seljak 1998; Abazajian \& Dodelson 2003).


While most techniques used in cosmology to constrain neutrino masses are based on the CMB or on galaxy clustering,
fewer studies involve the Ly$\alpha$ forest -- that is, the absorption lines 
in the spectra of high-redshift quasars, that are due to neutral hydrogen in the intervening  photoionized intergalactic medium (IGM).
Thanks to data from the SDSS (York et al. 2000), the statistical power  
of the  Ly$\alpha$ forest has greatly increased, so that it is now
emerging as a  very promising and unique
window into the high-redshift Universe, because it is at a redshift range inaccessible to other LSS probes and spans a wide interval in redshift. 
For this reason, it was recently possible, for instance, to detect for the first time the baryon acoustic oscillation (BAO) signal directly from the Ly$\alpha$ forest 
(Busca et al. 2013; Slosar et al. 2013). This will be even more so with future surveys, such as eBOSS (Comparat et al. 2013) and DESI (Schlegel et al. 2011).

The Ly$\alpha$ forest is particularly well suited to constrain neutrino masses, since 
massive neutrinos leave a redshift- and mass-dependent signature in the one-dimensional flux power spectrum
because the growth of cosmological structures on scales smaller than the neutrino free-streaming distance is suppressed. 
To detect this effect, careful modeling of the line-of-sight (LOS) power spectrum of the transmitted Ly$\alpha$ flux is required.
Pioneering  work along these lines has been carried out by  Croft et al. (1998, 2002), Zaldarriaga, Hui \& Tegmark (2001),  Viel et al. (2004, 2006, 2010), Seljak et al. (2005),  
McDonald et al. (2006), Seljak, Slosar \& McDonald (2006), and Kim \& Croft (2008). 
In particular, McDonald et al. (2006) and Seljak et al. (2006) used a sample of $3035$ moderate-resolution  
forest spectra from the SDSS to measure the one-dimensional 
flux power spectrum at $z =2.2 - 4.2$. They placed constraints on the linear matter power spectrum and on neutrino masses, 
while Viel et al. (2010) studied the impact of massive neutrinos in the  transmitted Ly$\alpha$ flux.
At present, the most precise measurement of the Ly$\alpha$ flux power spectrum comes from the Baryon Acoustic Spectroscopic Survey 
(BOSS; Dawson et al. 2013), with a sample of forest spectra almost 
 two orders of magnitude larger than in previous studies (Palanque-Delabrouille et al. 2013).
These Ly$\alpha$ forest measurements supplement those obtained from the population of luminous red galaxies (LRGs), and considerably 
extend the redshift range that can be studied.


The Ly$\alpha$ forest also offers one of the strongest reported constraints on neutrino mass when combined with WMAP 3-year CMB data 
(i.e. $\sum m_{\rm \nu}<0.17$ eV at 95\% CL; Seljak et al. 2006), but the constraint depends on the normalization of the observed LSS power spectrum 
relative to the CMB power spectrum; using recent data from the Planck satellite instead of those from WMAP, 
the previous constraints are weakened. Nevertheless, current neutrino mass limits are 
on the verge of distinguishing between a normal (one species with $m \sim 0.06$ eV) 
and inverted (two species with $m \sim0.06$ eV)  hierarchy, and
in the near future the degeneracy of neutrino masses
will be removed by combining cosmological results with atmospheric and solar neutrino
constraints. For example, the combination of Planck CMB data, WMAP 9-year CMB polarization data (Bennett et al. 2013),
and a measurement of BAO from BOSS, SDSS, WiggleZ, and the 6dF galaxy redshift survey (Jones et al. 2009) produces an upper limit of $\sum m_{\rm \nu} <0.23 $ (95\% CL),
while a more aggressive use of galaxy clustering into smaller scales and the nonlinear
clustering regime can lead to stringent constraints (Zhao et al. 2012; Riemer-S{\o}rensen et al. 2013).


However, the validity of the current limits on neutrino masses depends on the assumption that there are no systematic offsets
between estimates of the matter power spectrum obtained with different methods; according to
Viel et al. (2010), these uncertainties are not reflected in the quoted measurement errors.
To this end, one needs to gain a better understanding of the characteristic signatures of massive neutrinos in the power spectrum 
across different redshift slices, and be in control of the various systematics involved, especially at lower redshifts ($z=2 - 4$) and at small scales ($1 - 40 h^{-1}$Mpc) --
where the nonlinear evolution of density fluctuations 
for massive neutrinos is non-negligible. 
Particularly for the Ly$\alpha$ forest, 
constraints on neutrino masses are only limited by the systematic accuracy with which we can make these theoretical predictions. 
This is only possible through more and more sophisticated numerical simulations, where the full hydrodynamical treatment is performed
at scales where nonlinear effects become important for the neutrino component; so far, 
only a handful studies in the literature have addressed these aspects for the Ly$\alpha$ forest in some detail. 
Given that current and planned experiments such as BOSS, eBOSS and DESI will provide
excellent-quality data for the Ly$\alpha$ forest (see also the recent American 2013 report `Cosmic Frontier Vision', and in particular
Connolly et al. 2013), 
it is now timely to design and perform accurate numerical simulations capable of reproducing the effects of massive neutrinos.


The present study aims at filling this gap by  presenting a suite of state-of-the-art 
hydrodynamical simulations with cold dark matter, baryons and massive neutrinos, specifically 
targeted for modeling the low-density regions of the IGM as probed by the Ly$\alpha$ forest at high-redshift.
In addition to providing technical details on the simulations and on the improvements made with respect to pre-existing literature, we show here 
measurements of the simulated nonlinear three- and one-dimensional matter and flux power spectra, 
and characterize the statistics of the transmitted flux in the Ly$\alpha$ forest in presence of massive neutrinos. 
This is the first of a series of papers dedicated 
to quantify the effects of massive neutrinos in the Ly$\alpha$ forest across different redshift slices and at nonlinear scales.
In addition, we are planning to make the simulations available to the scientific community upon request; hence, the present work may serve as a
guide for a direct use of the simulations and of the products provided.


The layout of the paper is organized as follows:
in Section \ref{sec_modeling_lya}, we briefly outline the theory behind the modeling of the Ly$\alpha$ forest, along
with the most commonly used numerical techniques available.
In Section \ref{sec_implementing_neutrinos}, we focus on neutrino science and on the implementation of massive neutrinos in cosmological $N$-body simulations
and explain the method of our choice.
In Section \ref{sec_simulations_description}, we present our novel suite of hydrodynamical simulations and provide several
technical details on the code used for the run, initial conditions, optimization strategies and performance, along with various improvements and a description 
of the pipeline developed to extract the synthetic Ly$\alpha$ transmitted flux;
in the appendix, we also describe a sanity check we performed
to ensure that we correctly recover the limit of massless neutrinos.
In Section \ref{sec_first_results}, 
we present the first analysis of our suite of simulations, where in particular we
compute the three- and one-dimensional matter and flux power spectra, 
focusing on the imprint of massive 
neutrinos. 
We conclude in Section \ref{sec_conclusions}, 
where we summarize our main
achievements and explain how we will use the simulations presented here to constrain neutrino masses directly from the
Ly$\alpha$ forest, with improved sensitivity. 



\section{MODELING THE LYMAN-ALPHA FOREST} \label{sec_modeling_lya}


In this section we briefly summarize the basic theory of the Ly$\alpha$ forest as a cosmological probe and
the most commonly used numerical techniques for modeling 
the low-density regions of the IGM.
In particular, we focus
on the specific requirements necessary to accurately reconstruct the Ly$\alpha$ transmitted flux. 


\subsection{Ly$\alpha$ forest: overview and challenges}

The observational discovery of the Ly$\alpha$ forest traces back to Lynds (1971), although
the actual existence of an ionized IGM was already postulated back in the 1960s (Bahcall \& Salpeter 1965; Gunn \& Peterson 1965).
However, only some twenty years later was it realized 
that the numerous absorption features 
in the spectra of high-redshift quasars,
 bluewards of the redshifted resonant 
$1215.67$\r{A} emission line, 
directly trace the underlying dark matter fluctuations (Cen et al. 1994; Bi et al. 1995;
Zhang et al. 1995; Hernquist et al. 1996; Miralda-Escude et al. 1996; Bi \& Davidsen 1997; Hui, Gnedin \& Zhang 1997; Theuns et al. 1998).
Clearly, since hydrogen makes up most of the baryonic density of the Universe, the Ly$\alpha$
forest is also a direct tracer of the baryonic matter distribution 
over a wide range of scales and redshifts -- i.e. $k \sim 0.1 - 10~h{~\rm Mpc^{-1}}$; $1 \le z \le 6$.

Since then, considerable progress has been made toward a thorough understanding of the 
nature of these absorption features and of the properties of the IGM.
We now have observational evidence that 
 at high redshift the IGM contains the majority of baryons present in the Universe (Petitjean et al. 1993; Fukugita et al. 1998),
is highly ionized by the ultra-violet (UV) background produced by galaxies and quasars, and becomes increasingly neutral from $z=0$ to $z=7$ (Mortlock et al. 2011).
The overall physical picture that emerges is  relatively simple: 
the IGM probed by the Ly$\alpha$ forest consists of mildly nonlinear gas density fluctuations;
low column-density absorption lines trace the filaments of the cosmic web;
high column-density absorption lines trace the surrondings of galaxies;  
the gas traces the dark matter, and is photoionized and photoheated by the UV-background.
Although metals are present in the IGM (Cowie et al. 1995; Schaye et al. 2003; Aracil et al. 2004), stirring of the IGM due to feedback from
galaxies or active galactic nuclei (AGNs) does not significantly affect the vast majority of the baryons (Theuns et al. 2002; McDonald et al. 2005).
Photoionization heating and expansion cooling cause the gas density ($\rho$) and temperature ($T$) to be closely related, except
where mild shocks heat the gas (see Schaye et al. 2000), so that in low-density regions a simple redshift-dependent polytropic power-law temperature-density relation holds
(Katz, Weinberg \& Hernquist 1996; Hui \& Gnedin 1997):
\begin{equation}
T(z) = T_0(z) \Big ( {\rho \over \rho_0}   \Big ) ^{\gamma(z) -1},
\label{eq_T_rho}
\end{equation}
where $T_0$ and $\rho_0$ are the corresponding gas mean temperature and density, while
the parameter $\gamma$ depends on redshift, reionization history model, and  spectral shape of the UV background.
It is interesting to address the modifications to this simple relation caused by massive neutrinos,
and we return to this issue in Section \ref{sec_first_results}.

The gas of the IGM  is generally assumed to 
be in photoionization equilibrium with the UV background, and
it can be described by an optical depth $\tau(z)$ that depends on the evolving photoionization
rate (Peebles 1993). The optical depth for Ly$\alpha$ absorption is proportional to the neutral hydrogen density (Gunn \& Peterson 1965),
which -- since the gas is in photoionization equilibrium -- can also be expressed as
\begin{equation}
\tau = A \Big ( {\rho \over \rho_0}   \Big )^{\beta}, 
\end{equation}
where $\beta = 2.7 - 0.7 \gamma$ and $A$ depends on redshift, baryon density, temperature
at the mean density, Hubble constant, and photoionization rate.
While the optical depth is a tracer of the matter distribution on scales larger than the Jeans length of the photoionized IGM,
it is more conventional to use the mean transmitted flux $\bar{\cal{F}}$ instead, and define an
effective optical depth $\tau_{\rm eff}$ so that
\begin{equation}
\tau_{\rm eff} = - \ln \bar{\cal{F}}.
\end{equation}
The previous expression contains the uncertainties in the intensity of the UV background, the mean baryon density,
and other parameters that set the normalization of the relation between optical depth and density of the gas. 
Measurements of the mean transmission and of its evolution allow one to
constrain the basic cosmological parameters
(see Jenkins \& Ostriker 1991; Hernquist et al. 1996; Rauch et al. 1997; Rauch 1998; McDonald \& Miralda-Escude 2001).
The gas density is also closely related to that of the DM on large scales, while
on small scales the effects of thermal broadening and Jeans smoothing must be included. For more details on the
physics of  the IGM and its potential for cosmology, see Meiksin (2009).

Dynamical and thermal processes are essential in modeling the Ly$\alpha$ forest: 
therefore,
the effects of baryon  pressure,  nonlinear evolution of density perturbations, 
thermal and chemical evolution such as  adiabatic cooling due to the 
expansion of the Universe, UV background photoionization heating,  as well as Compton and recombination cooling need to
be taken into account. 
For instance, 
the Ly$\alpha$ flux power spectrum depends not only on the DM distribution, but also on the thermal state of the IGM, 
and on feedback effects due to star formation and 
AGNs. Hence, a full hydrodynamical modeling including effects of galaxy formation physics 
is necessary.
While there exist  several numerical challenges in simulating the Ly$\alpha$ forest, along
with a series of physical mechanisms still poorly understood, today we do have the computational capability of
carrying out full hydrodynamical treatments, as we perform in this work.
Despite hydrodynamic uncertainties, hierarchical models of structure formation are now capable of
reproducing almost all the aspects of the Ly$\alpha$ forest.
We also note that once the spectrum is modeled as a continuous phenomenon, there
is no need to resolve every single feature (Weinberg et al. 1999, 2003), so that 
the forest can be studied with relatively moderate resolution spectra.


\subsection{Hydrodynamical simulations in a nutshell}

The rapid progress made in our theoretical understanding of the Ly$\alpha$ forest
is mainly due to the improved ability to simulate 
all the physical effects that
impact the IGM 
more and more realistically -- thanks to state-of-the-art  computational facilities. 
In fact, while the forest has been traditionally studied using a variety of analytical
techniques such as the {Zel'dovich} approximation (Doroshkevich \& Shandarin 1977; McGill 1990; Hui, Gnedin \& Zhang  1997;  Matarrese \& Mohayaee 2002),
 the lognormal approximation (Hamilton 1985; Coles \& Jones 1991; Bi 1993; Bouchet et al. 1993; Kofman et al. 1994; Gnedin \& Hui 1996; Bi \& Davidsen 1997;
Viel et al. 2002), or semi-analytic models (Balian \& Schaeffer 1989; Bernardeau \& Schaeffer 1992, 1999; Valageas, Schaeffer \& Silk 1999; Pichon et al. 2001),
it is only with hydrodynamical simulations
that the interplay between gravity and gas pressure on the structure of the photoionized IGM
can be modeled self-consistently -- so that 
most of the observed properties of the Ly$\alpha$ forest are successfully reproduced and 
the uncertainties in the theoretical modeling overcome. 

Traditionally, cosmological hydro-simulations come in two basic flavors: 
smoothed particle hydrodynamics (SPH; Gingold \& Monaghan 1977; Lucy 1977), and grid-based methods;
there are also more sophisticated combinations of the two categories.
The SPH technique -- adopted in this study -- uses 
particles to represent the baryonic fluid, employs an artificial viscosity to simulate shocks (Springel \& Hernquist 2002), 
and it is Lagrangian in nature: this implies that the resolution is concentrated in regions of high-density.
On the other hand, grid-based methods use a grid of cells to represent the gas properties, which may
provide a superior resolution
of the low-density regions of the IGM and more accurate treatments of shocks, but
at a higher computational cost. 
For more details on hydrodynamical techniques relevant to this study, see Katz et al. (1996).

Both approaches have been successfully used to model the Ly$\alpha$ forest at low- and high-redshift  
and to obtain quantitative estimates of the clustering amplitude and constraints on cosmological
and astrophysical parameters. The number of dedicated studies has increased
 in recent years, since the forest is emerging as a key probe of the  hydrogen reionization epoch.
A long, but still incomplete list of relevant numerical works includes 
Gnedin \& Hui (1996, 1998), Croft et al. (1998, 1999, 2002), Hui et al. (2001), 
McDonald et al. (2000, 2001, 2005), Meiksin \& White (2001), 
Gnedin \& Hamilton (2002), 
Zaldarriaga et al. (2001, 2003), 
Seljak et al. (2003), Bolton \& Haehnelt (2007), 
 Viel et al. (2003, 2004, 2010, 2012), 
 Crain et al. (2009), 
Bolton \& Becker (2009), Schaye et al. (2010). 

The importance of having full hydrodynamical simulations cannot be stressed enough.
To provide an example, McDonald et al. (2005) used hydrodynamical simulations extended with hydro-particle-mesh  (HPM) realizations to 
analyze the SDSS Ly$\alpha$ forest power spectrum and infer the corresponding linear theory power spectrum.
However, their HPM simulations -- calibrated
by a limited number of hydrodynamical runs -- were found to be discrepant by up to $20\%$  when compared with full hydrodynamical simulations, 
with respect to the statistical properties of the Ly$\alpha$ flux distribution  (Viel et al. 2006). 
Hence, while using approximate numerical calculations is certainly attractive
because computationally less demanding, a complete hydrodynamical treatment is mandatory to reach
the precision that data are now beginning to show. 

Before moving on to the treatment of massive neutrinos, 
 we stress that the development of progressively more sophisticated numerical simulations 
 is an area of rapid progress -- particularly crucial for a  realistic modeling
 of the Ly$\alpha$ forest.  With increasing computational power, 
 there is currently less motivation to 
use approximation methods -- although more work is needed to
 understand a multitude of complex baryonic processes.
This is even more so when massive neutrinos are included in the picture, 
and the scope of the present work is to add more knowledge in this direction. 

\begin{figure}
\centering
\includegraphics[angle=0,width=0.48\textwidth]{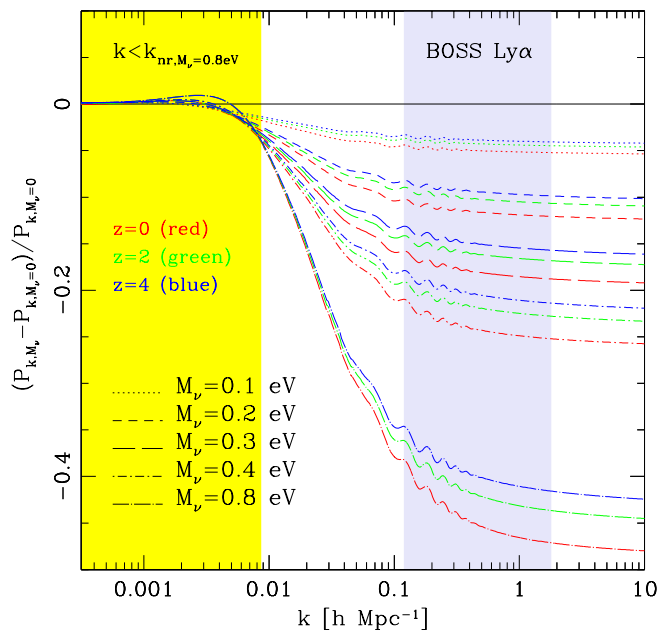}
\caption{Linear 
theory predictions
for the matter power spectra with massive neutrinos, normalized by
the corresponding case of massless neutrinos.
The cosmological parameters considered are those used for our simulations (see Section \ref{sec_simulations_description}); the
neutrino mass range is indicated in the figure. 
Different colors (and similar line styles) show the evolution in redshift for $z=0, 2, 4$, respectively, as a function of  the neutrino mass.
The yellow area corresponds to values of $k< k_{\rm nr, M_{\nu}=0.8~eV} $, 
where a linear description for the neutrino evolution is
sufficient; the gray zone highlights the range of $k$ approximately covered by the one-dimensional flux power spectrum 
obtained from the Ly$\alpha$ BOSS survey.}
\label{fig_neutrino_linear_theory_A}
\end{figure}



\section{IMPLEMENTING MASSIVE NEUTRINOS} \label{sec_implementing_neutrinos}

In this section we first provide a synthetic overview of the effects of massive neutrinos in cosmology -- focusing on the Ly$\alpha$ forest;
in particular, we present the expected linear predictions for the matter power spectra in presence of massive neutrinos, with the set of cosmological parameters
adopted in our simulations.  
We then briefly describe
how neutrinos are implemented.
In Section \ref{sec_first_results},
the linear predictions shown here are compared with nonlinear measurements obtained from our simulations.


\subsection{Revival of neutrino science}

The impact of massive neutrinos on the CMB and LSS 
was investigated long ago
(see for example Bond, Efstathiou \& Silk 1980; Klypin et al. 1993; Ma \& Bertschinger 1995; Dodelson et al. 1996; Hu, Eisenstein \& Tegmark 1998;
Hu \& Dodelson 2002; Abazajian et al. 2005; Hannestad 2005; Seljak et al. 2006),
and with a renewed interest quite recently (e.g. Saito et al. 2008, 2009; Wong 2008; 
  Brandbyge et al. 2008; Brandbyge \& Hannestad 2009;  Viel et al 2010;  Marulli et al. 2011; Bird et al 2012;  Carbone et al. 2012; Hou et al. 2012;  
  Lesgourgues \& Pastor 2012).
The renewed interest  is mainly driven 
by the  large amount of cosmological data available today, 
which allow placing competitive limits on the neutrino mass-scale and hierarchy. 
For instance, simply with the improvement of
a factor of two from Seljak et al. (2006), one should be
able to distinguish between a normal hierarchy and 
an inverted one -- a fact within reach in the very near future, given high-quality upcoming surveys such as eBOSS and DESI. 

The effects of cosmological neutrinos on the evolution of density perturbations in the linear regime 
is well understood. 
 In what follows, we only discuss a few general aspects of cosmological neutrinos
  relevant for the Ly$\alpha$ forest, 
and refer to Lesgourgues \& Pastor (2006, 2012) for a more exhaustive treatment.

Neutrinos decouple from the cosmic plasma before the electron-positron annihilation (around $\sim 1$ MeV),
resulting in a subsequent neutrino temperature  $T_{\nu}$ that is lower than the photon temperature $T_{\gamma}$, namely
 \begin{equation}
 T_{\nu} = (4/11)^{1/3} T_{\gamma},
 \end{equation}
and a number density $n_{\rm \nu}$ lower than the photon number density:
\begin{equation}
n_{\rm \nu} = N_{\rm eff} \Big (  {3 \over 4} \Big ) \Big ( {4 \over 11 } \Big ) n_{\rm \gamma},
\end{equation}
where $n_{\rm \gamma}$ is the density of the CMB photons, and 
the factor 3/4 comes from the difference between the Fermi-Dirac and Bose-Einstein statistics.
Moreover, they  behave as additional radiation while ultra-relativistic, traveling at the speed of light
 with a free-streaming length equal to the Hubble radius,   and as an additional CDM
component when they become non-relativistic.  
Subsequently, massive neutrinos affect structure formation by free-streaming and by delaying 
matter domination. 
These effects can be parameterized by their ultimate fractional contribution to the matter density:
\begin{equation}
f_{\rm \nu} = \Omega_{\rm \nu}/\Omega_{\rm m}, ~ \Omega_{\rm \nu} h^2 = {M_{\rm \nu} \over 93.14 ~{\rm eV}},
\end{equation}
where $h$ is the present value of the Hubble constant in units of 100 km~$s^{-1}{\rm Mpc}^{-1}$,
$M_{\rm \nu}=\sum m_{\rm \nu}$ is the sum of the neutrino masses of the three species considered,
and $\Omega_{\rm m}$ is the matter energy density in terms of the critical density.

\begin{figure*}
\centering
\includegraphics[angle=0,width=0.90\textwidth]{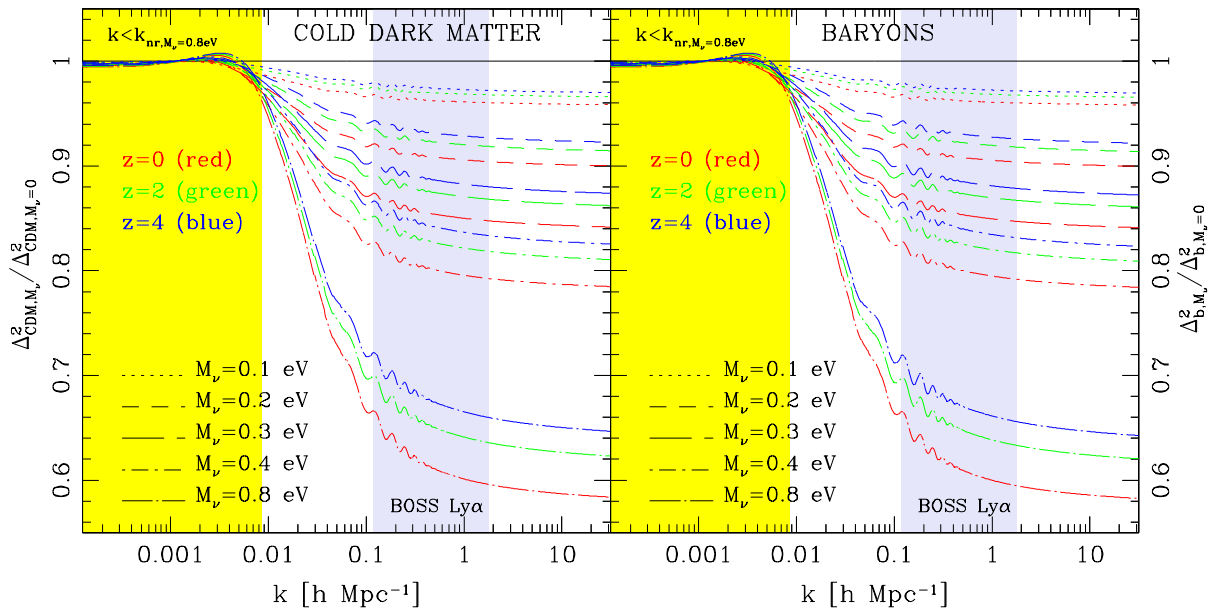}
\caption{Dimensionless linear power spectra per component  in presence of massive neutrinos, 
normalized by the corresponding case of massless neutrinos -- as defined in the main text. The left panel shows the linear evolution of the CDM 
component, while the right panel displays the corresponding baryonic evolution. The linear evolution of the two components is very similar.
Line styles, redshifts, colored areas, and neutrino mass ranges are same as in the previous figure.}
\label{fig_neutrino_linear_theory_B}
\end{figure*}

Neutrinos in the mass range $0.05~{\rm eV} \le m_{\rm \nu} \le 1.5~{\rm eV}$ become non-relativistic
in the redshift interval $3000 \ge z \ge 100$, approximately
 around $z_{\rm nr} \sim 2000 ~(m_{\rm \nu}$/1eV) -- during the matter domination era; for the given mass-intervals considered in this study, 
 all our runs started well in the non-relativistic regime.
When neutrinos are non-relativistic, there is a minimum wavenumber
\begin{equation}
k_{\rm nr} \sim 0.018 ~\Omega_{\rm  m}^{1/2} \Big [ {m_{\rm \nu} \over 1~{\rm eV}} \Big ]^{1/2} h~{\rm Mpc^{-1}}
\label{eq_k_nr}
\end{equation}
above which the physical effect produced by their free-streaming damps small-scale density fluctuations, while
modes with $k < k_{\rm nr}$ evolve according to linear theory. 
The free-streaming leads to a suppression of power on small scales;
with increasing neutrino mass, this suppression becomes stronger
and its shape and amplitude depend mainly on the total mass, but only weakly on redshift (Bond, Efstathiou \& Silk 1980).
At scales $k>0.1$ the suppression is constant, while at $0.01 < k< 0.1$ it gradually decreases to
zero -- with $k$ expressed in units of $h~{\rm Mpc^{-1}}$. When $k \ll 0.01$ (very large scales), the influence 
of neutrinos in the matter power spectrum becomes negligible.
All these effects are clearly seen in 
Figure \ref{fig_neutrino_linear_theory_A}, where we show the linear theory predictions
for the matter power spectra, which include massive neutrinos ($P_{\rm k, M_{\nu}}$), normalized by
the corresponding case of massless neutrinos ($P_{\rm k, M_{\nu}=0}$).
The cosmological parameters adopted are those used for our simulations and reported in Section \ref{sec_simulations_description};
we consider the following neutrino masses: $M_{\nu} = 0.1, 0.2, 0.3, 0.4, 0.8$ eV.
With different colors but similar line styles, we also show the evolution in redshift for three significant intervals, namely 
$z=0$ (red), $z=2$ (green), and $z=4$ (blue). All the various linear predictions were computed with the CAMB code (Lewis, Challinor \& Lasenby 2000).
The yellow area in the figure corresponds to values of $k$ lower than $k_{\rm nr}$  for $M_{\nu}=0.8$ eV (i.e., the most massive case considered here) obtained
 from (\ref{eq_k_nr}),
below which a linear description for the neutrino evolution is
sufficient.
For masses $M_{\nu}<0.8$ eV, the corresponding $k_{\rm nr, M_{\nu}}$ values are lower than $k_{\rm nr, M_{\nu}=0.8~eV}$.
The gray area in the same figure shows the $k$-range approximately covered by the BOSS survey,
relatively to the one-dimensional Ly$\alpha$ forest power spectrum. As can be clearly seen, our primary range of interest lies  well outside the
zone in which a linear description would be sufficient -- for the neutrino masses considered in this study; hence, a full nonlinear treatment of the neutrino component is mandatory. 
In Section \ref{sec_first_results}, we
compare these linear predictions with the corresponding nonlinear evolutions as a function of neutrino mass 
and quantify the departures from linearity; we also determine at which $k$ these departures are maximized.

Figure \ref{fig_neutrino_linear_theory_B} presents the dimensionless linear power spectra per component when massive neutrinos are
included, normalized by the corresponding zero-neutrino-mass case. The general convention used in this paper
sets $\Delta_{\rm i}^2 = k^3 P_{\rm i}(k)/2 \pi^2$, where
the subscript $i$ specifies the component considered. 
In detail, the left panel shows the CDM linear evolution, while the 
right panel displays the evolution of the baryonic component; neutrino mass ranges, redshifts, and line styles are the
same as in Figure \ref{fig_neutrino_linear_theory_A}. Evidently, the linear evolution of the two components is very similar and closely coupled,
with slight departures at increasing redshifts.
Note also the remarkable suppression of power (about $40\%$), caused solely by a $6\%$ component.

\begin{figure}
\centering
\includegraphics[angle=0,width=0.45\textwidth]{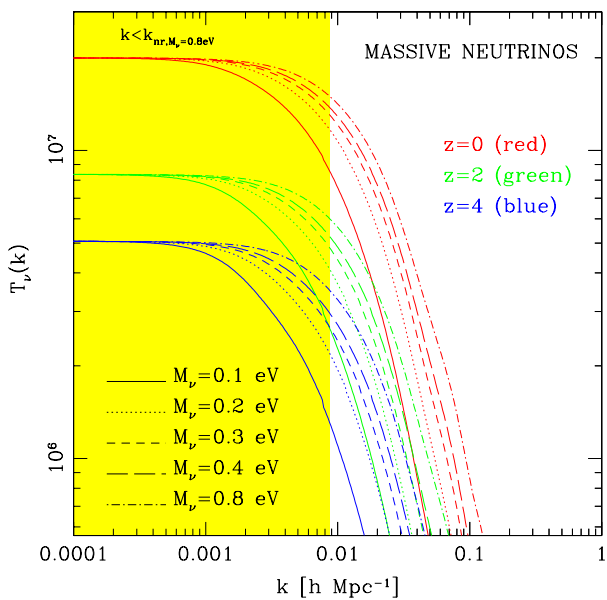}
\caption{Neutrino linear transfer functions $T_{\rm \nu}(k)$ 
for the same mass and redshift ranges considered in the previous figure; the normalization is arbitrary. 
The yellow area is same as in Figure \ref{fig_neutrino_linear_theory_A}.}
\label{fig_neutrino_linear_theory_C}
\end{figure}

Finally, Figure \ref{fig_neutrino_linear_theory_C} shows the neutrino linear transfer functions $T_{\rm \nu}(k)$
for the same mass and redshift ranges as considered before,  with arbitrary normalization.

It is of considerable  interest to
investigate how these effects propagate in the nonlinear regime, not only at the level of the three-dimensional
matter power spectrum, but also
for the one-dimensional Ly$\alpha$ flux power spectrum: we address these questions in
Section \ref{sec_first_results}.


\subsection{Particle implementation of massive neutrinos}

Implementing massive neutrinos in cosmological $N$-body simulations
is a delicate subject. Firstly, neutrinos can be treated either as a fluid or as an ensemble of particles.
Secondly, one may describe their evolution with linear theory or
perform a full nonlinear treatment; clearly, the second option comes 
with a series of numerical challenges, \textit{in primis} the problem of shot noise
introduced by the high thermal velocities of the neutrino component.

Several attempts along these lines have already been made in the literature, even long ago
(e.g., White, Frenk \& Davis 1983; Klypin et al. 1993; Ma \& Bertschinger 1994).
More recently, Brandbyge et al. (2008)
described a simple method for including the effect of massive neutrinos in large-scale $N$-body simulations,
using a hybrid TreePM approach, but neglecting all the hydrodynamics; their findings already 
showed that the suppression of power due to the presence of massive neutrinos is 
increased by nonlinear effects.  
Subsequently, Brandbyge \& Hannestad (2009) modeled
neutrinos as a fluid with a grid method, and pointed out the 
 relative benefits and drawbacks of implementing the effects of neutrinos in the form of particles 
versus a grid-based implementation.
In their code, the gravitational force due to neutrinos is calculated using the linearly evolved density distribution
of the neutrinos in Fourier space. Obviously, this technique 
eliminates
the  Poisson noise at small scales introduced by an alternative particle representation, which results in 
higher accuracy in regions where the effect of the nonlinear neutrino evolution is mild.
With this approach a series of computational problems are avoided or drastically reduced, such as
memory and CPU time consumption  -- as one does not need to store
neutrino positions and velocities. 
In another study,  Brandbyge \& Hannestad (2010)
combined grid- and particle-based methods with a hybrid technique 
to achieve good accuracy at small and large scales while keeping the CPU consumption under control:
neutrinos are first discretized on a grid, and subsequently part of the grid is
converted into $N$-body particles,  when the thermal motion of neutrinos decreases to a few times the flow velocities in the simulation.
Instead,  Ali-Ha{\"i}moud \& Bird (2013) used a different technique:
the CDM component is obtained via $N$-body computations, while the smooth neutrino component is evaluated from that background 
by solving the Boltzmann equation linearized with respect to the neutrino overdensity. 

In the present work, we choose a more direct and computationally intensive approach --  following Viel et al. (2010): 
neutrinos are modeled as an additional type of particle in the $N$-body setup (on top of gas and DM),
and  a full hydrodynamical treatment is carried out, well-inside the nonlinear regime -- including the
effects of baryonic physics which affect the IGM.
In particular, we make no approximations for the evolution of the neutrino component, nor 
interchange between grid- and particle-based implementations to save CPU time or speed up the computations.
The adopted implementation technique is primarily driven by our main goal to
accurately reproduce all the main features of the Ly$\alpha$ forest, at the quality level of BOSS or  future 
deep Ly$\alpha$ surveys. 
As evident from Figure \ref{fig_neutrino_linear_theory_A} (i.e., yellow versus gray areas), the
one-dimensional Ly$\alpha$ forest data provided by BOSS lies in a $k$-range where
nonlinear evolution of cosmological neutrinos cannot be neglected: hence, any attempt
to speed-up calculations by using approximate linear solutions for the neutrino component would
compromise our ability to accurately reproduce all the features of the forest.
To this end, Viel et al.  (2010) previously compared particle and grid neutrino representations and found
that their difference in terms
of power spectra are mainly driven by the fact that the nonlinear evolution at small scales is not properly reproduced by the grid method;
they also argued that on scales relevant for the Ly$\alpha$ forest it provides higher accuracy 
 to account for the nonlinear evolution rather than limiting the description to the linear case, 
 despite the effect of the Poisson contribution on the neutrino power spectrum introduced by the
particle-based modeling. This fact alone would be sufficient to justify our 
choice of the particle-based implementation for neutrinos.
In addition, we are not limited by computational time or memory  
because we have access to state-of-the-art computational facilities to perform a complete hydrodynamical treatment -- as we
describe next. 



\section{OUR SIMULATIONS} \label{sec_simulations_description}

In this section we present our new suite of hydrodynamical simulations with massive
neutrinos and provide several technical details on the codes used for the runs, the performance,
and the various optimization strategies. We also briefly describe the workflow pipeline and the post-processing procedure
developed to extract the line of sight (LOS) and particle samples to accurately model the Ly$\alpha$ transmitted flux. 


\subsection{Suite of simulations with massive neutrinos}

\begin{table}[t]
\centering
\caption{Basic parameters of our simulations, common to all the runs -- if not specified otherwise.}
\doublerulesep2.0pt
\renewcommand\arraystretch{1.5}
\begin{tabular}{cc} 
\hline \hline  
Parameter &  Value \\
\hline
$\sigma_8 (z=0)$          &                   0.83  \\
$n_{\rm s}$  &                  0.96  \\
$H_0$ [km s$^{-1}$Mpc$^{-1}$]              &                        67.5\\
$\Omega_{\rm m}$ &                 0.31 \\
$\Omega_{\rm b}$  &               0.044   \\
$\Omega_{\rm \Lambda}$   &             0.69    \\
$T_0 (z=3)[K]$ &             15000    \\
$\gamma(z=3)$   &             1.3    \\
Starting redshift    &               30         \\
\hline
\label{tab_param_sims}
\end{tabular}
\end{table}

\begin{table*}[t]
\begin{center}
\centering
\caption{List of our simulation suite (group I) -- best-guess (BG) and neutrino (NU) runs$^\ast$.}
\doublerulesep 2.0pt
\renewcommand\arraystretch{1.5}
\begin{tabular}{cccccc} 
\hline \hline
 Simulation set &   $M_{\rm \nu}$ [eV]  &   $\sigma_8(z=0)$  & Boxes [Mpc/h] & $N_{\rm p}^{1/3}$ & Mean particle separation  [Mpc/h]  \\ 
\hline
BG a/b/c & 0          & 0.830             &     100/25/25  & 768/768/192 & 0.1302/0.0325/0.1302 \\
NUBG a/b/c & 0.01  &      0.830             &    100/25/25  & 768/768/192 & 0.1302/0.0325/0.1302 \\
NU01 a/b/c & 0.1         &       0.830        &    100/25/25  & 768/768/192 &0.1302/0.0325/0.1302 \\
NU01-norm a/b/c & 0.1    &     0.810                 &    100/25/25  & 768/768/192 &0.1302/0.0325/0.1302\\
NU02 a/b/c & 0.2     &          0.830            &    100/25/25  & 768/768/192 &0.1302/0.0325/0.1302\\
NU03 a/b/c & 0.3       &      0.830         &    100/25/25  & 768/768/192 &0.1302/0.0325/0.1302\\
NU04 a/b/c & 0.4            &      0.830      &    100/25/25  & 768/768/192 &0.1302/0.0325/0.1302\\
NU04-norm a/b/c & 0.4        & 0.733             &    100/25/25  & 768/768/192 &0.1302/0.0325/0.1302\\
NU08 a/b/c & 0.8     &     0.830         &    100/25/25  & 768/768/192 &0.1302/0.0325/0.1302 \\
NU08-norm a/b/c & 0.8       & 0.644               &    100/25/25  & 768/768/192 &0.1302/0.0325/0.1302 \\
\hline
\label{tab_suite_sims_A}
\end{tabular}
\end{center}
\end{table*}

\begin{table*}
\begin{center}
\centering
\caption{List of our simulation suite (group II) -- neutrino cross-terms{$^\ast$}.}
\doublerulesep 2.0pt
\renewcommand\arraystretch{1.5}
\begin{tabular}{ccccccccccc} 
\hline \hline
 Simulation set &   $M_{\rm \nu}$ [eV]  &   $\sigma_8(z=0)$  & Boxes [Mpc/h] & $N_{\rm p}^{1/3}$ & $\gamma$ & $H_0$ & $n_{\rm s}$ & $\Omega_{\rm m}$ & $T_0$ \\ 
\hline
$\gamma +$NU08 a/b/c & 0.8   &      0.83                  &     100/25/25 & 768/768/192 & 1.6&67.5 &0.96 &0.31 &15000 \\
$H_0+$NU08 a/b/c & 0.8        & 0.83                  &    100/25/25 & 768/768/192 & 1.3&72.5 &0.96 &0.31 &15000 \\
$n_{\rm s}+$NU08 a/b/c & 0.8         & 0.83                  &    100/25/25 & 768/768/192 & 1.3&67.5 &1.01 &0.31 &15000 \\
$\Omega_{\rm m}+$NU08 a/b/c & 0.8           & 0.83                  &    100/25/25 & 768/768/192 &1.3&67.5 &0.96 &0.36 &15000 \\
$\sigma_8+$NU08 a/b/c & 0.8           &          0.88      &    100/25/25 & 768/768/192 & 1.3&67.5 &0.96 &0.31 &15000 \\
$T_0+$NU08 a/b/c & 0.8         & 0.83                  &    100/25/25 & 768/768/192 & 1.3&67.5 &0.96 &0.31 &21000 \\
\hline
\label{tab_suite_sims_B}
\end{tabular}
\end{center}
{\tiny {$^\ast$} {\rm a/b/c} indicate the different box size and number of particles in the simulation.}
\end{table*}

We performed a total of 48 hydrodynamical simulations, both with varying neutrino mass and fixed cosmological and astrophysical parameters (group I), or with
a fixed neutrino mass and slight variations in the basic cosmological and astrophysical parameters   (group II)  around what we indicate as the `best-guess' run -- 
this is the reference simulation set without massive neutrinos (but a massless neutrino component) and a cosmology compatible with the latest Planck (2013) results.
The basic parameters common to all the realizations are reported in Table \ref{tab_param_sims}.

For a given neutrino mass, we always performed a set of three simulations with different box sizes and number of particles (their combinations determine
the lowest and highest $k$-modes that can be resolved), which are appropriate for the quality of BOSS; 
specifically,  we adopted a box size of $100~h^{-1}{\rm Mpc}$ for large-scale power  with a number of particles per component $N_{\rm p}=768^3$ 
(simulations `a' in Tables \ref{tab_suite_sims_A} and \ref{tab_suite_sims_B}), and
a box size of $25~h^{-1}{\rm Mpc}$ for small-scale power, in this case with $N_{\rm p}=768^3$ or $192^3$, respectively (simulations `b'
and `c' in Tables \ref{tab_suite_sims_A} and \ref{tab_suite_sims_B}).
Extensive convergence and resolution  tests in support of our settings
have been carried out in Borde et al. (2014) -- but see also Section \ref{sec_convergence_tests}. 
In particular, the reason behind our specific choice is the ability to match the sensitivity of the BOSS quasar catalog (P${\rm \hat{a}}$ris et al. 2012) from Data Release 9 (Ahn et al. 2012),
and is also related to the application of the splicing technique proposed by McDonald (2003), which allows
correcting the larger box size simulation for the lack of resolution
and the small box  for the lack of nonlinear coupling between the highest and lowest $k$-modes; in this way,
we are  able to achieve an equivalent resolution of $3 \times 3072^3 \simeq 87$ billion particles in a $(100~h^{-1} {\rm Mpc} )^3$ box size 
-- optimal also for eBOSS and DESI -- without the need of running a
single but computationally prohibitive numerical simulation. 

When we included massive neutrinos we always kept $\Omega_{\Lambda} + \Omega_{\rm m}$ fixed to give a flat geometry
(with $\Omega_{\rm m}=\Omega_{\rm b}+\Omega_{\rm \nu}+\Omega_{\rm CDM}$) and varied
the additional massive neutrino component $\Omega_{\rm \nu}$  to the detriment of $\Omega_{\rm CDM}$. 
Moreover, most of our runs were  tuned to have $\sigma_8=0.83$ at $z=0$ by construction, which is the observed Planck (2013)
value. 
However, to characterize the effect of massive neutrinos with respect to the case of massless neutrinos,
we also ran simulations with the initial spectral amplitude $A_{\rm s}$ fixed as in  the best-guess, and therefore
with values of $\sigma_8$ changing across redshifts; these additional simulations are termed normalized and are used  here to
quantify the impact of massive neutrinos on the matter power spectrum;
in models with massive neutrinos, the power is suppressed on scales 
smaller than the free-streaming scale when the normalization is fixed, as explained previously.

\begin{figure*}

\includegraphics[angle=0,width=0.307\textwidth]{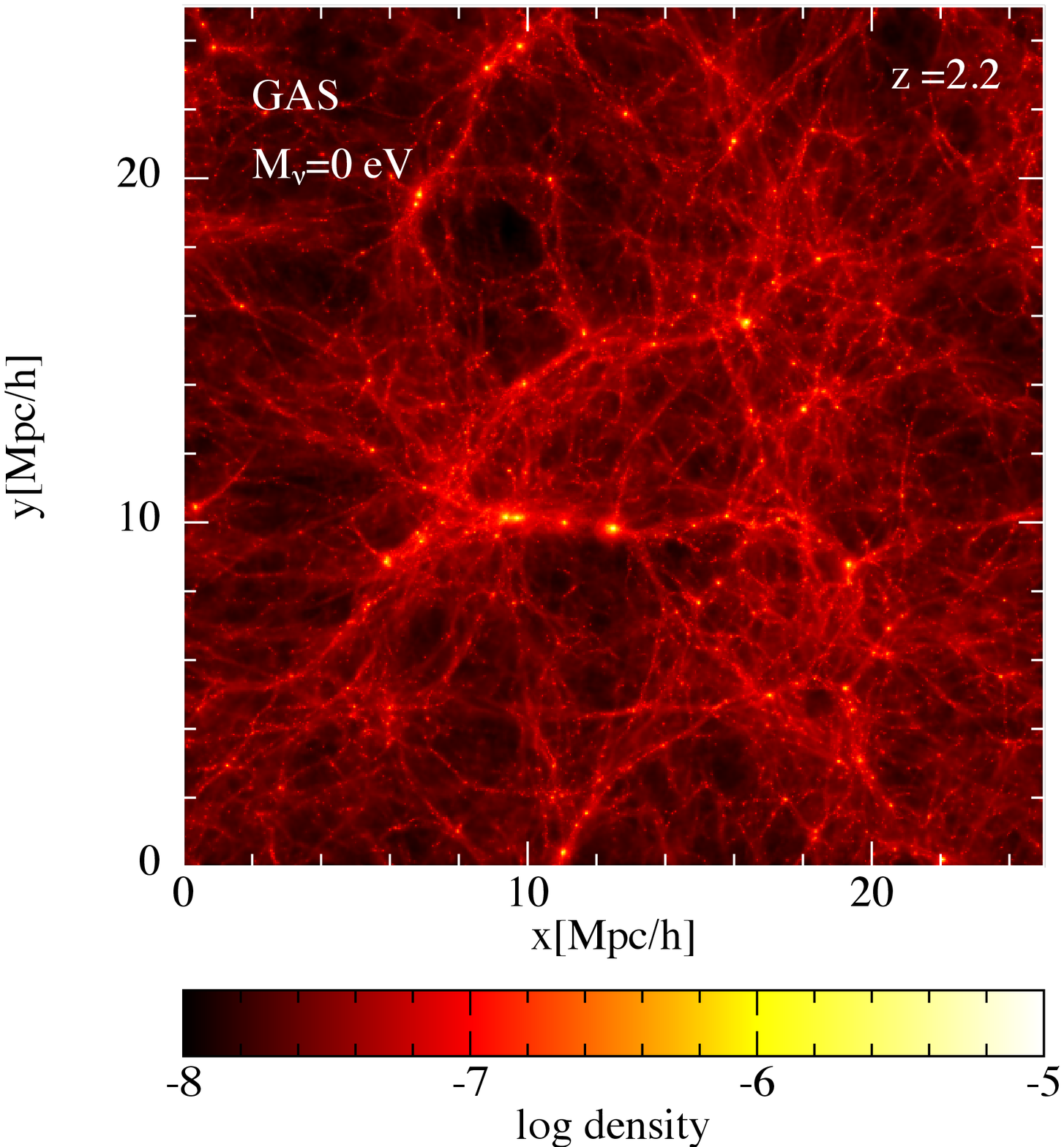}
\includegraphics[angle=0,width=0.307\textwidth]{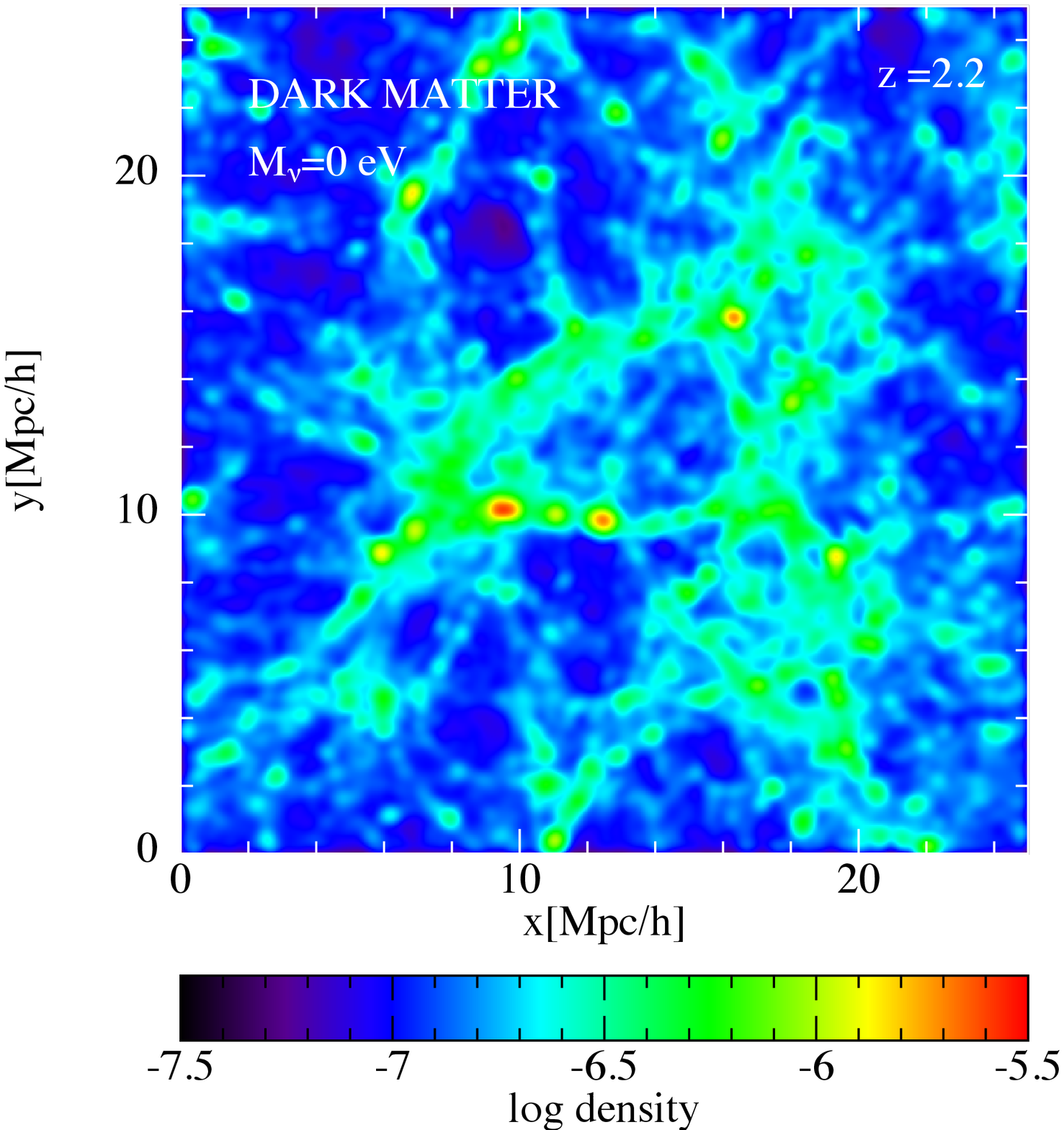} 

\includegraphics[angle=0,width=0.307\textwidth]{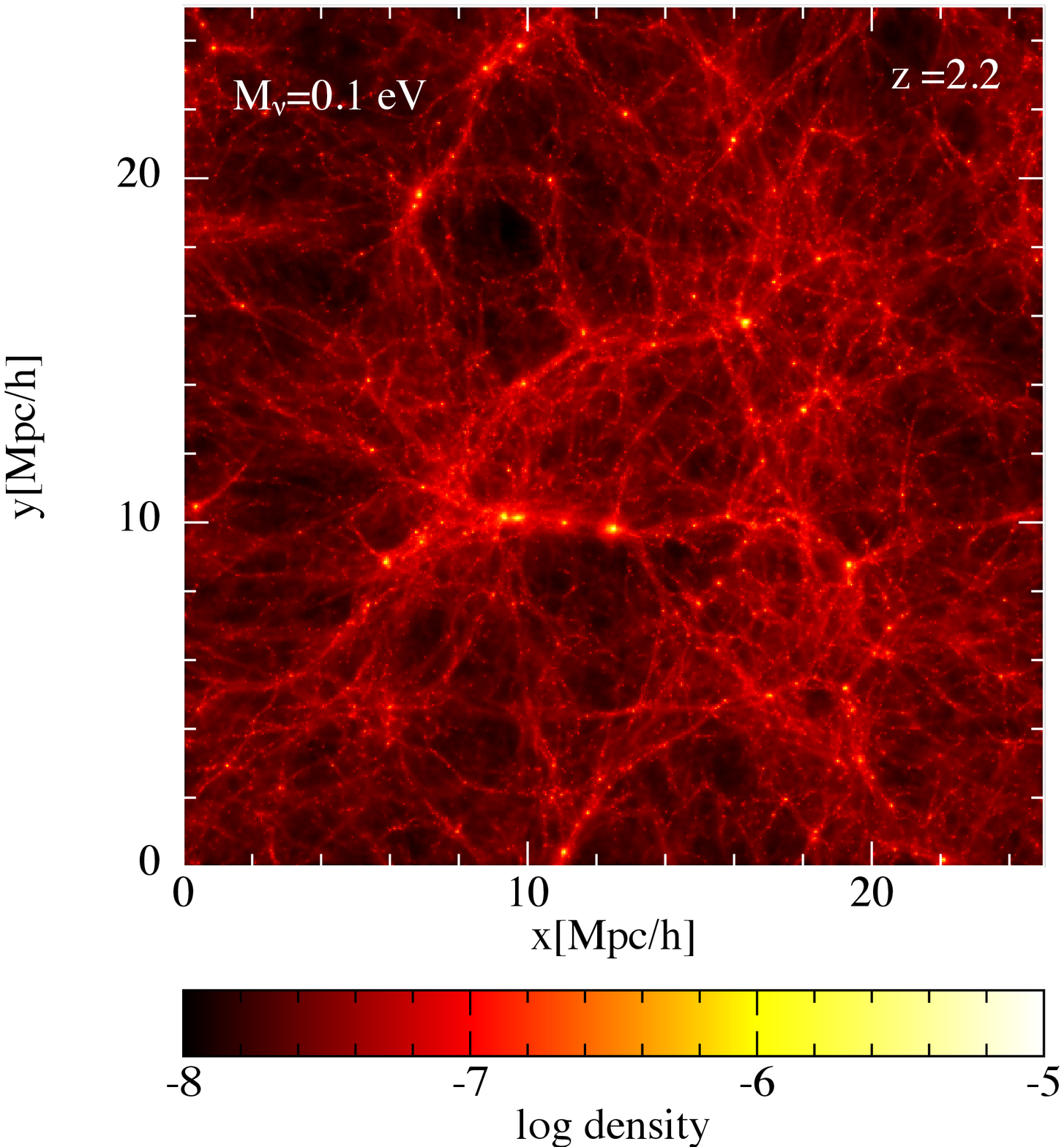} 
\includegraphics[angle=0,width=0.307\textwidth]{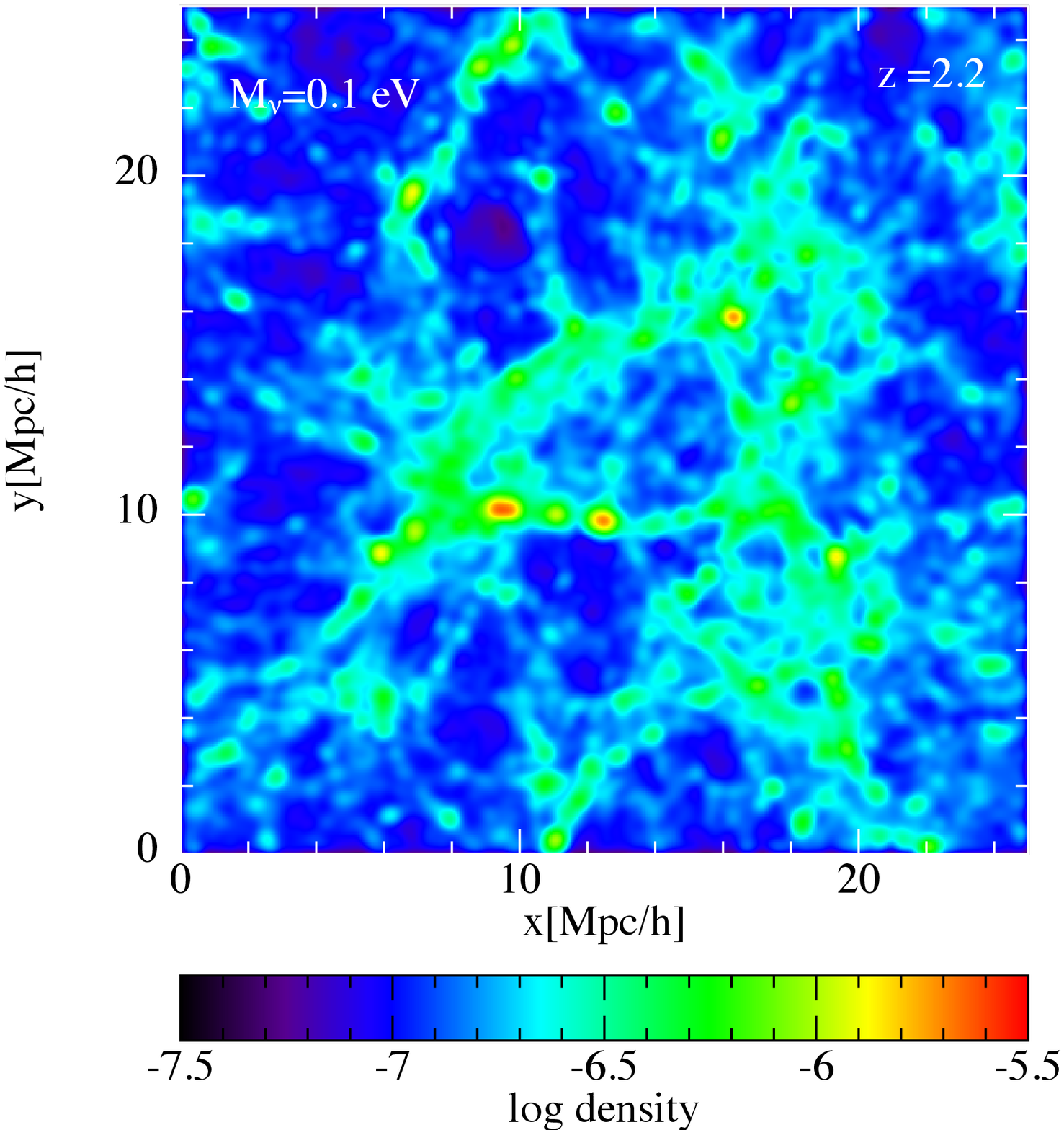} 
\includegraphics[angle=0,width=0.307\textwidth]{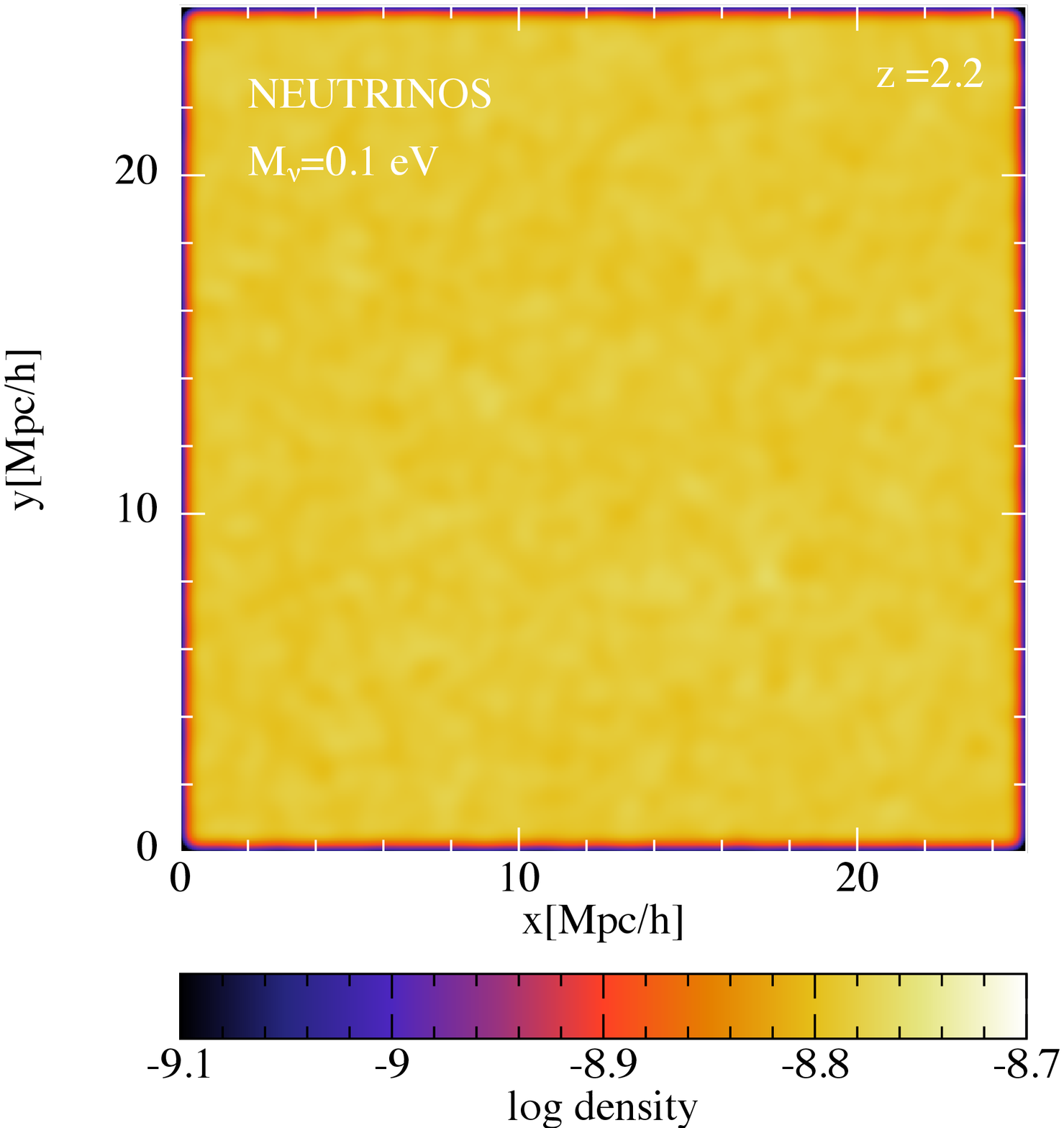}

\includegraphics[angle=0,width=0.307\textwidth]{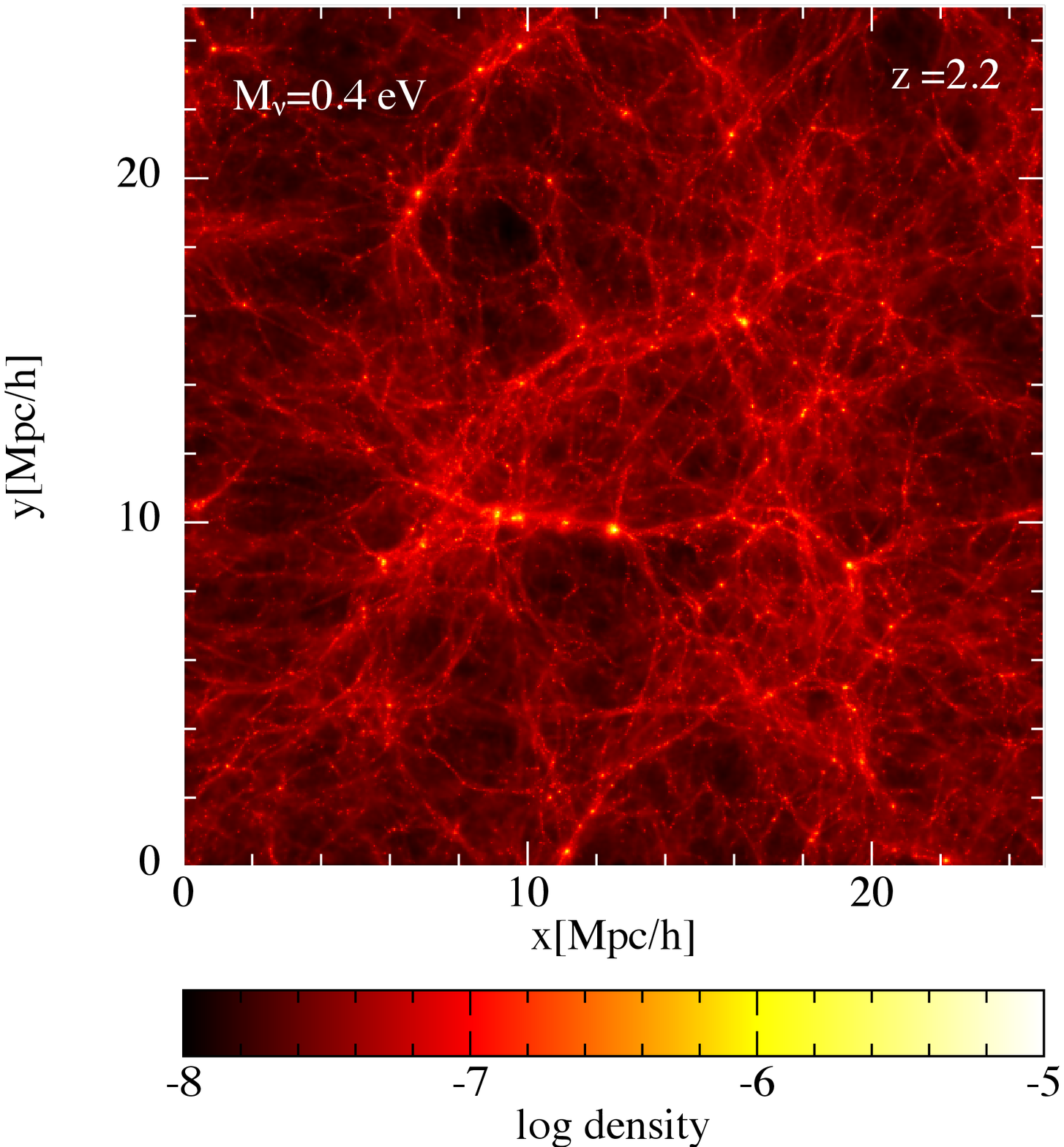} 
\includegraphics[angle=0,width=0.307\textwidth]{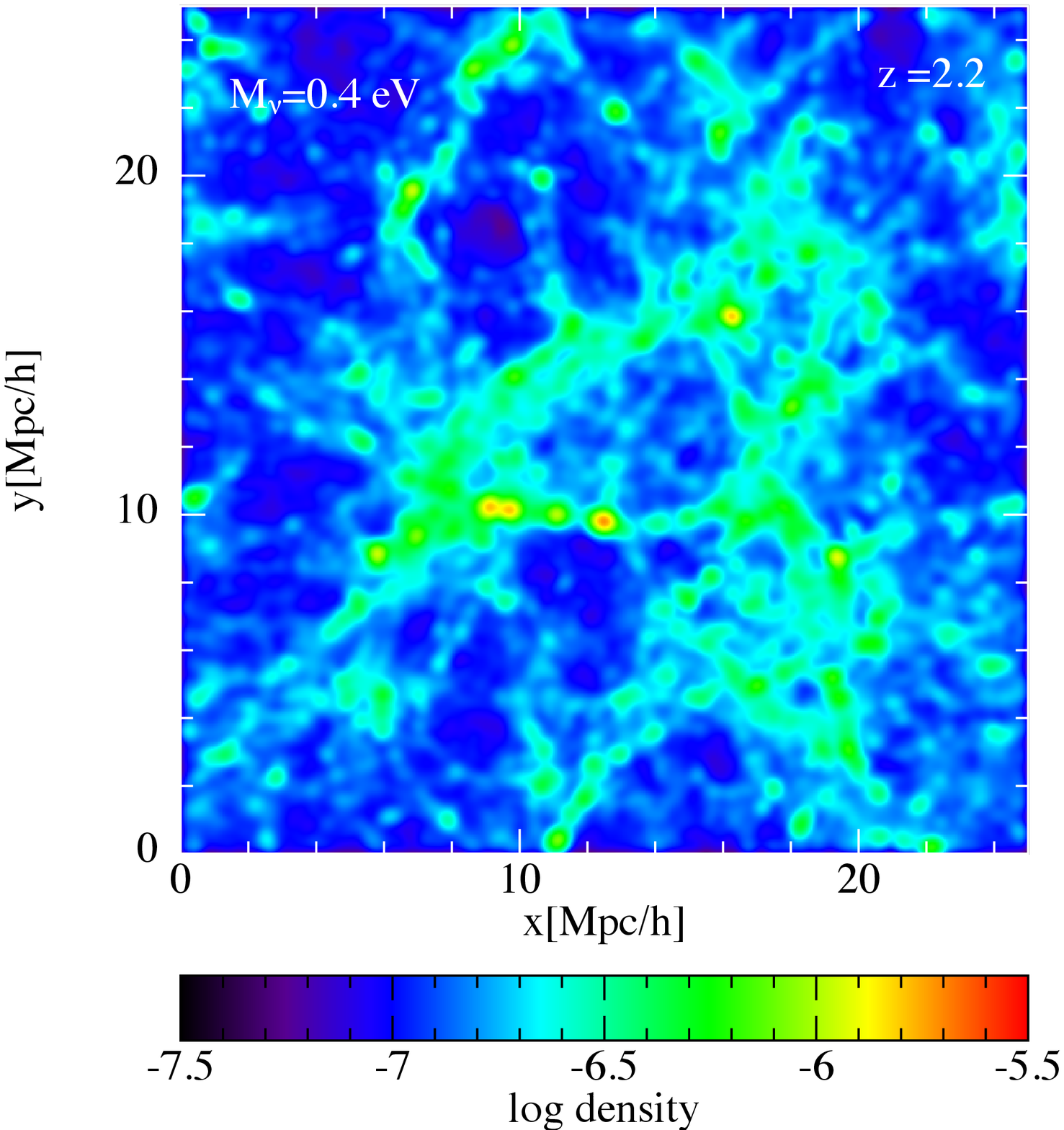} 
\includegraphics[angle=0,width=0.307\textwidth]{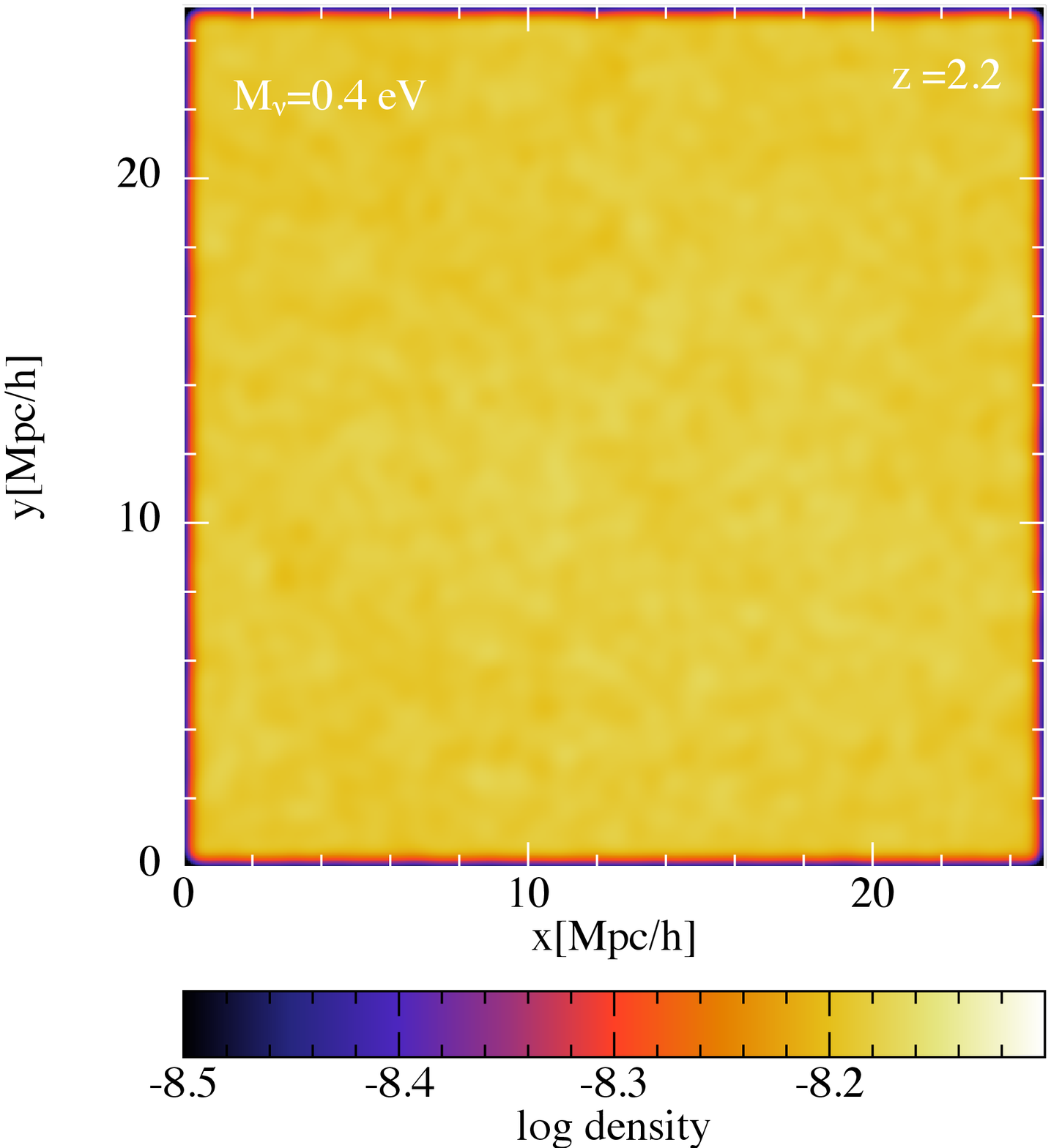}

\includegraphics[angle=0,width=0.307\textwidth]{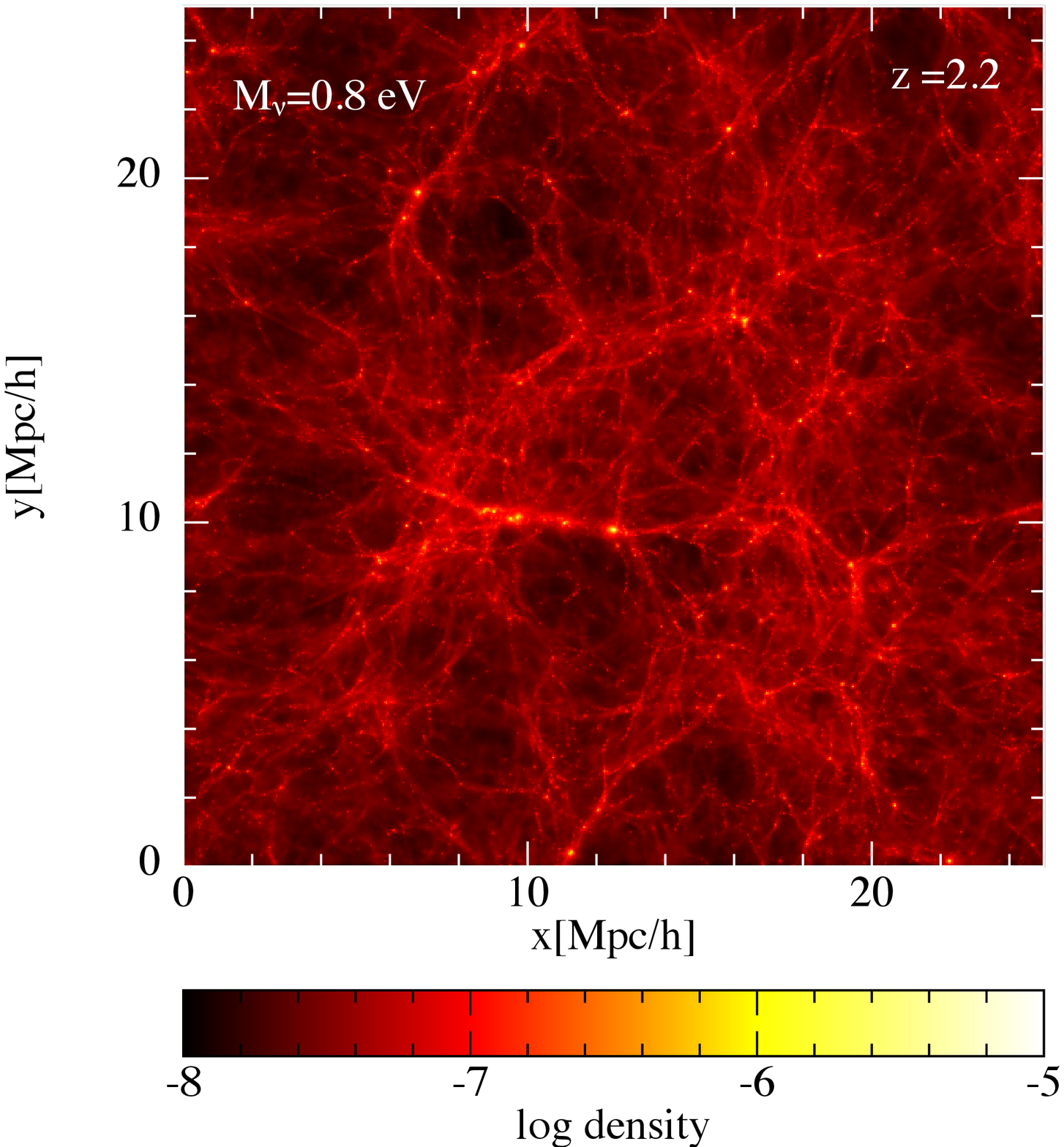}
\includegraphics[angle=0,width=0.307\textwidth]{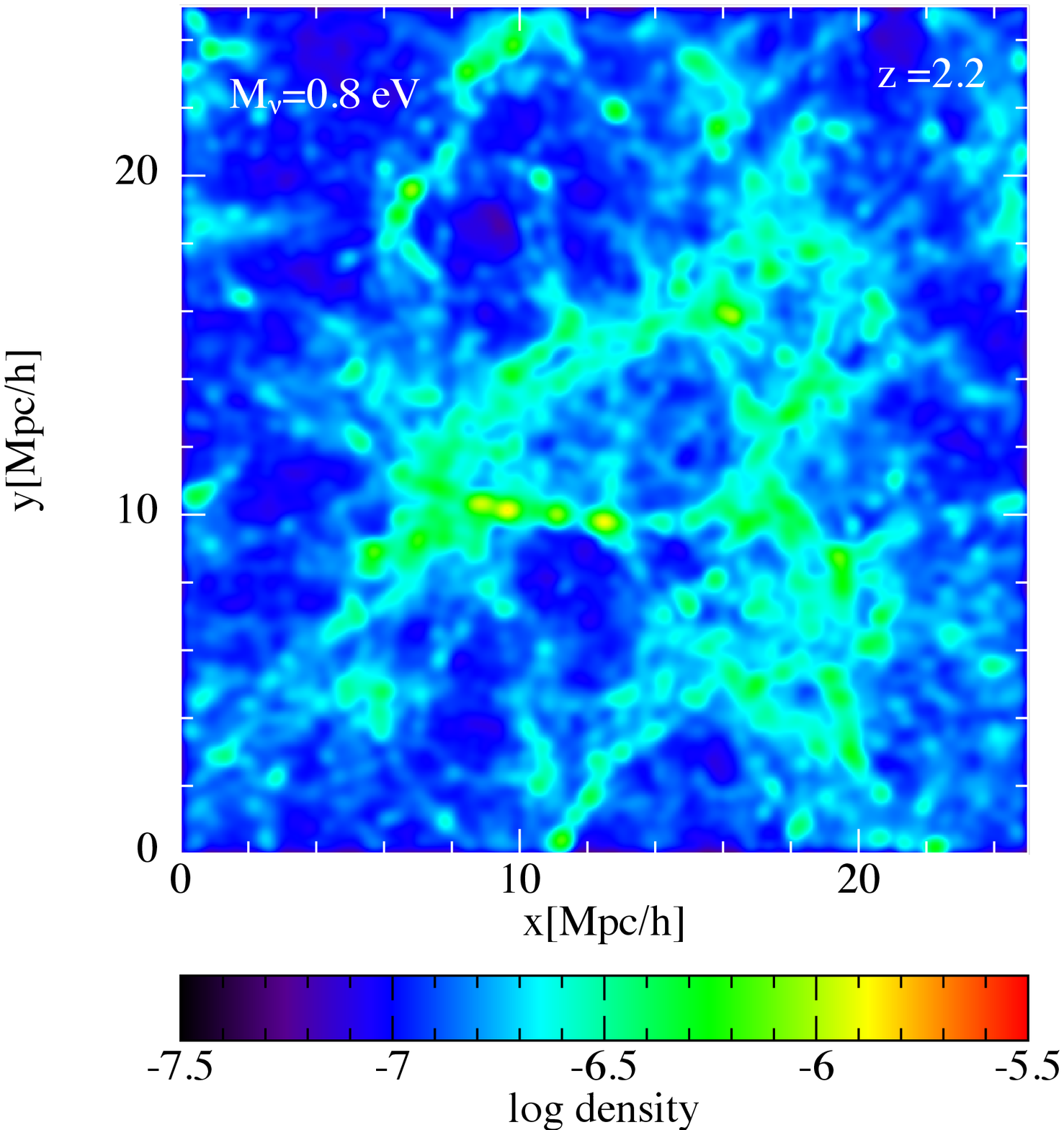} 
\includegraphics[angle=0,width=0.307\textwidth]{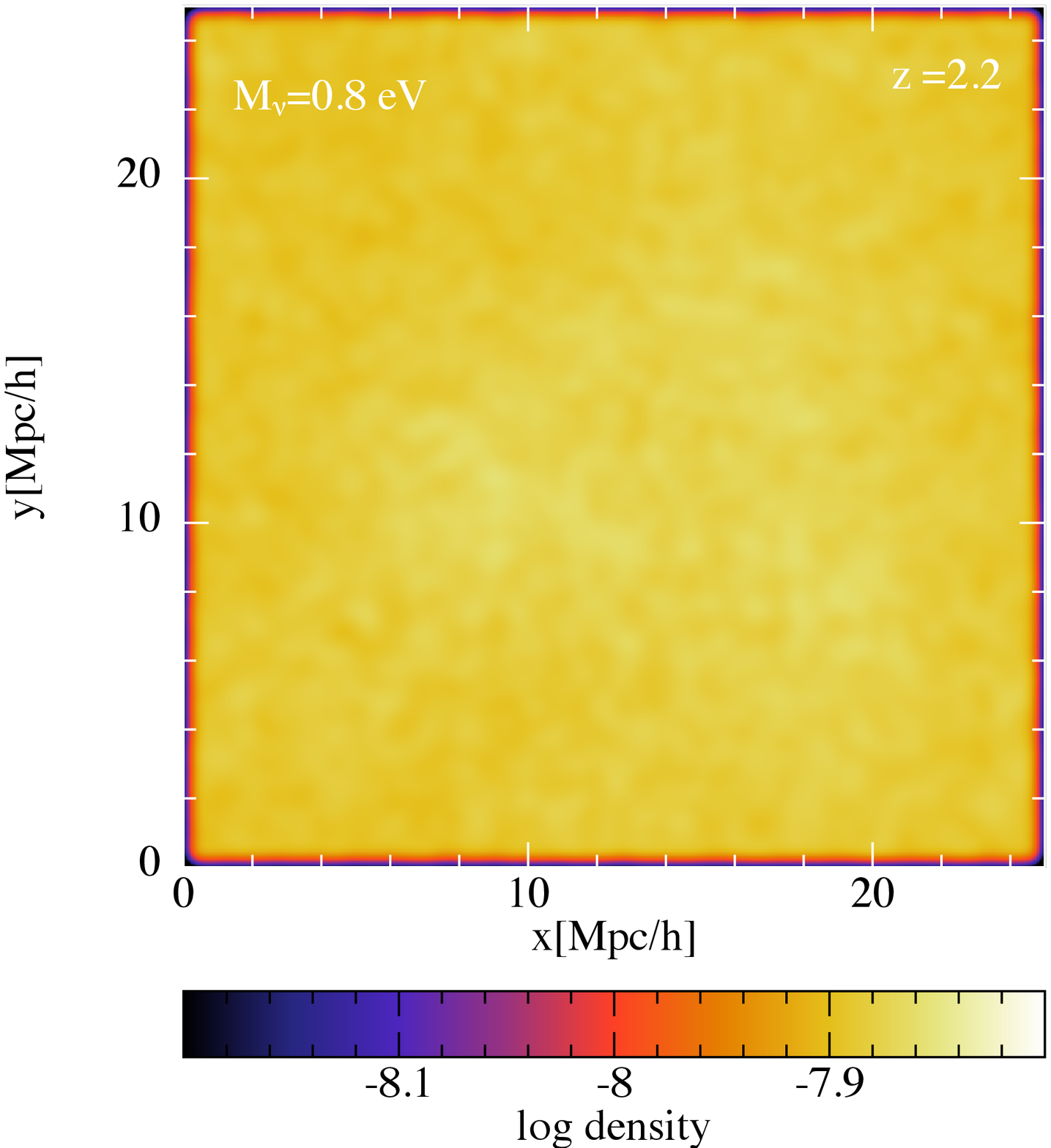}

\caption{Visual examples of snapshots at $z=2.2$ from simulations with a box size of
 $25~h^{-1}{\rm Mpc}$ 
and a resolution of $N_{\rm p}=192^3$ particles per type. 
The upper top panels are full projections of the density field 
in the $x$ and $y$ directions across $z$ from our best-guess reference simulation without massive neutrinos (but with a massless neutrino component),
while in descending order the other panels are for $M_{\rm \nu}=0.1,0.4$, and $0.8~{\rm eV}$. 
Gas (left panels), dark matter (central panels), and neutrino (right panels) components -- when present -- are shown.
The axis scales are
 in ${\rm Mpc}/h$.
 The various plots are 
smoothed with a cubic spline kernel, and both the DM and neutrino components are treated in the same way as the gas.
See the text for more details.}
    \label{fig_sims_visualization_A}
\end{figure*}

\begin{figure*}
\includegraphics[angle=0,width=0.307\textwidth]{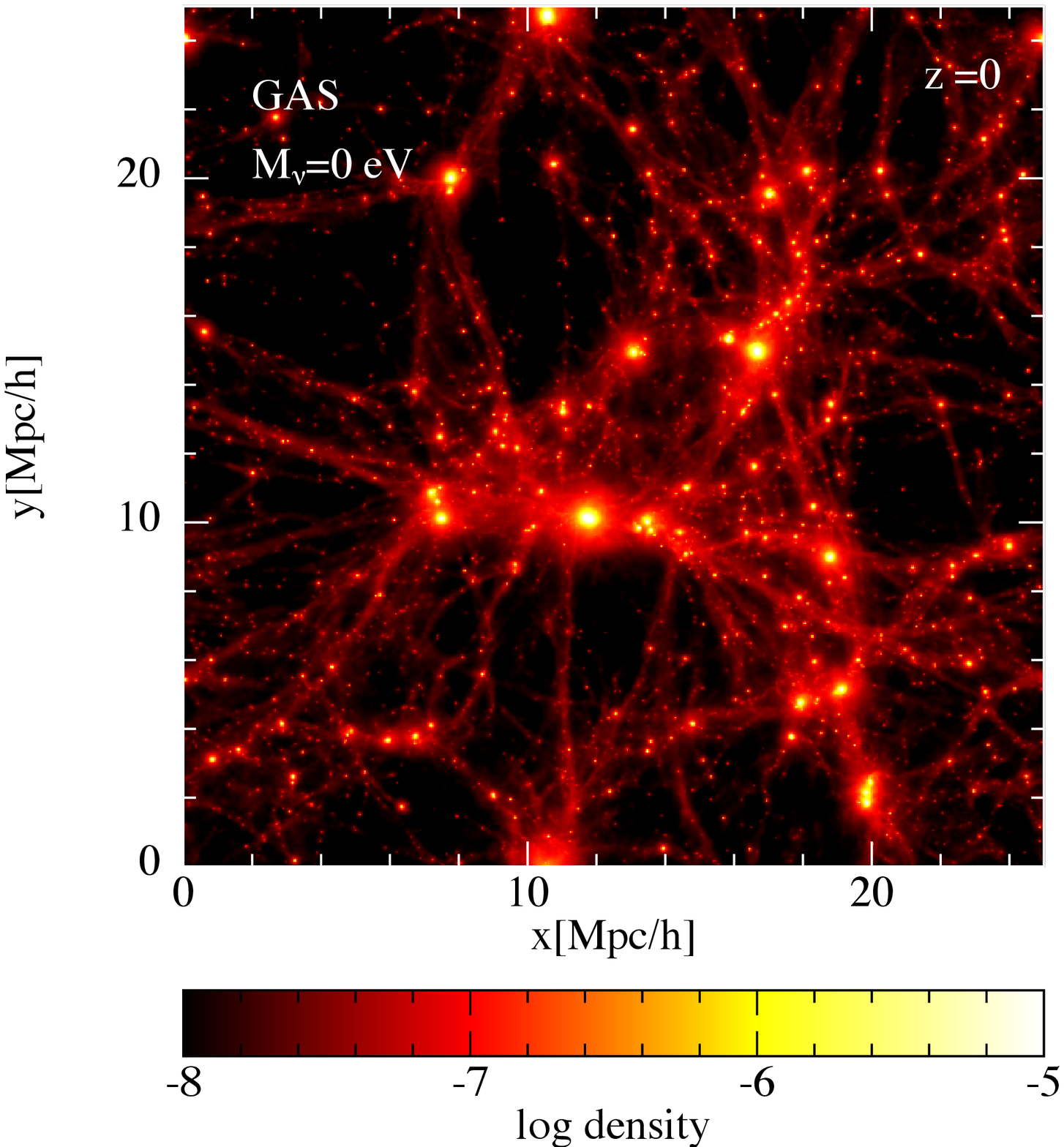}
\includegraphics[angle=0,width=0.307\textwidth]{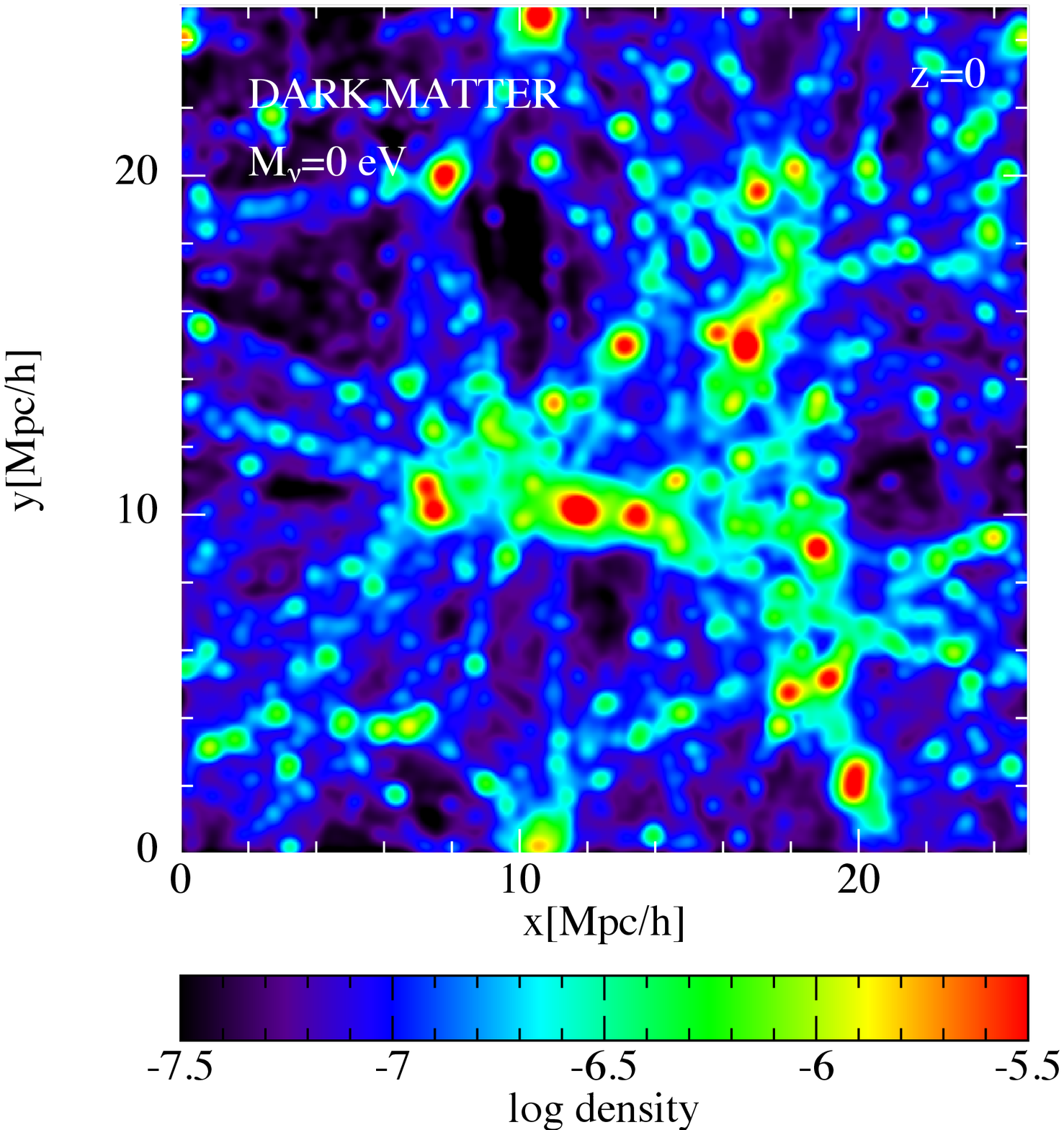} \\

\includegraphics[angle=0,width=0.307\textwidth]{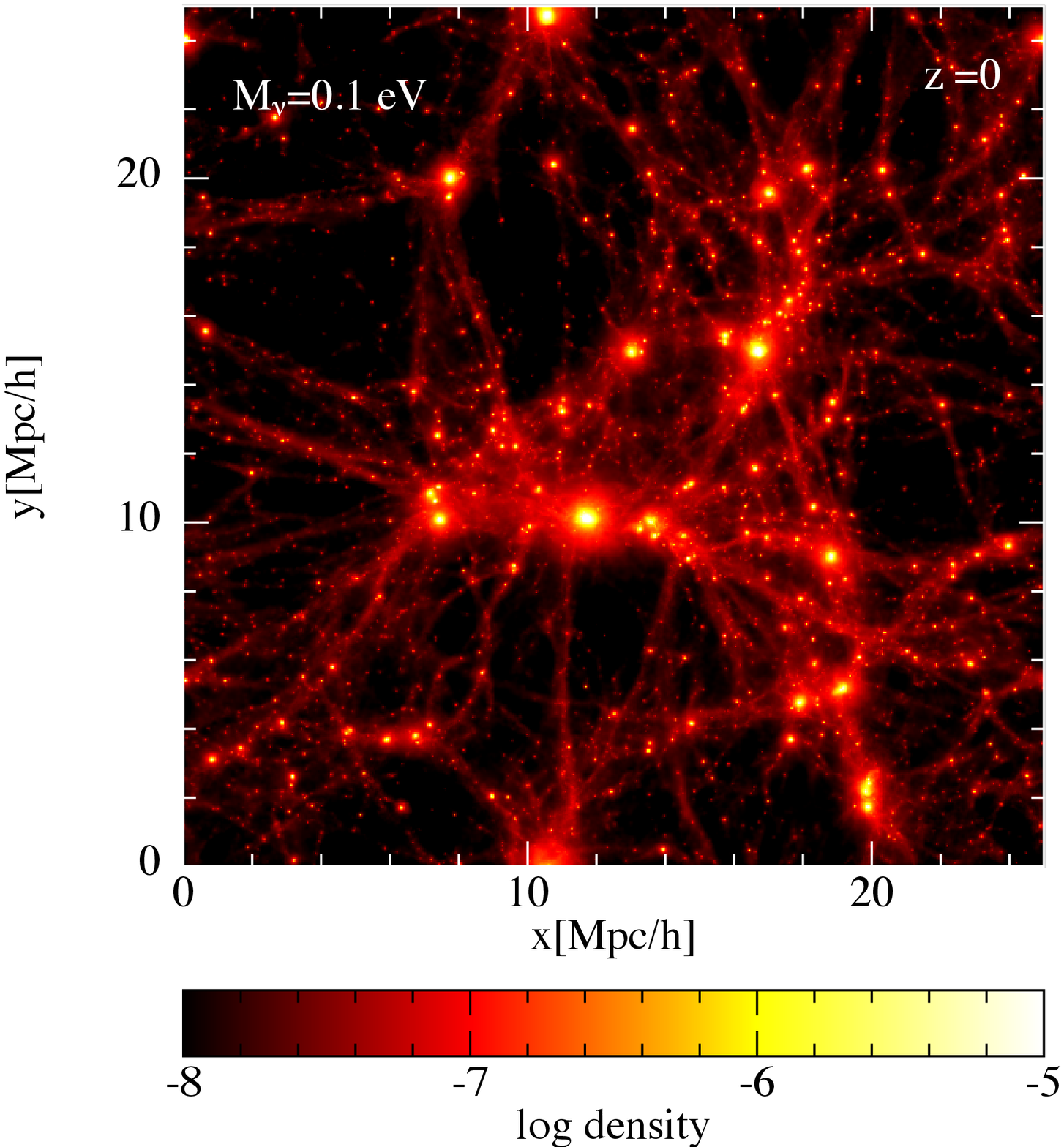}
\includegraphics[angle=0,width=0.307\textwidth]{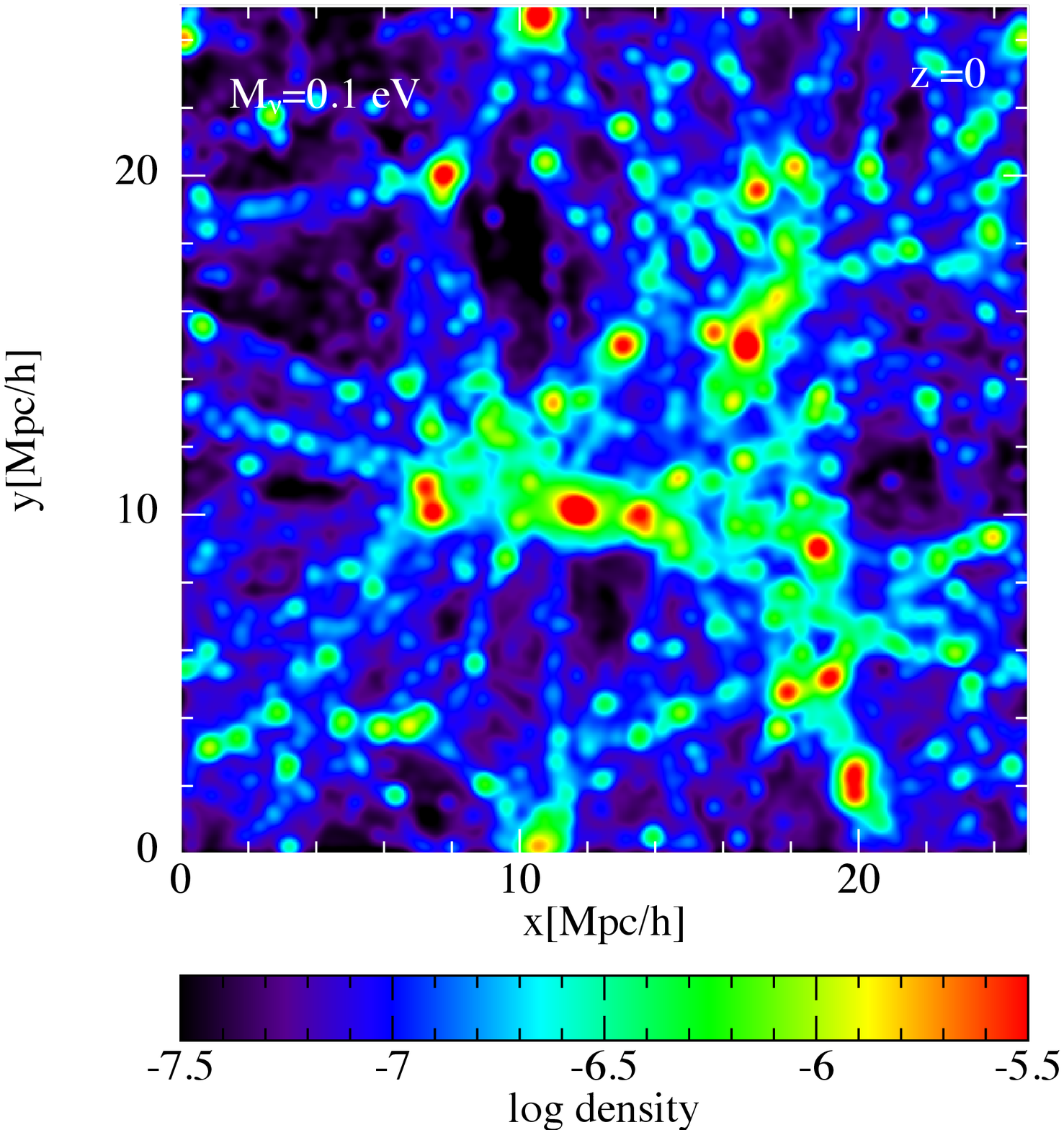} 
\includegraphics[angle=0,width=0.307\textwidth]{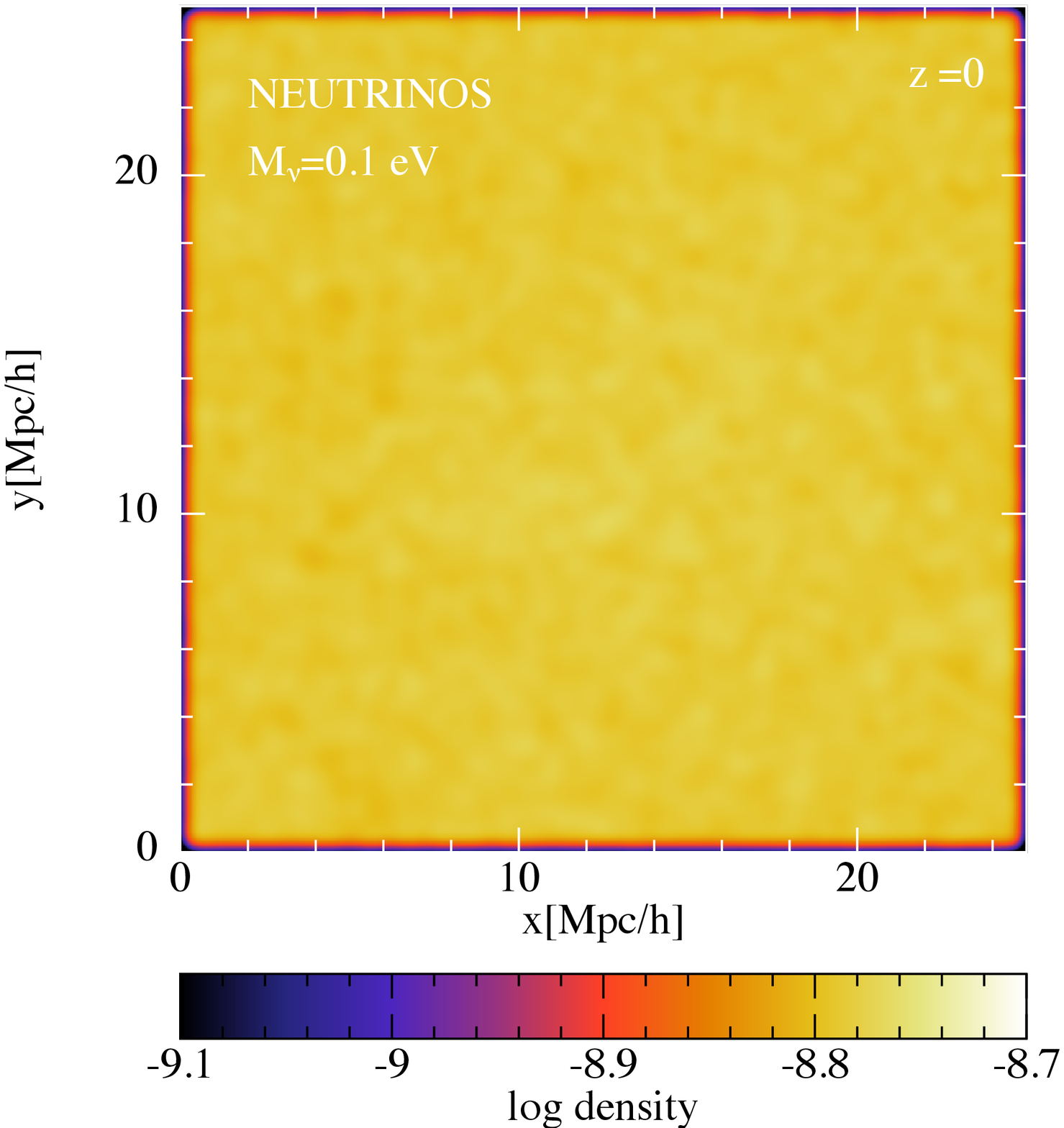}

\includegraphics[angle=0,width=0.307\textwidth]{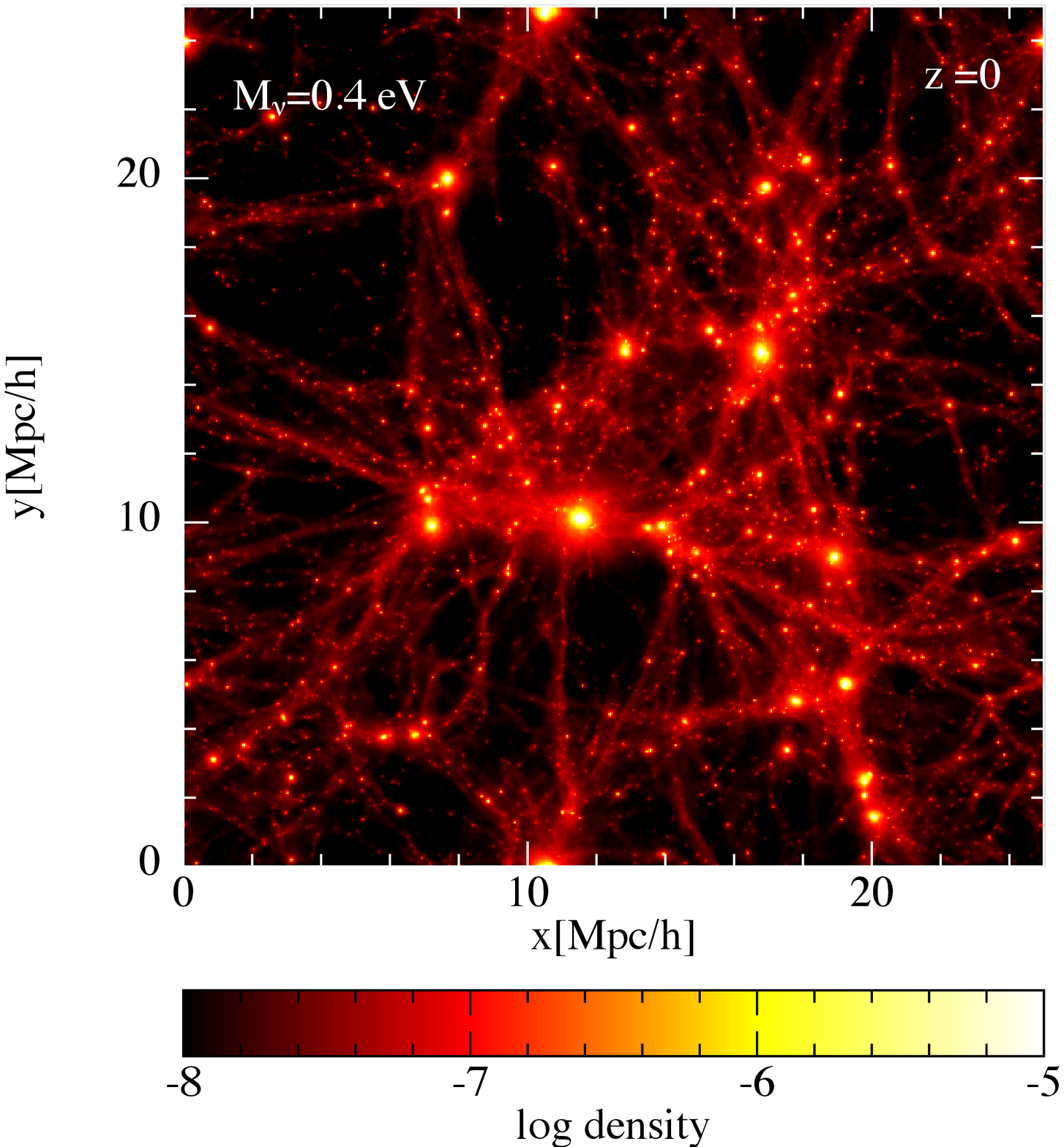}
\includegraphics[angle=0,width=0.307\textwidth]{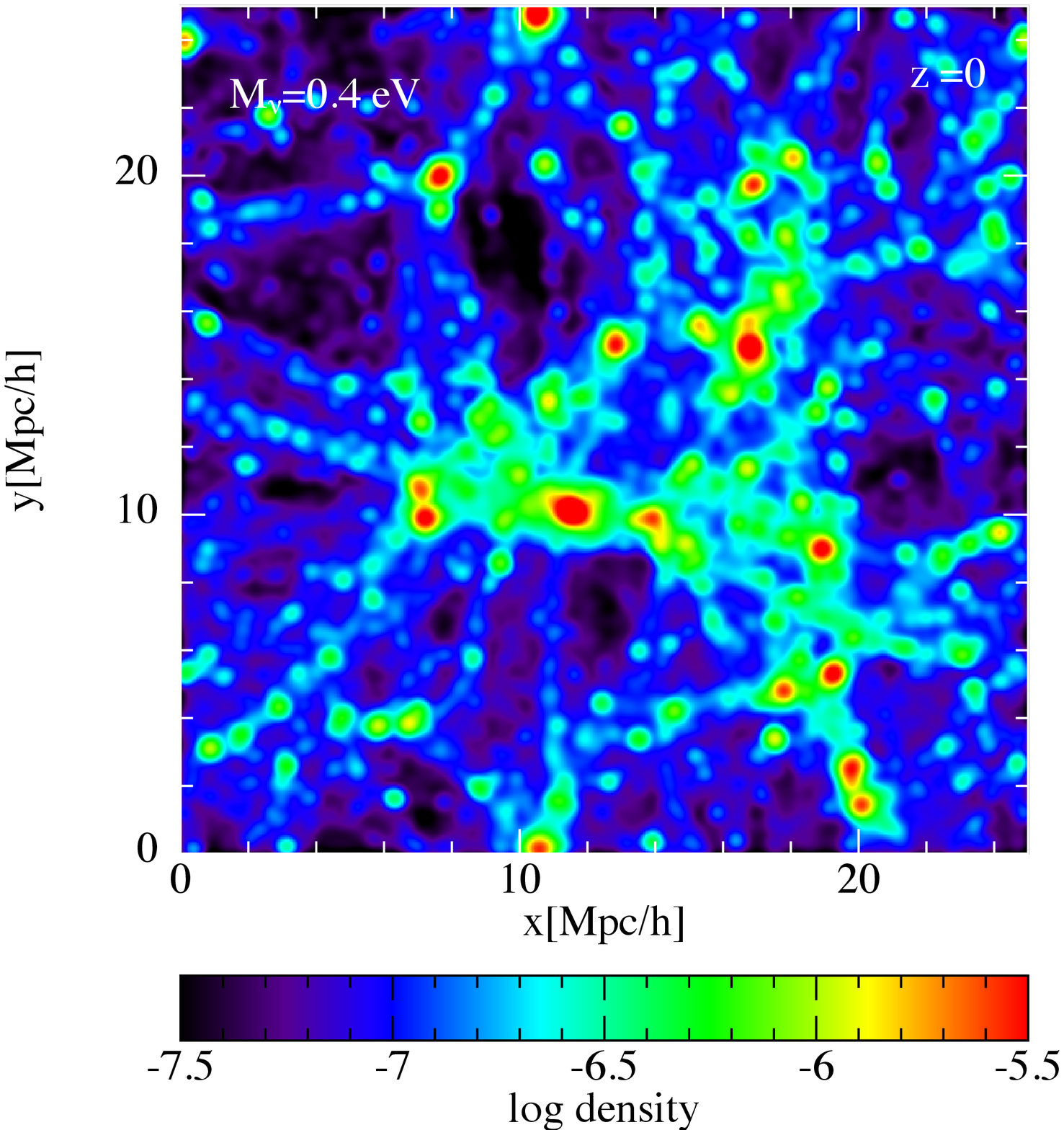} 
\includegraphics[angle=0,width=0.307\textwidth]{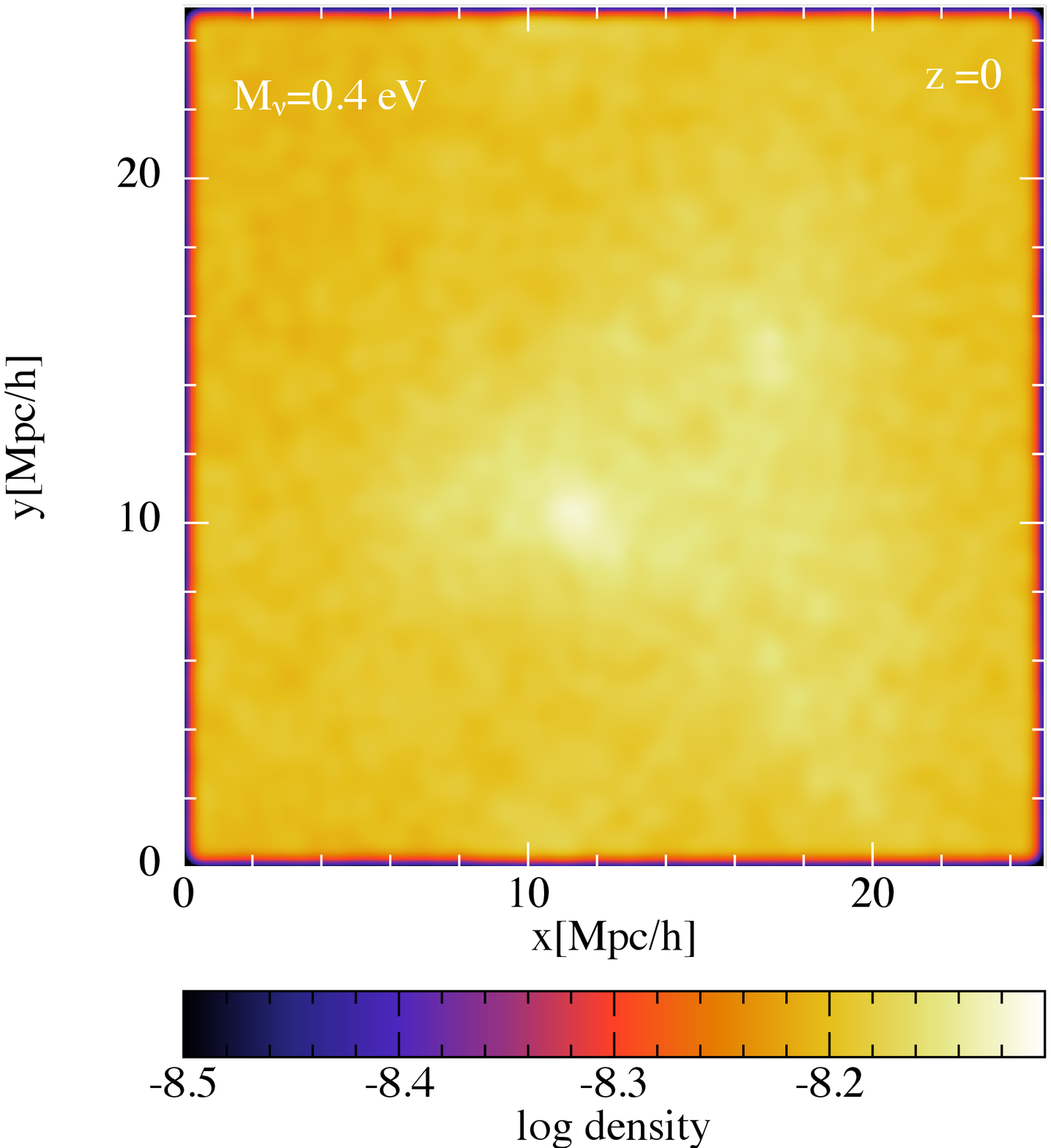}

\includegraphics[angle=0,width=0.307\textwidth]{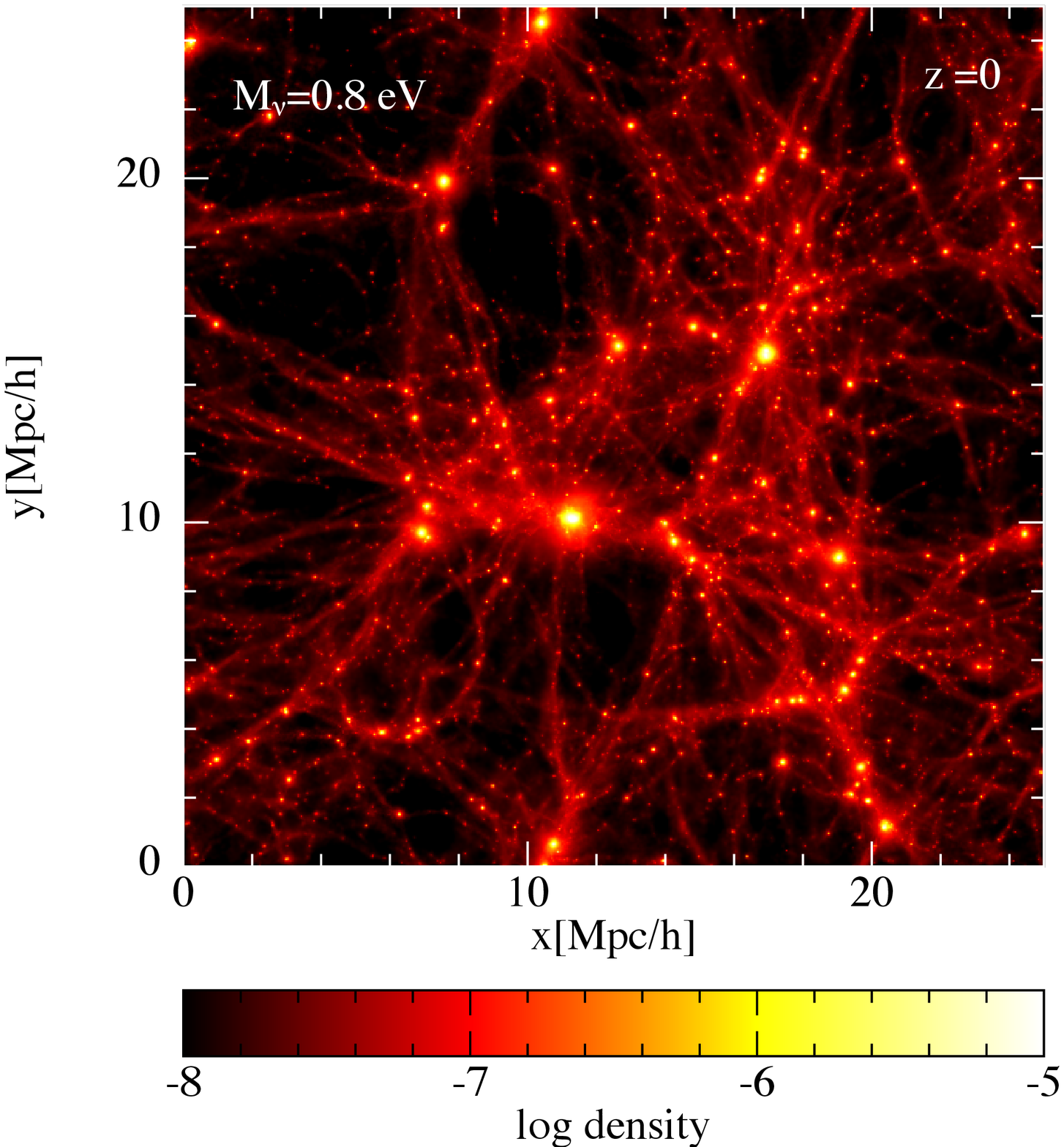}
\includegraphics[angle=0,width=0.307\textwidth]{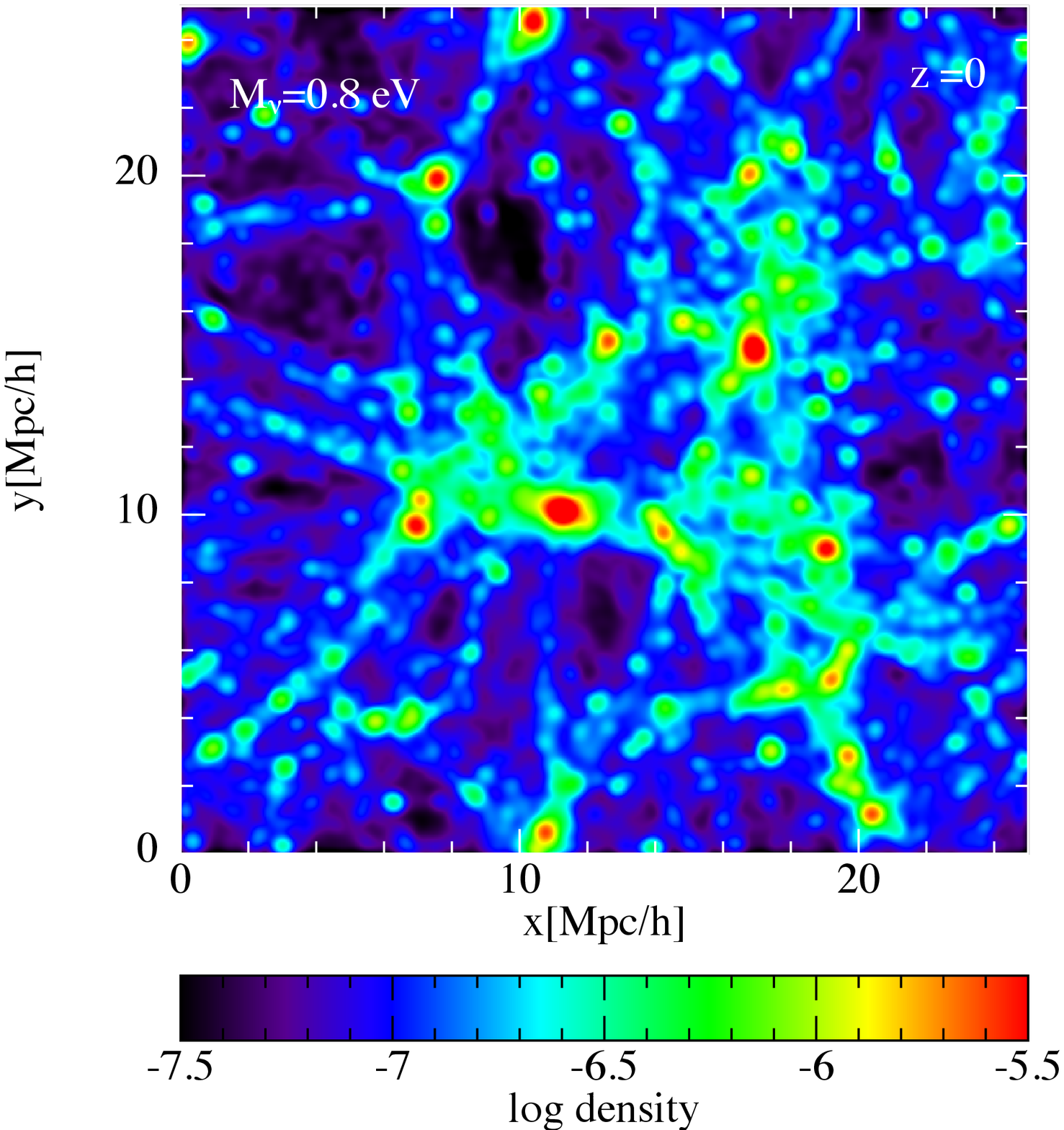} 
\includegraphics[angle=0,width=0.307\textwidth]{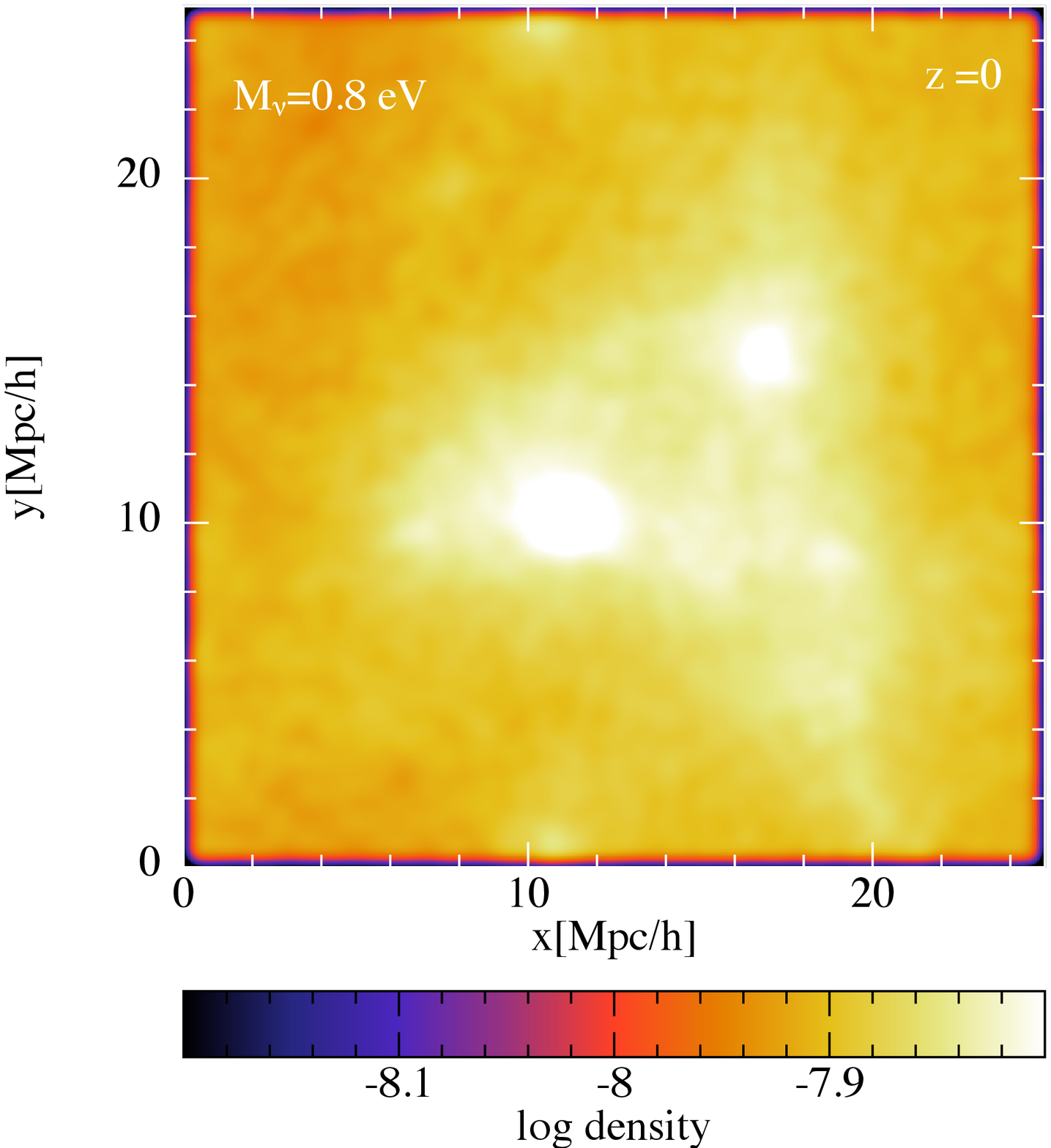}

\caption{Same as in the previous figure, but now at $z=0$.}
    \label{fig_sims_visualization_B}
\end{figure*}

Aside from the best-guess run, which only has a massless neutrino component, all our other
simulations contain three degenerate species of massive neutrinos implemented as a single particle-type, where
$M_{\rm \nu}=0.1,0.2,0.3,0.4$, and $0.8$ eV.
To ensure that the various realizations correctly  converge when $M_{\rm \nu}=0~{\rm eV}$, we also
ran a simulation set with a very low neutrino mass, i.e. $M_{\rm \nu}=0.01~{\rm eV}$ -- indicated as NU best-guess (see the appendix for a sanity check test). 
In addition, we performed a series of realizations with the neutrino mass fixed to be $M_{\nu}=0.8~{\rm eV}$, and 
slightly varied the basic cosmological and astrophysical parameters around the best-guess reference.
Specifically, we considered variations of $\pm 0.05$ in the amplitude of the matter power spectrum $\sigma_8$, in the spectral index of the primordial density fluctuations $n_{\rm s}$,
and in the matter density content $\Omega_{\rm m}$, while we varied the Hubble constant $H_0$ by $\pm 5$;
regarding astrophysical parameters, we altered both $T_0$ and $\gamma$, the former by $\pm 7000$ and the latter by $\pm 0.3$.
The suite of simulations with best-guess cosmological and astrophysical parameters and varying neutrino mass (group I) -- 
including runs with different normalizations --
is summarized in Table \ref{tab_suite_sims_A};
the  realizations indicated as neutrino cross-terms  (group II), in which we
kept the neutrino mass fixed to be $M_{\rm \nu}=0.8~{\rm eV}$ but varied cosmology and astrophysics around the best-guess,
are listed in Table \ref{tab_suite_sims_B}.
To this end, we note that the reason for producing cross-terms is motivated by  the multidimensional parameter estimation procedure 
outlined in Viel et al. (2010); in a  forthcoming study, we will apply  
this technique to constrain cosmological parameters
and neutrino masses from the Ly$\alpha$ forest by combining results from these simulations and BOSS Ly$\alpha$ data.

All our runs started at $z = 30$, with initial conditions  having the same random seed and based on
the second-order
Lagrangian perturbation theory (2LPT; Crocce et al. 2006)
instead of on the Zel'dovich approximation.
Snapshots were produced at regular intervals in redshift  between $z=4.6 - 2.2$, with $\Delta z=0.2$; for a few runs, we also reached $z=0$. 
We provide 
visual examples of our snapshot outputs at $z=2.2$ and $z=0$  in Figures  \ref{fig_sims_visualization_A}
and  \ref{fig_sims_visualization_B} for the
gas (left panels), dark matter (central panels), and neutrino (right panels) components -- when present.
The upper top panels are projections of the density field 
along the $x$ and $y$ directions (and across $z$) from our best-guess reference simulation, which only contains 
massless neutrinos, with a box size of $25~h^{-1}{\rm Mpc}$
and a relatively low resolution $N_{\rm p}=192^3$ particles per type;
in descending order, the other panels are for $M_{\rm \nu}=0.1,0.4$, and $0.8~{\rm eV}$. 
The axis scales are
 in $h^{-1}{\rm Mpc}$. 
 Note that for the neutrino component the density scale is kept fixed only for a given neutrino mass, 
 but changes for different $M_{\rm \nu}$ values.
The various plots were
smoothed with a cubic spline kernel, and both DM and neutrinos were treated like the gas.
It is nontrivial to visualize the neutrino
component, especially because of shot noise -- in essence,
for a very small neutrino mass, the overall effect is similar to that of random noise, 
whereas structures  start to appear at increasing $M_{\rm \nu}$ and decreasing redshifts.

In all our simulations, 
the gas was assumed to be of primordial composition with a helium mass fraction of $Y=0.24$.
Metals and evolution of elementary abundances were neglected.
As in Viel et al. (2010), we used a simplified criterion for star formation: all gas particles whose overdensity with respect to the mean 
is above 1000 and whose temperature is lower than $10^5$K  were turned immediately
into star particles.
This criterion, while having negligible effects on the Ly$\alpha$ flux statistics, speeds the calculations up considerably --
see Viel et al (2006, 2009), where effects  of adopting this simplified strategy 
 were estimated to be about $0.2 \%$ in the Ly$\alpha$ statistics,  compared with a more elaborate  multiphase model.

The various simulations were performed with periodic boundary conditions and an equal number of dark matter, gas, and neutrino particles.
We employed the entropy formulation of SPH proposed by Springel \& Hernquist (2002).
Gas in the simulations was photoionized and heated by a spatially uniform 
ionizing background.  This background was applied in the optically thin limit 
and was switched on at $z=9$.  The resulting reference thermal history in our 
simulations is consistent with the recent temperature measurements of Becker 
et al. (2011), assuming a slope for the temperature-density relation of 
$\gamma=1.3$.  We furthermore explored a variety of different thermal histories 
around this reference, parameterized by $T_0$ and $\gamma$, which allowed us  to span a 
plausible range for these two parameters within the observational 
uncertainties.  We achieved this by rescaling the amplitude and density 
dependence of the photoionization heating rates in the simulation (e.g. Becker et al. 2011).
Details on the software developed
for this study are provided next.


\subsection{Codes, optimization, and performance}

The basic code used for our simulations is Gadget-3 (Springel et al. 2001; Springel 2005), 
supplemented by CAMB (Lewis, Challinor \& Lasenby 2000), and a modified version of 2LPT (Crocce et al. 2006) to determine
the initial conditions. 

In particular,  Gadget-3 (GAlaxies with Dark matter and Gas intEracT) is a massively parallel tree-SPH code
 for collisionless and gasdynamical cosmological simulations. 
Gravitational interactions are computed with a hierarchical multipole expansion via the standard $N$-body method, and gas-dynamics is followed with 
SPH having fully adaptive smoothing lengths, so that energy and entropy are conserved; collisionless DM and gas are both represented by particles. 
The gravitational force computation uses a hierarchical multipole expansion, optionally in the form of a tree-PM algorithm: 
short-range forces are treated with the tree method, and long-range forces with Fourier techniques.
For our realizations, we set the number of mesh cells of the PM grid  equal to the number of particles. 

With respect to its original version, Gadget underwent a series of improvements and optimizations over several years 
to maximize the work-load balance and the 
efficiency in memory consumption and communication bandwidth. 
The high-level optimization of the code is obtained via a new parallelization strategy, based on a space 
decomposition achieved with a space-filling curve (i.e. the Peano-Hilbert decomposition). This fact guarantees a 
force independent of the processor number. 

Several other physical processes have also been implemented in Gadget-3, from radiative cooling/heating
physics to nonstandard DM dynamics, star formation, and feedback.
However, in our case
feedback options were disabled and galactic winds neglected, 
as suggested by  the results of Bolton et al. (2008), who found that winds have a negligible effect on the Ly$\alpha$ forest.

Along the lines of Viel et al. (2010),  Bird et al. (2012), and Villaescusa-Navarro et al. (2013a,b),
Gadget-3 has been modified to simulate the evolution of the neutrino density distribution.
In particular, neutrinos are treated as a separate collisionless fluid and are implemented as an additional particle-type on top of gas and DM
(see Section \ref{sec_implementing_neutrinos}).
To save computational time,  the small-scale neutrino clustering is neglected, and their short-range gravitational tree
force in the TreePM scheme is not computed.
Hence, the spatial resolution for the neutrino component is on order of the grid  resolution
used for the PM force calculation. 
We also note that the time-step used by the code is set by the DM alone, and is not affected by the neutrino component. 

Lines of sight and particle samples were obtained from Gadget-3 snapshots with an extraction procedure briefly described in the next section;  we
also developed additional software to handle the post-processing phase.
We ran all our parallel codes on the thin nodes of the  Curie supercomputer, owned by GENCI and operated in the TGCC by CEA --
the first French Tier0 system open to scientists through the French participation in the PRACE research infrastructure.

 \begin{figure*}
\centering
\includegraphics[angle=0,width=0.47\textwidth]{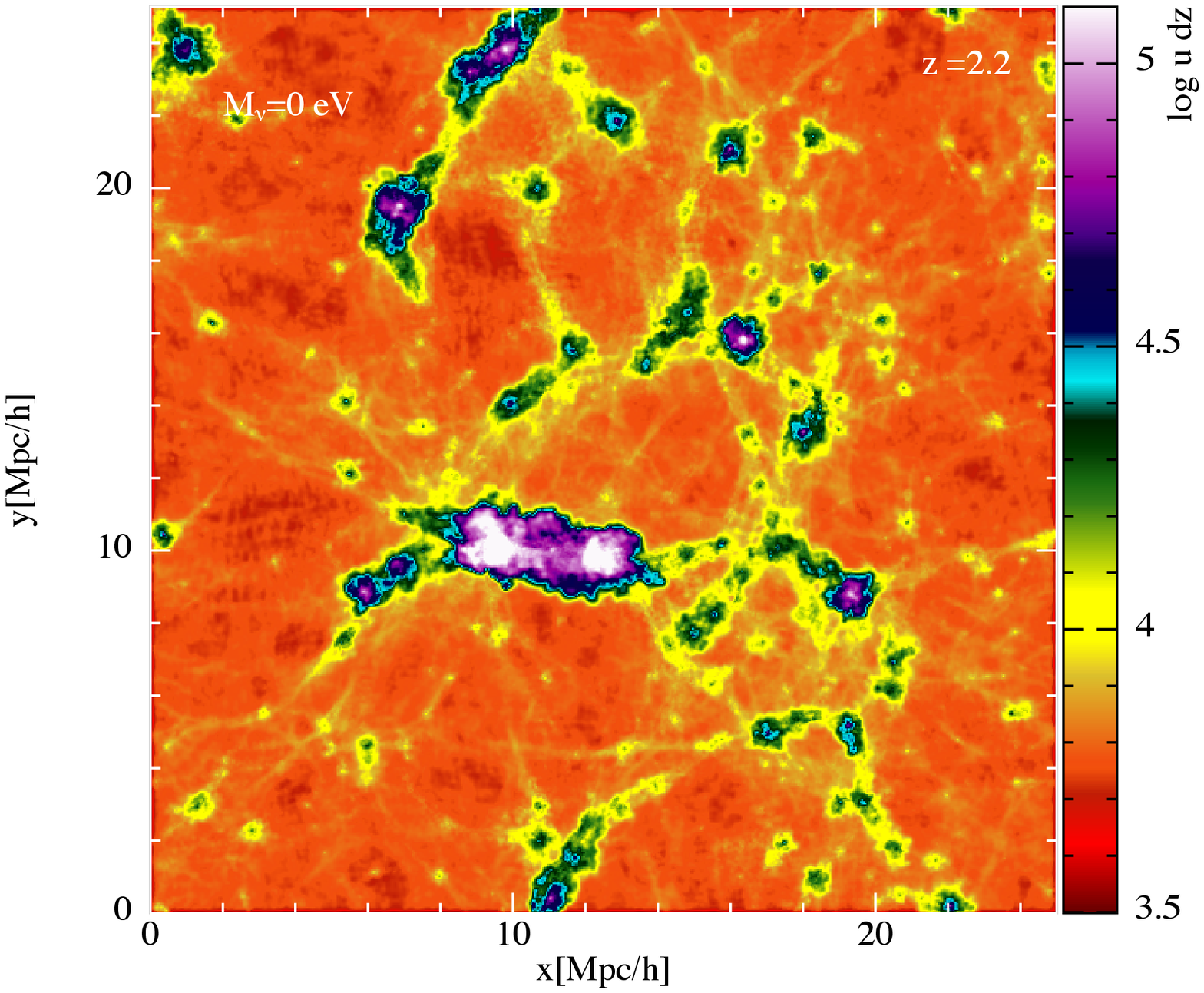}
\includegraphics[angle=0,width=0.47\textwidth]{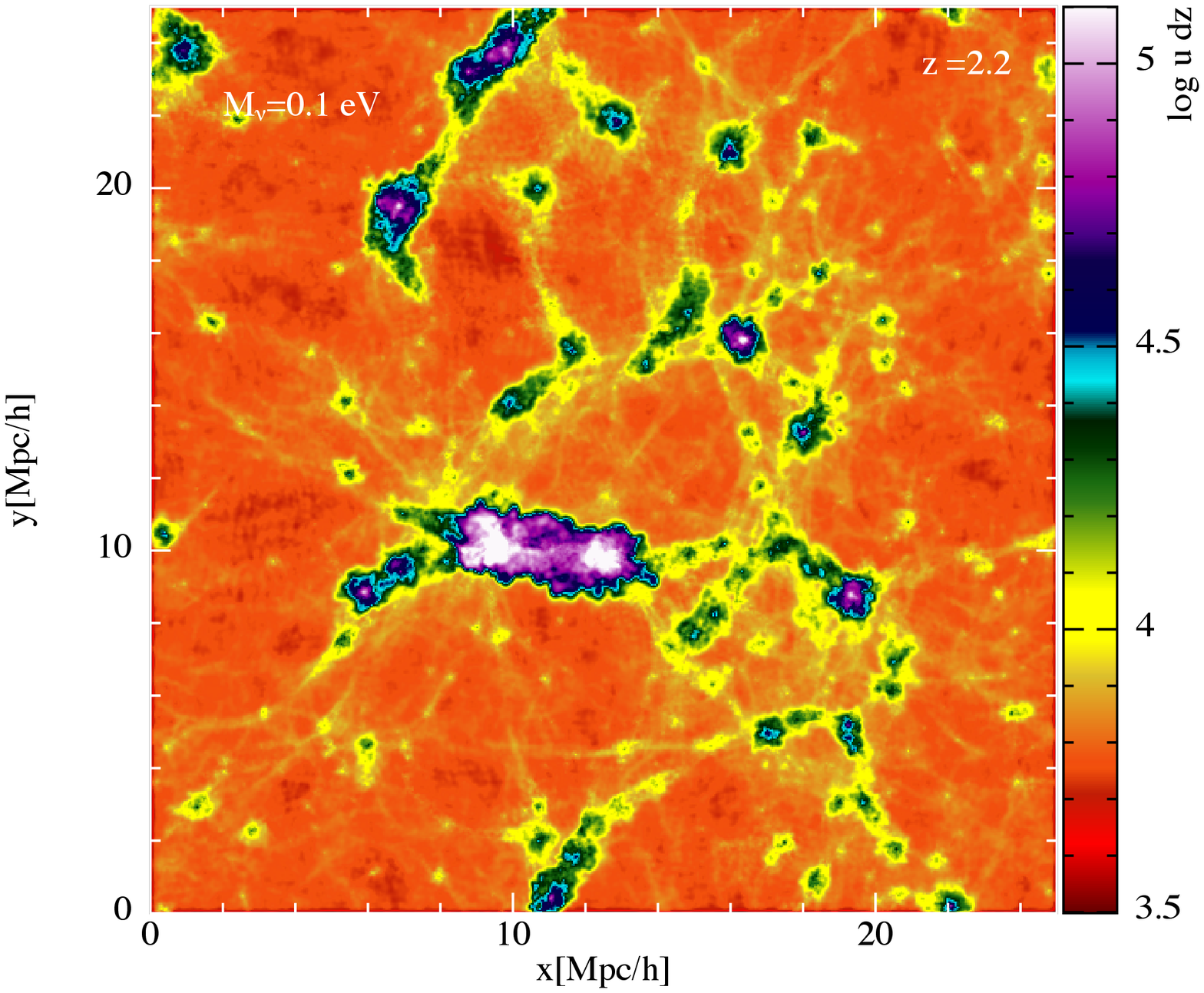}
\includegraphics[angle=0,width=0.47\textwidth]{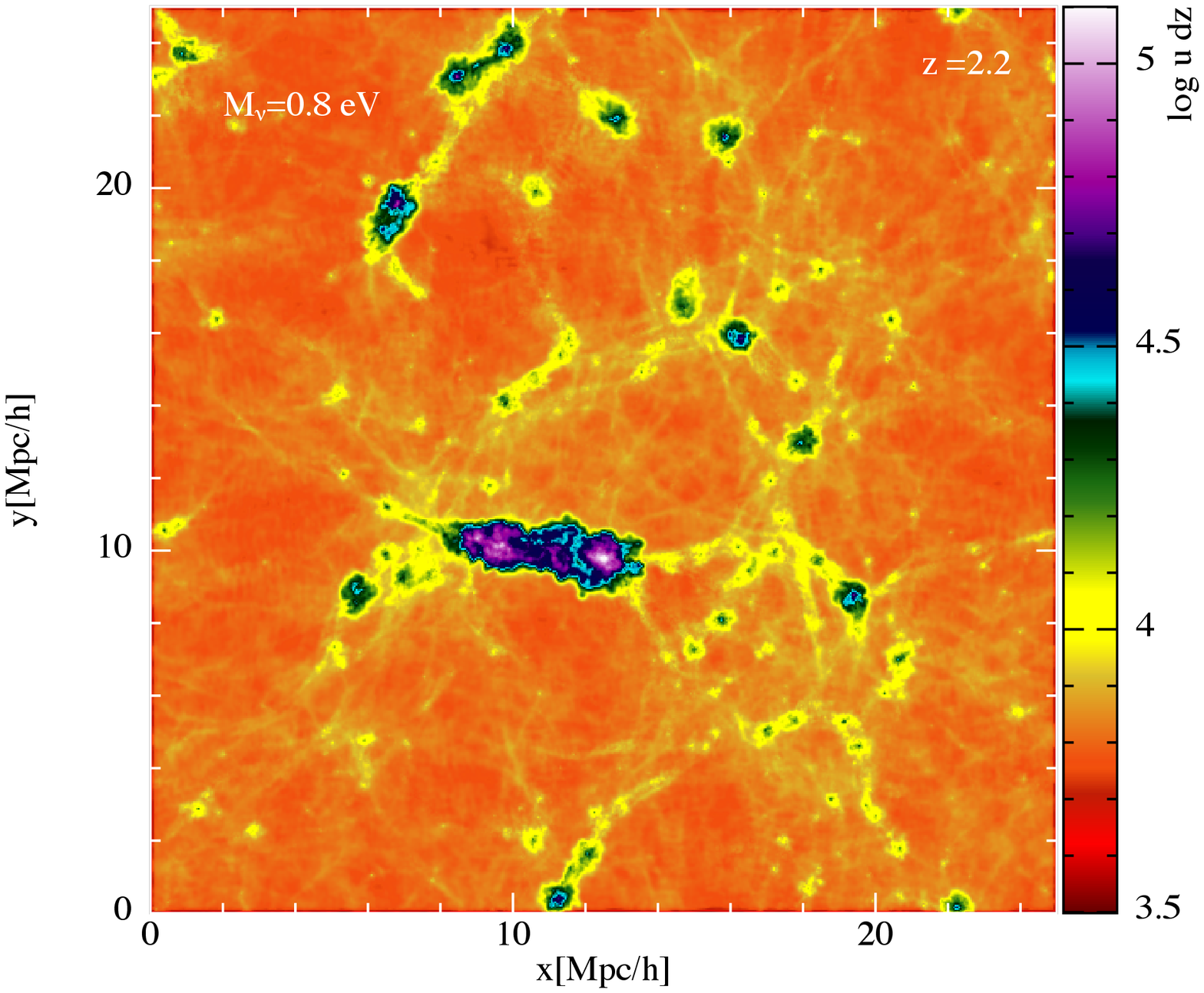}
\includegraphics[angle=0,width=0.47\textwidth]{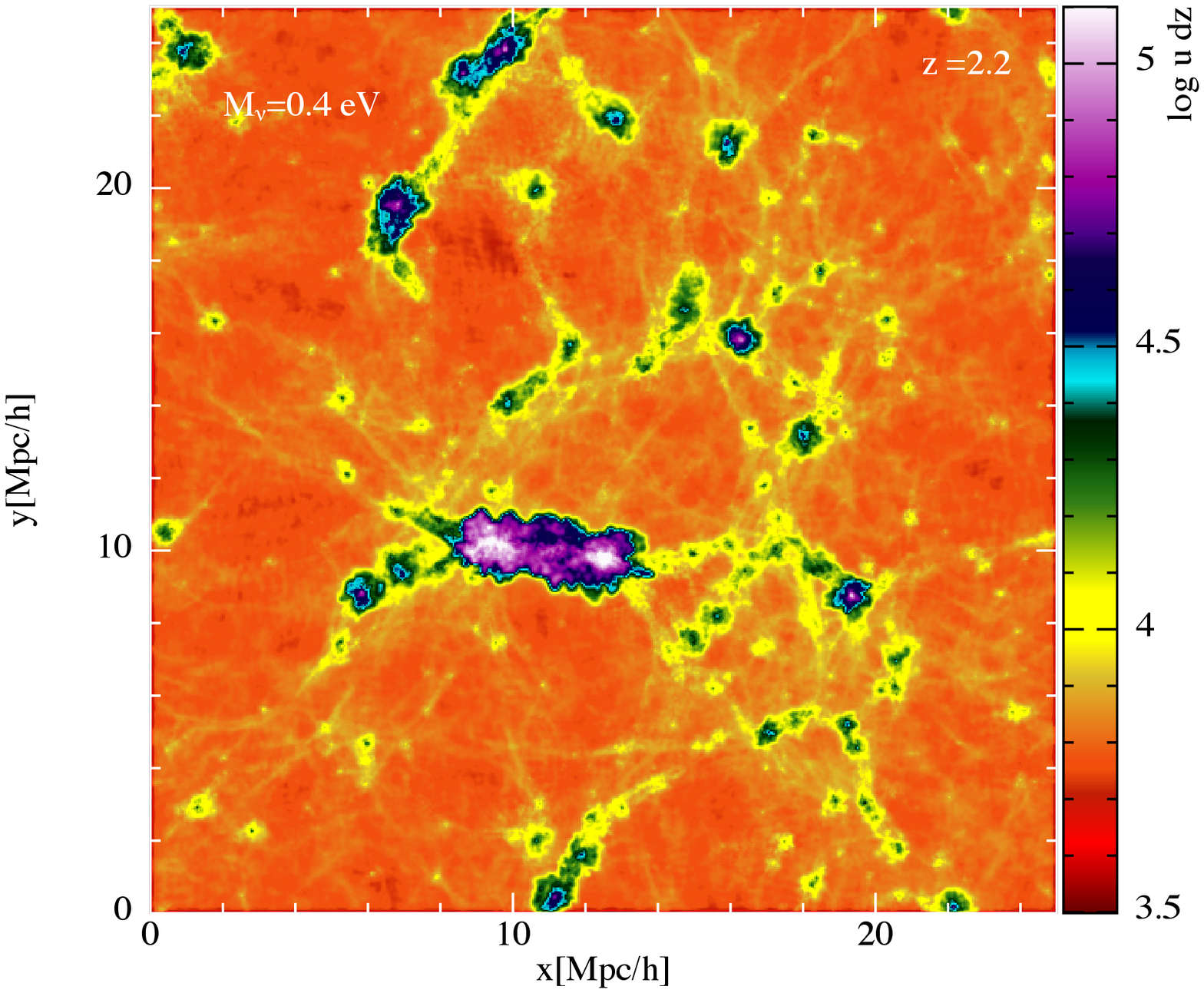}
\caption{Slice of the internal energy of the gas
from simulation snapshots at $z=2.2$ when the box size is $25~h^{-1}{\rm Mpc}$ and the resolution is $N_{\rm p}=192^3$/type.
The upper left panel is from a simulation with massless neutrinos, and in clockwise direction the values of the neutrino mass increase as 
$M_{\rm \nu}=0.1, 0.4$, and $0.8$ eV.  
Changes in the thermal state of the gas are relevant for the power-law
$T_0-\gamma$ relation (eq. \ref{eq_T_rho}).}
\label{fig_visualization_internal_energy}
\end{figure*}


\subsection{Pipeline and post-processing}

A typical snapshot from Gadget-3 at a given redshift contains information about positions and velocities for all the components (gas, DM, neutrinos, stars),
in addition to specific information about the SPH treatment of the gas (i.e., internal energy, density, hydrogen, and electron
fraction and smoothing length).
The snapshot
goes through an elaborate pipeline to obtain
an averaged flux power spectrum and compute the
temperature-density relation (cf. eq. \ref{eq_T_rho}).
To characterize the Ly$\alpha$ flux statistics, 
10,000 randomly placed simulated quasar sightlines were drawn through the simulation box.
Given our largest $100~h^{-1}{\rm Mpc}$ box size, 
this implies an average spacing between sightlines of $10~h^{-1}$kpc 
  --  far smaller than the scale probed by the Ly$\alpha$ forest.
To generate the flux power spectrum, the absorption due to each SPH particle near the sightline was calculated
from the positions, velocities, densities, and temperatures of all the SPH particles at a given
redshift -- following the procedure
described in Theuns et al. (1998) using the SPH formalism;
this provides a number of simulated quasar spectra that were smoothed with a three-dimensional cubic spline kernel.
As done in Borde et al. (2014), each spectrum was rescaled by a constant so that the mean flux across all spectra and absorption bins 
matched that observed mean flux at redshift $z$ (Miralda-Escud{\'e} et al.1996;  Kim et al. 2007; Meiksin 2009).
In particular, we
fixed the photoionization rate by
requiring the effective optical depth at each 
redshift to follow the empirical power law $\tau_{\rm eff}(z) =  \tau_{\rm A} (1+z) ^{\tau_{\rm S}}$, with $\tau_{\rm A}=0.0025$ and $\tau_{\rm S}=3.7$. 
The normalization was performed a posteriori, since finding and 
fixing the appropriate photoionization rate a priori for each of the simulations would be more computationally demanding.  
However, the rescaling of the optical depths is possible
and routinely done, because
simply changing the intensity of the UV background at a fixed redshift without
changing the reionization history does not vary the temperature of the gas
for an optically thin IGM in ionization equilibrium; the
instantaneous temperature only depends on the spectral shape of the UV background and
the gas density.  This has been demonstrated analytically (e.g., equation
2.16 in Theuns 2005). The rescaling coefficients,  which were determined  independently 
for every redshifts, were found to be between $-20\%$ and $+20\%$. 
On the other hand, changing the reionization history would instead modify the integrated thermal history
and hence the amount of Jeans smoothing in the IGM -- although   
the impact of varying the hydrogen reionization history on the Ly$\alpha$ forest at $2<z<4$ is relatively modest
(e.g., Viel et al. 2005, 2006, 2009; Becker et al. 2011). 
After performing the normalization procedure,
the mean over all the rescaled spectra was used as the extracted flux
power spectrum for the box. 
Finally, the splicing technique of McDonald (2003) was applied to increase the
effective resolution (see also Borde et al. 2014 for more details on the splicing method).




\section{FIRST RESULTS} \label{sec_first_results}

In this section we present the first results from the analysis of our suite of hydrodynamical simulations.
In particular, after briefly mentioning convergence and resolution tests and showing a few visualization examples, we
compute the three- and one-dimensional matter- and flux-power spectra and characterize the one-dimensional statistics of the
Ly$\alpha$ transmitted flux  in presence of massive neutrinos. 


\begin{figure*}
\includegraphics[angle=0,width=0.325\textwidth]{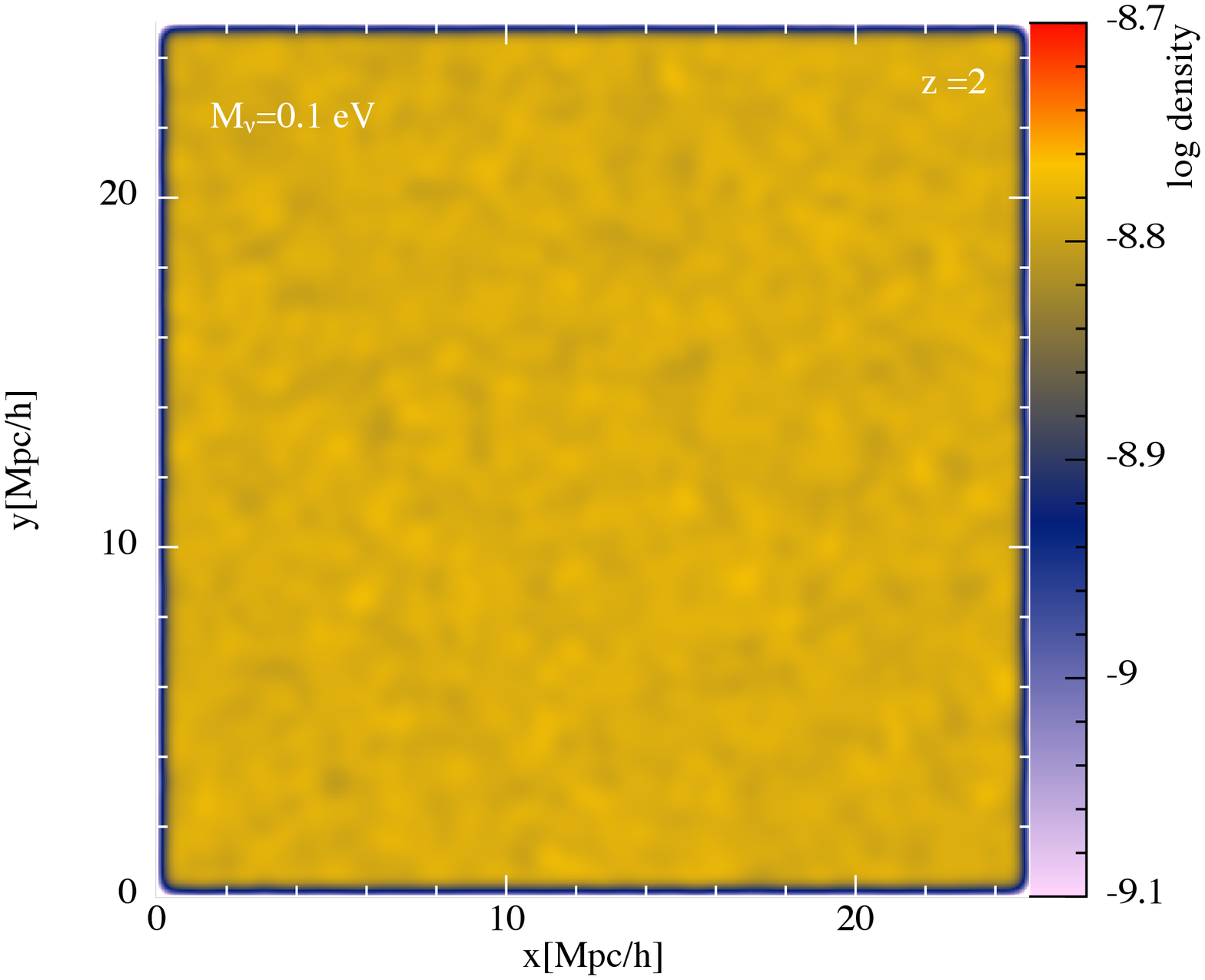}
\includegraphics[angle=0,width=0.325\textwidth]{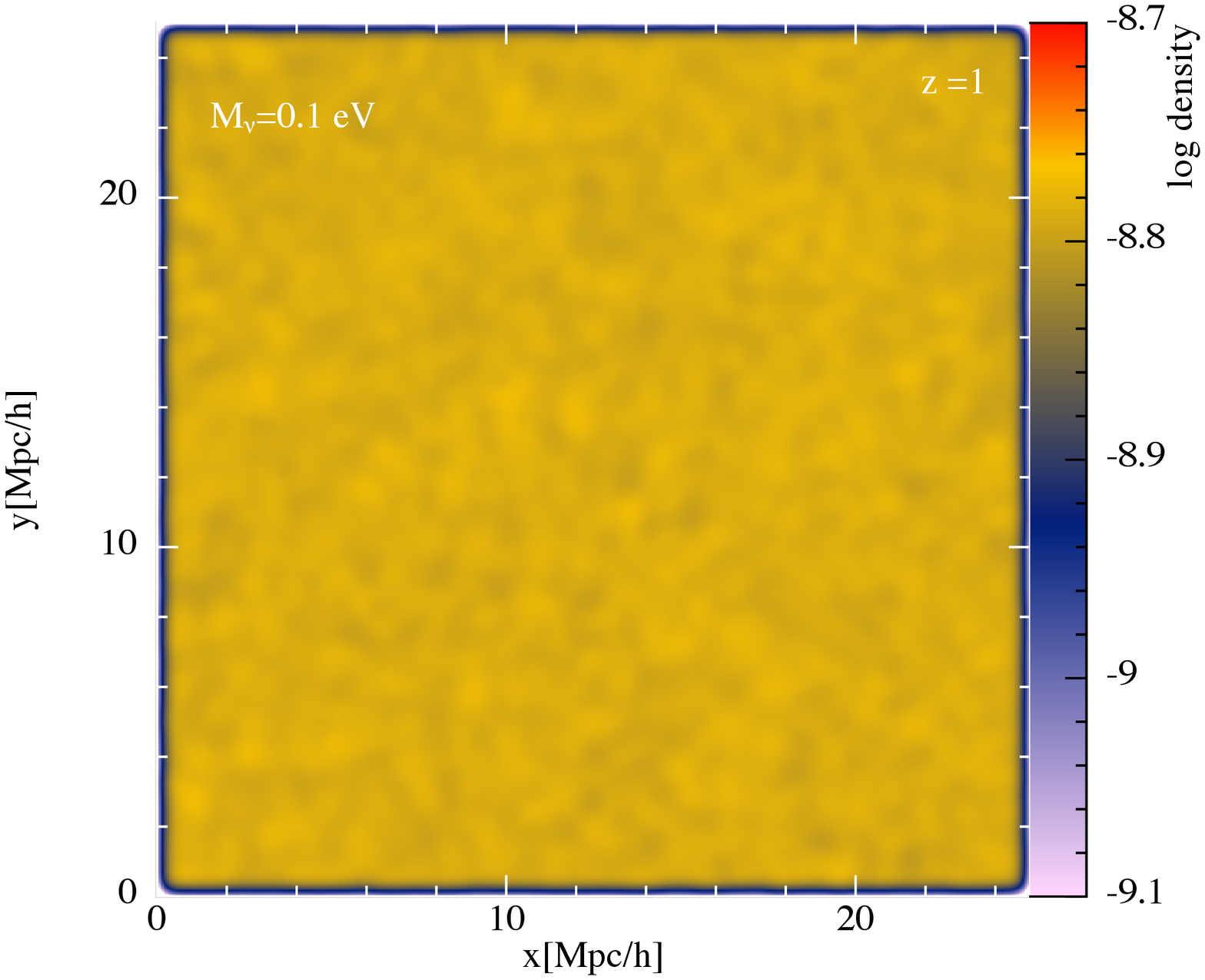}
\includegraphics[angle=0,width=0.325\textwidth]{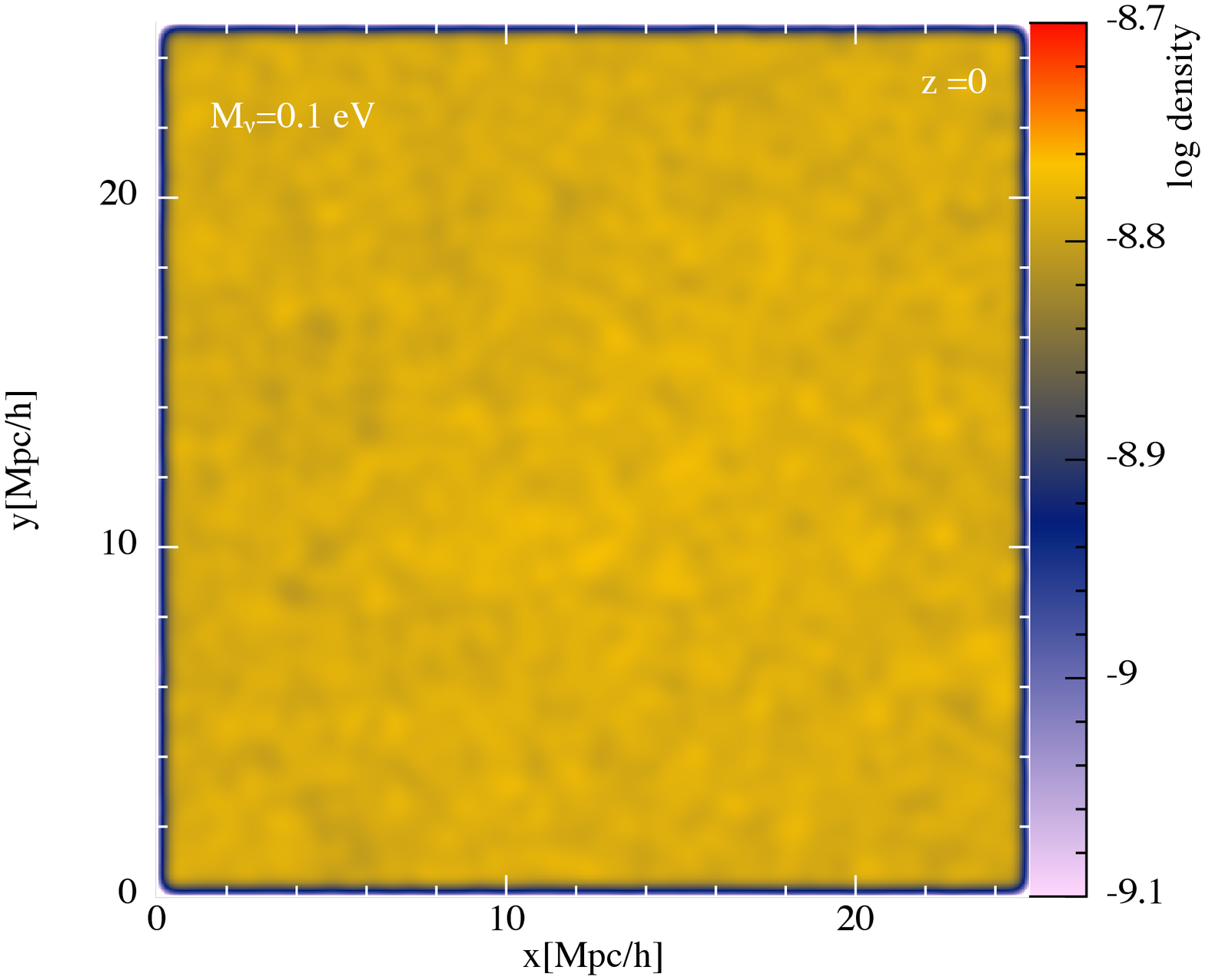}

\includegraphics[angle=0,width=0.325\textwidth]{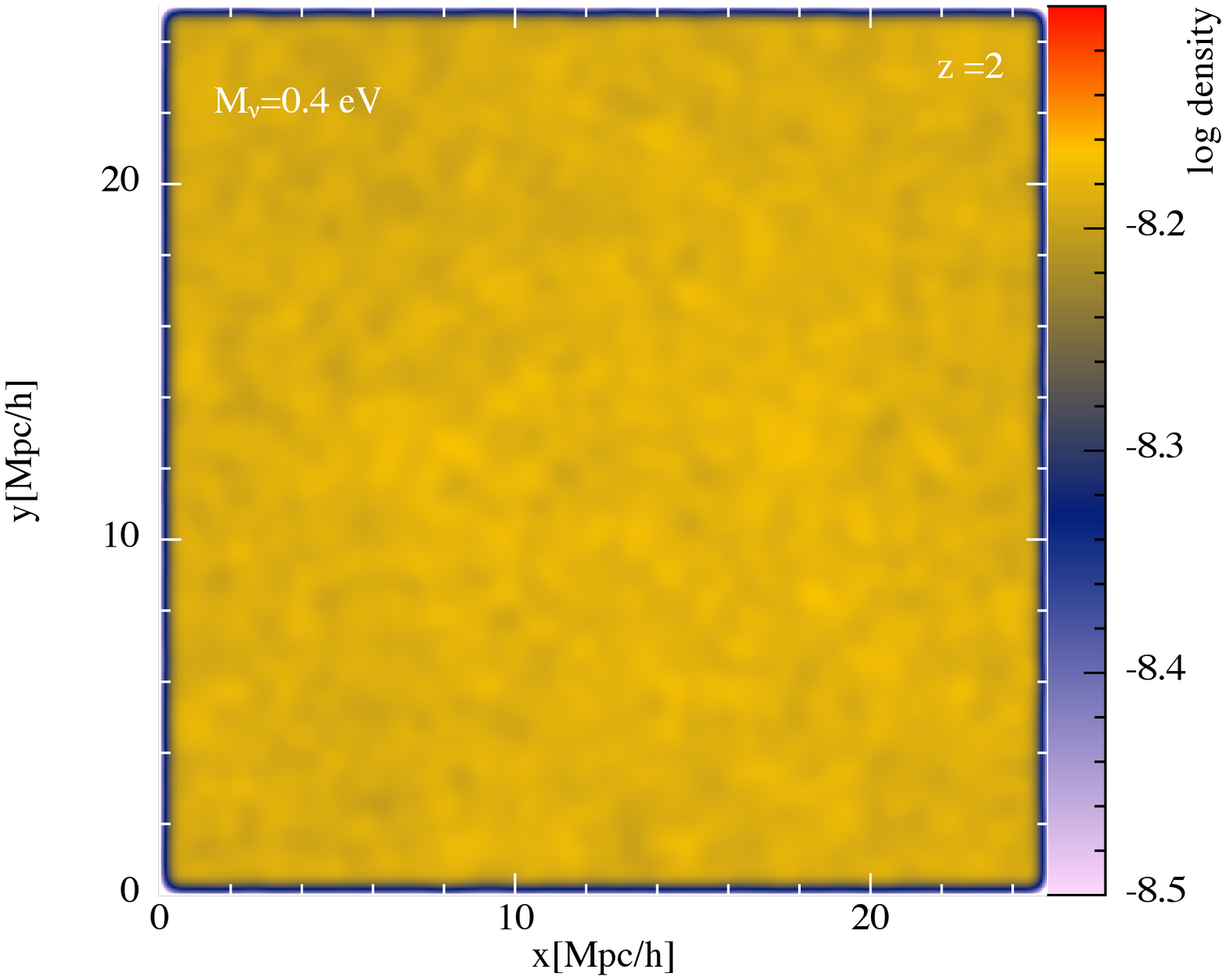}
\includegraphics[angle=0,width=0.325\textwidth]{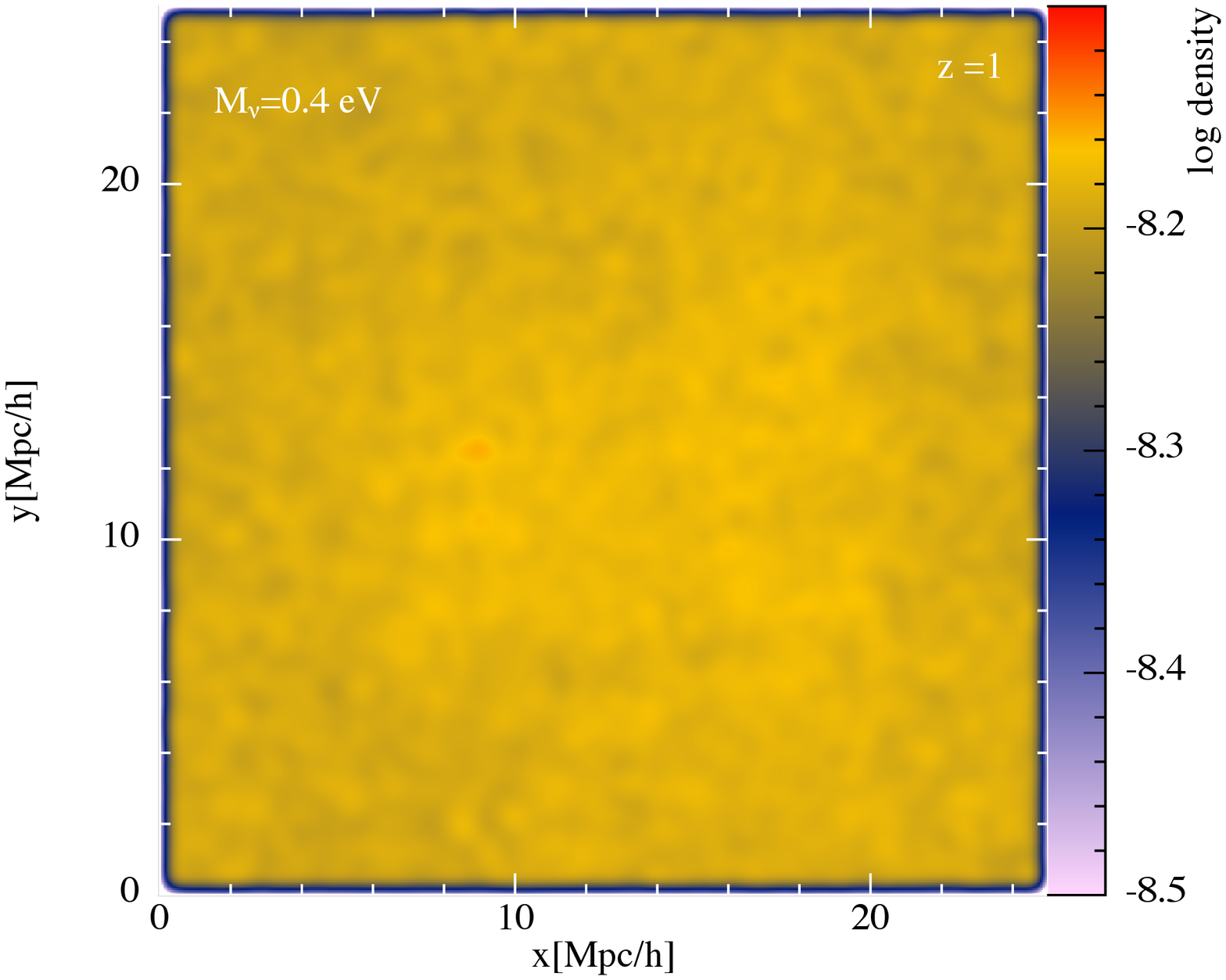}
\includegraphics[angle=0,width=0.325\textwidth]{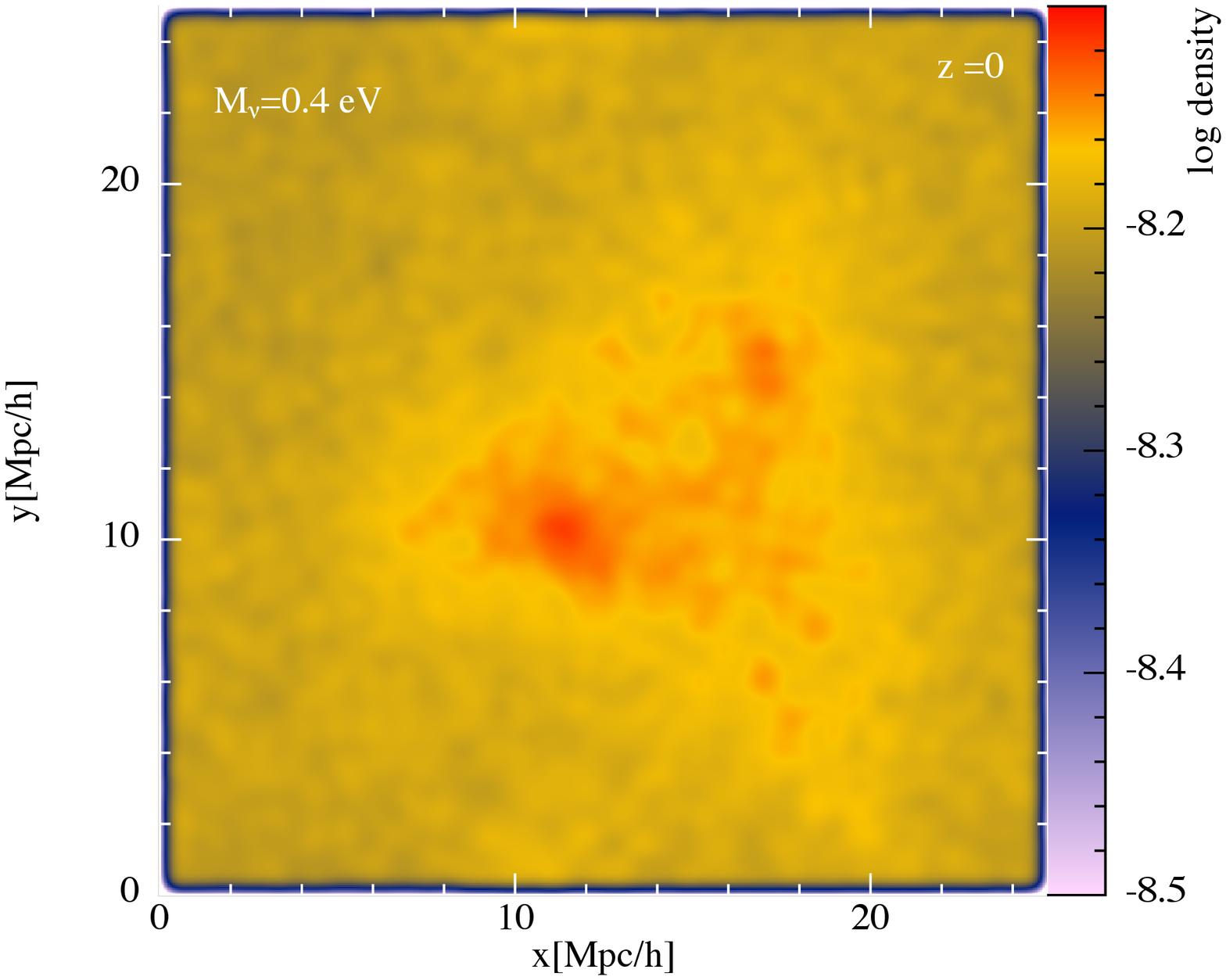}

\includegraphics[angle=0,width=0.325\textwidth]{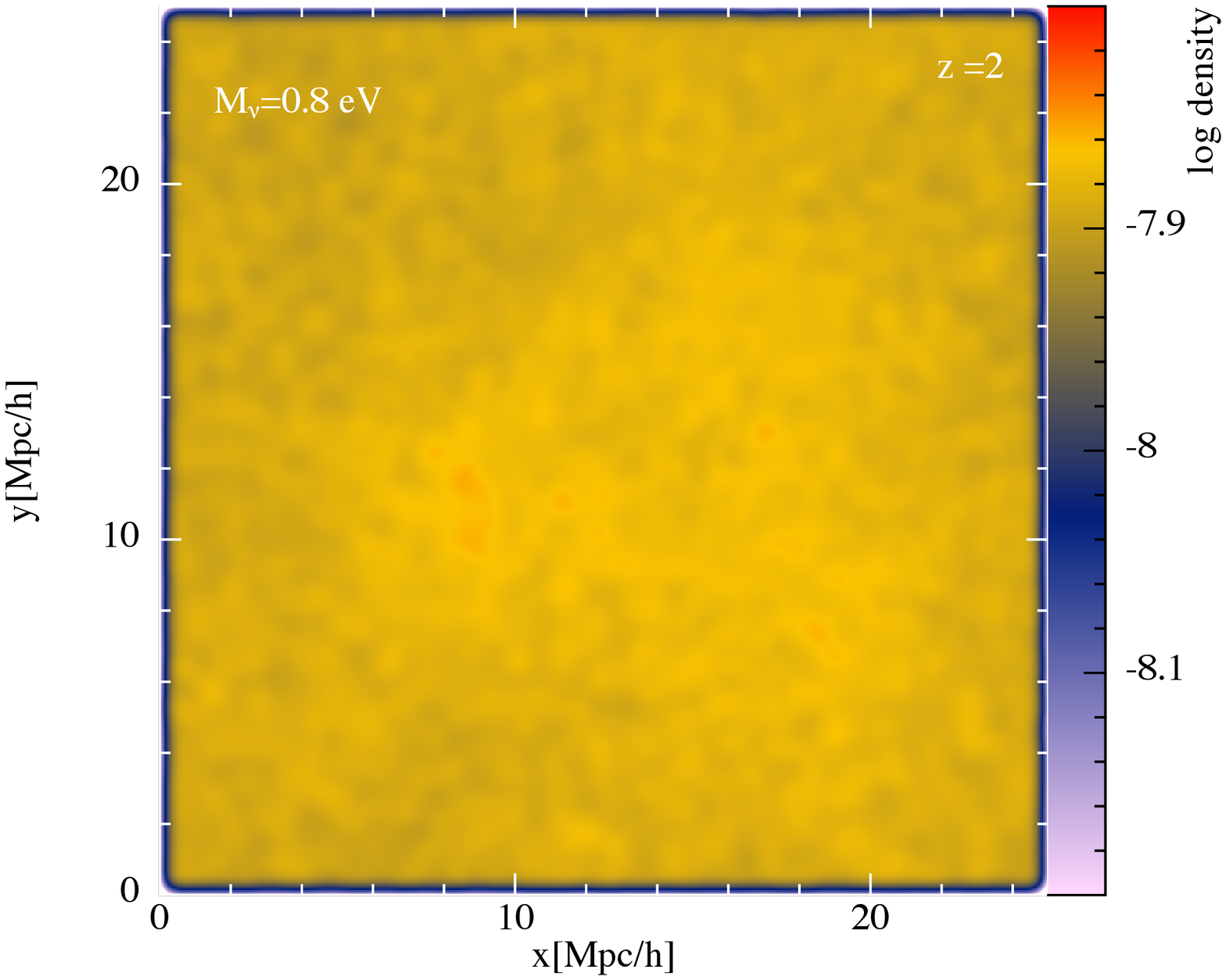}
\includegraphics[angle=0,width=0.325\textwidth]{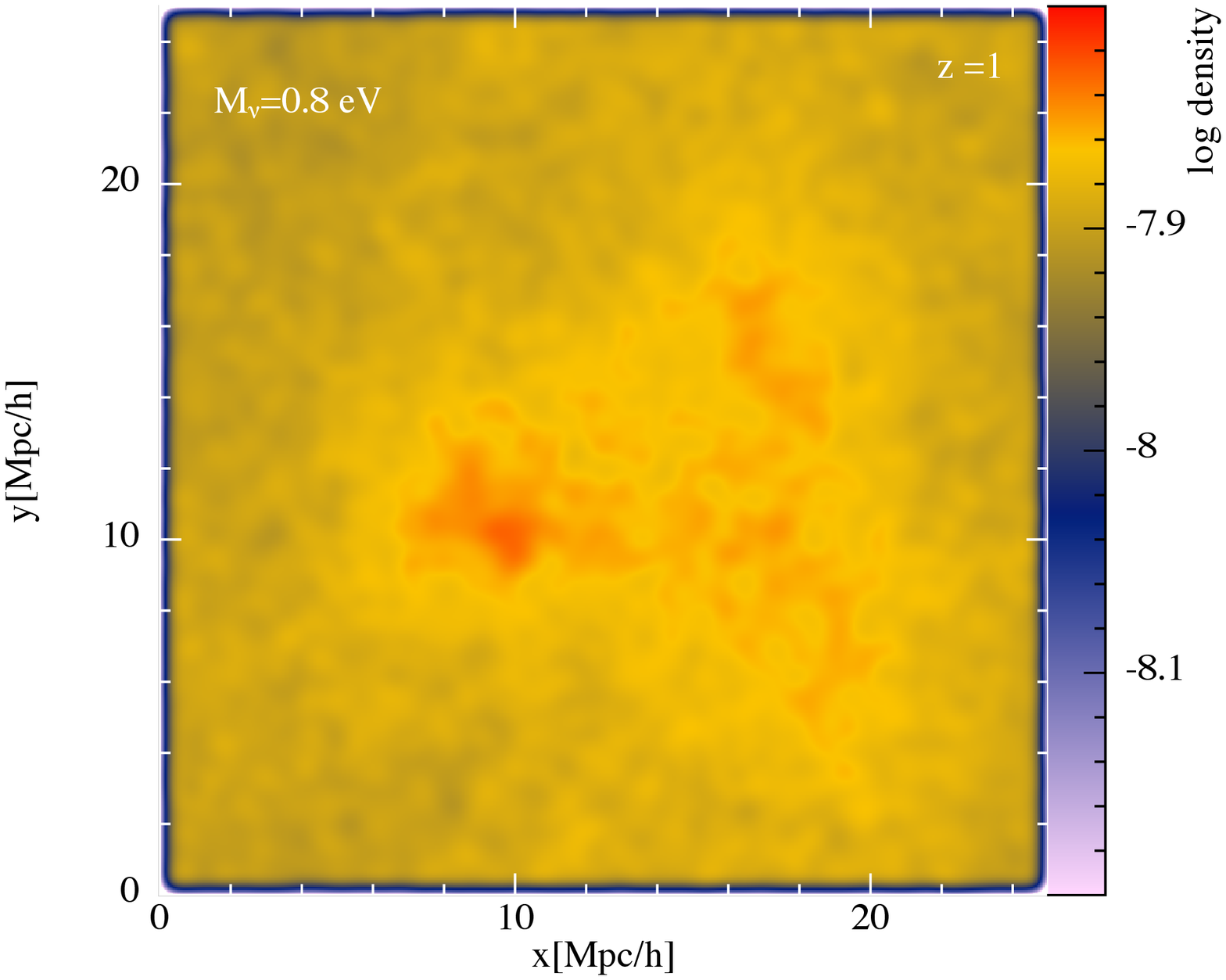}
\includegraphics[angle=0,width=0.325\textwidth]{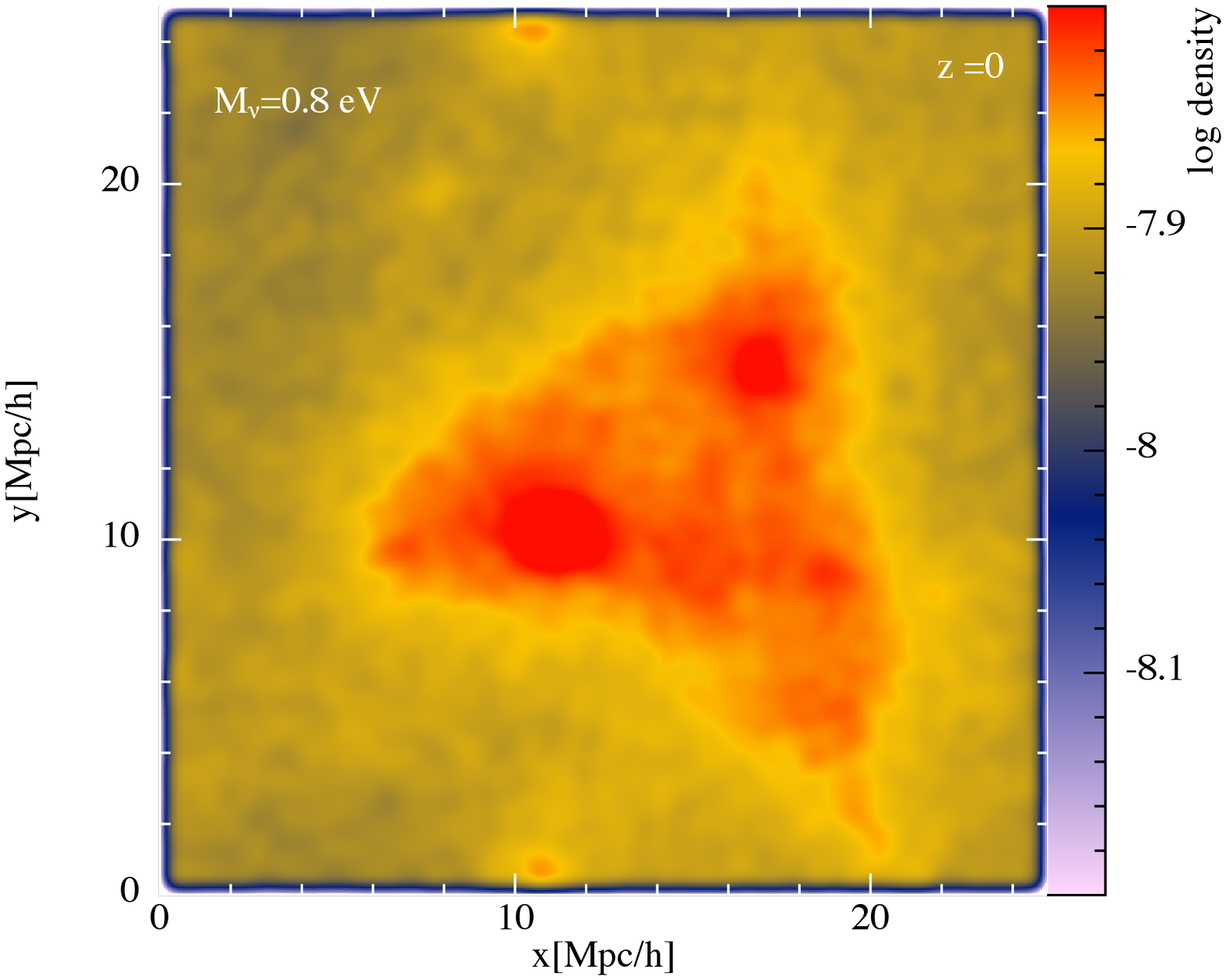}

\caption{Density evolution of the neutrino component at $z=2$ (left panels), $z=1$ (central panels), and $z=0$ (right panels),
when $M_{\rm \nu}=0.1$ eV (top), $M_{\rm \nu}=0.4$ eV (central), and $M_{\rm \nu}=0.8$ eV (bottom).
All the simulations have a box size of $25~h^{-1}{\rm Mpc}$ and resolution $N_{\rm p}=192^3$/type.
Because they are free-streaming, 
the effect of neutrinos is similar to a Gaussian noise component for very small masses, but as the mass increases, clustering effects are noticeable
 and are more 
pronounced for larger neutrino masses and lower redshifts.}
\label{fig_visualization_neutrinos}
\end{figure*}

\begin{figure*}
\centering
\includegraphics[angle=0,width=0.95\textwidth]{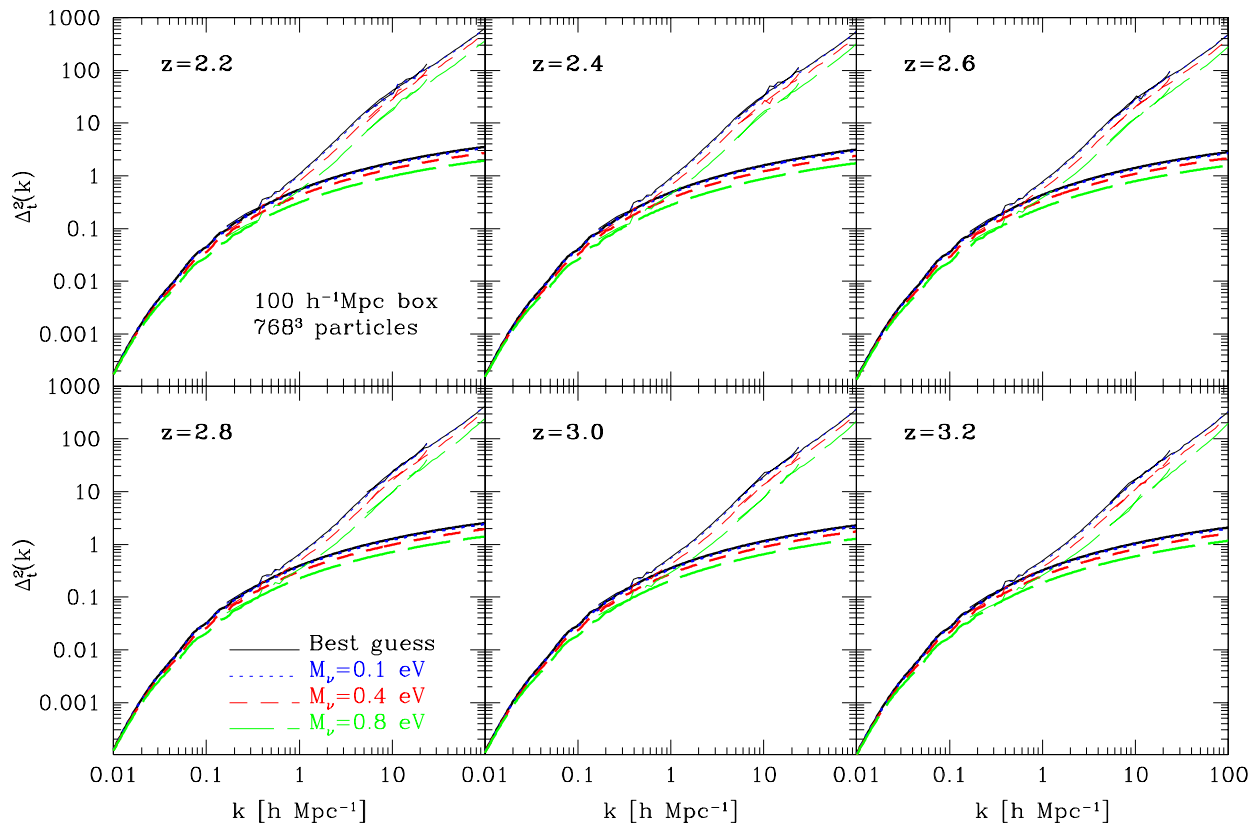}
\caption{Linear (thick lines) and nonlinear (thin lines) evolution of the 
three-dimensional total matter power spectrum computed from 
the best-guess realization (black lines) and from runs with massive neutrinos with $M_{\rm \nu}=0.1, 0.4$, and $0.8$ eV (dotted, dashed and long-dashed lines).
Six intervals in redshifts are considered, from $z=2.2$ to $z=3.2$, with a spacing of $\Delta z=0.2$.}
\label{fig_3d_matter_ps_comparisons_A}
\end{figure*}

\begin{figure}
\centering
\includegraphics[angle=0,width=0.34\textwidth]{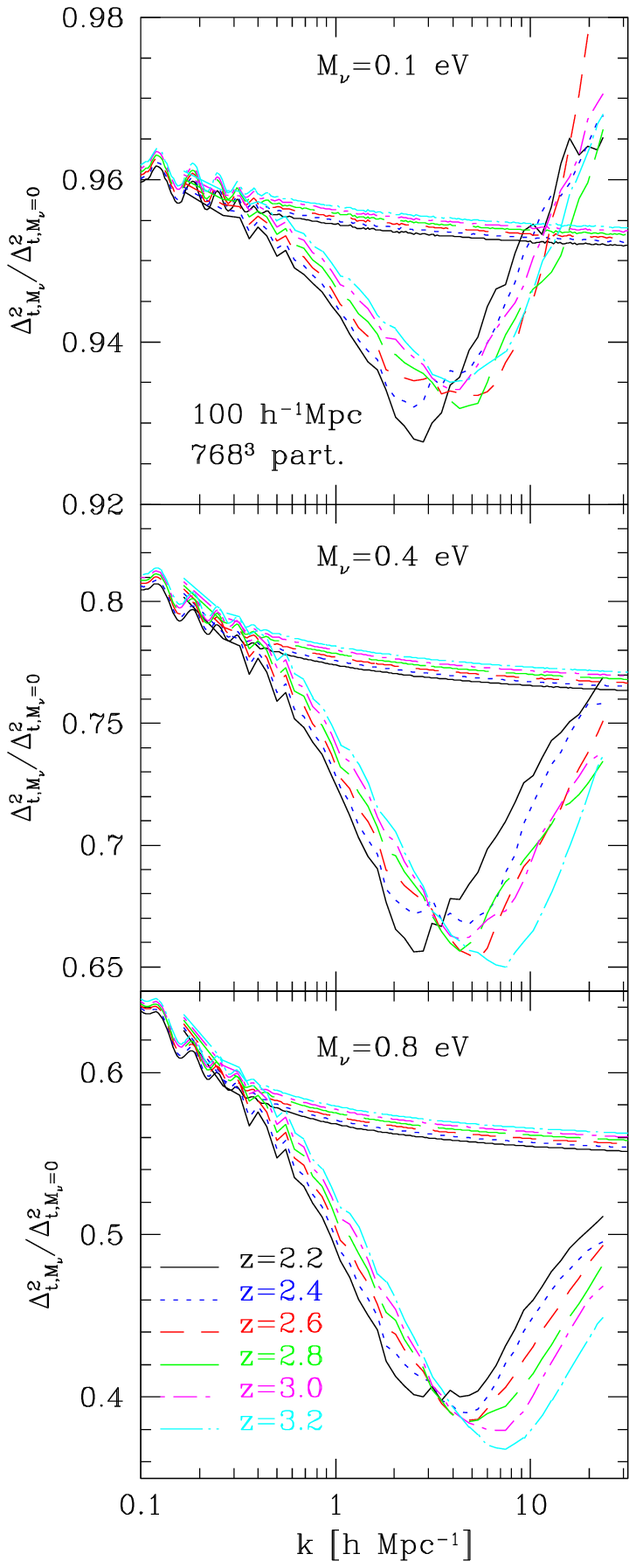}
\caption{Three-dimensional matter power spectra with massive neutrinos, normalized
by their corresponding   massless neutrino power spectra at the same resolution.
The top panel represents $M_{\rm \nu}=0.1~{\rm eV,}$ the middle panel
  denotes $M_{\rm \nu}=0.4~{\rm eV,}$ and the bottom one shows $M_{\rm \nu}=0.8~{\rm eV.}$
The almost
straight lines denote linear theory expectations.
Note the characteristic scale- and redshift-dependent suppression of the 3D power caused by the neutrino  free-streaming.}
\label{fig_3d_matter_ps_comparisons_B}
\end{figure}


\subsection{Convergence and resolution tests} \label{sec_convergence_tests}

Accurately modeling the Ly$\alpha$ flux power spectrum and 
achieving numerical convergence for the Ly$\alpha$ forest 
is challenging 
because most of the signal comes from poorly resolved 
underdense regions.
In addition, current data at high redshift 
are noisier than those at low-$z$, which increases the sample variance in the simulation box.
Hence, checks for convergence and resolution are important, and one needs to find an optimal compromise
between the box size of the simulation, the total number of particles used in the runs, and the overall CPU consumption.
Clearly, convergence requirements will always depend on the physical
process under consideration, as well as on the precision of the
observational data with which the simulations are compared. 
To this end, extensive tests on convergence and resolution -- along the lines of 
Theuns et al. (1998), Bryan et al. (1999), Regan, Haehnelt \& Viel (2007) and 
Bolton \& Becker (2009) -- have
been carried out in Borde et al. (2014). Their results have motivated the choices of
box sizes and resolutions in this work, and the overall strategy of using a set of three simulations 
and  applying the splicing technique (instead of performing 
a single but computationally too demanding run), which allows for a substantial decrease of 
modeling errors because of the improved particle resolution. 
Given our setting choices, numerical convergence is safely reached; however,   
since we also added the neutrino component 
as another type of particle and performed a complete hydrodynamical treatment,
our simulation workload was heavier than simpler realizations with only gas and DM 
by about $20\%$ -- when the number of particle per species was kept equal.

In closing this section, we note that
while the tests conducted in Borde et al. (2014) did not consider
massive neutrinos, their stringent results about convergence and resolution requirements
are readily applicable to 
our case. This is simply because including massive neutrinos is essentially equivalent
(with very minor effects, at least in terms of convergence and resolution)
to a neutrino-less situation with a slightly different value of the parameter $\sigma_8$ -- see
for example Viel et al. (2010), where the degeneracy $\sigma_8-M_{\rm \nu}$ is
discussed in some detail. 
In addition, observational uncertainties on the BOSS power spectrum are at a level
that is less stringent than the requirements imposed in Borde et al. (2014).

Regarding the splicing technique, 
neutrinos are expected to introduce a smooth suppression in terms of matter power spectrum, and the splicing technique 
is able to capture this effect in the range of redshifts and wavenumbers we are interested in. 
In fact, the splicing technique can detect smooth variations of amplitude in matter power across the scales 
(although, very likely, this will no longer be the case for warm dark matter where the induced 
suppression is abrupt and stronger than in the neutrino case).


\subsection{Visualizations}

Massive neutrinos induce changes in the thermal state of the gas and in
the LSS clustering of DM.
Differences are even visually perceptable for relatively large neutrino masses, for example, in the distribution of the internal energy of the gas (and hence of its temperature), 
when compared with simulations with massless neutrinos.
Figure \ref{fig_visualization_internal_energy} provides an example:  in the various panels, we show a slice of the internal energy of the gas 
from simulation snapshots at $z=2.2$, when the box size is $25~h^{-1}{\rm Mpc}$ and the resolution is $N_{\rm p}=192^3$/type; 
the upper left panel is from a simulation with massless neutrinos, and in clockwise direction the values of the neutrino mass increase as 
$M_{\rm \nu}=0.1, 0.4$, and $0.8$ eV.  
Changes in the thermal state of the gas are
particularly relevant for the power-law
$T_0-\gamma$ relation (cf. eq. \ref{eq_T_rho}), which is
thought to arise  from the 
competition between photoheating and cooling due to the adiabatic expansion of the Universe,  
following reionization.
The evolution of this relation has been measured in the data and depends on the reionization history and the hardness of the UV background 
(Schaye et al. 2000; Ricotti et al. 2000; McDonald et al. 2001; Rollinde et al. 2013),
although in reality  the picture is more complicated -- because of radiative transfer effects during the epoch of HeII reionization.
Nevertheless,  the temperature at the 
characteristic overdensity probed by the Ly$\alpha$ is now 
quite well measured (e.g. Becker et al. 2011).  The main uncertainty that 
remains is the slope ($\gamma$) of the $T_0-\rho$ relation:
this is still poorly measured and translates into an 
uncertainty on $T_0$ (at mean density). 
Hence, a more accurate modeling of the thermal state of the gas is required to 
reduce uncertainties in the thermal state of the IGM -- when massive neutrinos are also present.

Figure \ref{fig_visualization_neutrinos} shows the density evolution of the neutrino component from 
simulations with 
 $25~h^{-1}{\rm Mpc}$ box sizes and resolution $N_{\rm p}=192^3$/type, 
at $z=2$ (left panels), $z=1$ (central panels), and $z=0$ (right panels)
as a function of the neutrino mass; top panels show $M_{\rm \nu}=0.1$ eV, 
intermediate panels represent $M_{\rm \nu}=0.4$ eV, and the bottom panels
$M_{\rm \nu}=0.8$ eV.
The axis scales are
 in $h^{-1}{\rm Mpc}$. 
 Note again that for the neutrino component the density scale is kept fixed only for a given neutrino mass, 
 while it changes across different $M_{\rm \nu}$ values.
The distribution of the neutrino density has been 
smoothed  with a cubic spline kernel to eliminate spurious Poisson noise at the smallest scales to
obtain genuine cosmological density fluctuations
of the neutrinos that occur only on large scales -- because of their free-streaming. 
According to Viel et al. (2010), typical neutrino fluctuations at the largest scales are about $10 \%$ around the mean, while for gas and DM the 
fluctuations are usually much stronger.
Moreover,  the growth of structures is less evolved in the simulation with neutrinos (i.e., the voids are less empty) since their suppressed 
clustering slows down the growth of the perturbations in the
overall matter density, and this in turn affects the properties of the gas and DM.

Clearly, one of the main consequences of the particle-based implementation of massive
neutrinos is the presence of shot noise. 
To this end,  Viel et al. (2010) have conducted an extensive computation of shot noise (see their Section 3.4 and their Figures 9 and 17) 
and considered the effect of varying the number of neutrino particles both on the matter and flux power spectra.
Their findings suggest that 
doubling the neutrino
particles for each spatial dimension shifts the
Poisson contribution to the matter power spectrum by a factor of roughly two to smaller scales. 
Hence, it would be desirable for Ly$\alpha$ studies to increase the number of neutrino particles 
to decrease 
the Poisson contribution to the matter power spectrum and
sample the neutrino 
power spectrum properly on scales between 0.1 and 2 $h~{\rm Mpc^{-1}}$.
However, Viel et al. (2010) also pointed out that 
increasing the number of neutrino particles
by a factor of eight can be done but required a factor of $\sim2$ more in
CPU time. Therefore, one has to balance the demand 
in terms of parallel computing resources with the desired resolution.
Fortunately, although the neutrino power spectrum is affected by shot noise
at the smallest scales,  the impact on the matter power spectrum, and thus on the one-dimensional flux power spectrum (which is
the main quantity we are after), is still very small because the neutrinos constitute a very small fraction
of the overall matter density. Hence, in  our regime of interest -- which is 
analogous to that of Viel et al. (2010) -- a single neutrino
particle per CDM particle is sufficient.


\subsection{Three-dimensional matter power spectra} \label{sec_3d_matter_ps}

Next, we consider the set of  neutrino simulations with $768^3$ particles per type and a box size of $100~h^{-1}$Mpc,
with the same spectral amplitude as the corresponding best-guess run (i.e., the normalized simulations).
Values of $\sigma_8$ at $z=0$ for these realizations are provided in Table \ref{tab_suite_sims_A}.
From these runs, we compute the three-dimensional total matter power spectra and compare results
with linear predictions. 

In Figure \ref{fig_3d_matter_ps_comparisons_A}, we show 
results of this comparison. 
Black lines denote the best-guess realization, and 
 dotted, dashed, and long-dashed lines are used for 
 runs with massive neutrinos with  $M_{\rm \nu}=0.1, 0.4$, and $0.8$ eV.
Six intervals in redshifts were considered, from $z=2.2$ to $z=3.2$, with a spacing $\Delta z=0.2$.
The linear evolution (thick lines) was computed from CAMB as
explained in Section \ref{sec_implementing_neutrinos}, 
while the nonlinear power spectra (thin lines) were obtained from Gadget-3 snapshots.
As can be directly inferred from the various panels, 
the free-streaming of neutrinos results in a 
suppression of the power spectrum of the total matter 
distribution at scales probed by the Ly$\alpha$ forest data,  which is higher 
than the linear theory prediction by about $\sim 5 \%$ ($\sim 9 \%$) at scales $k\sim 1~h~{\rm Mpc^{-1}}$
when $M_{\rm \nu}=0.4$ eV ($M_{\rm \nu}=0.8$ eV) and is strongly redshift dependent. 
The effects of free-streaming of neutrinos on the matter power spectrum have been discussed in detail in
Viel et al. (2010):  we here confirm their findings of 
a mass- and redshift-dependence suppression of the power spectrum at small scales, which is
more significant with increasing neutrino mass.
At large scales, linear and nonlinear evolution in the power spectrum are
similar, as already pointed out in Figure \ref{fig_neutrino_linear_theory_A}, where
we argued that a linear description for the neutrino component is sufficient
inside the yellow area (when  $k <k_{\rm n, M_{\nu}=0.8~eV}$).
 

\begin{figure*}
\centering
\includegraphics[angle=0,width=0.95\textwidth]{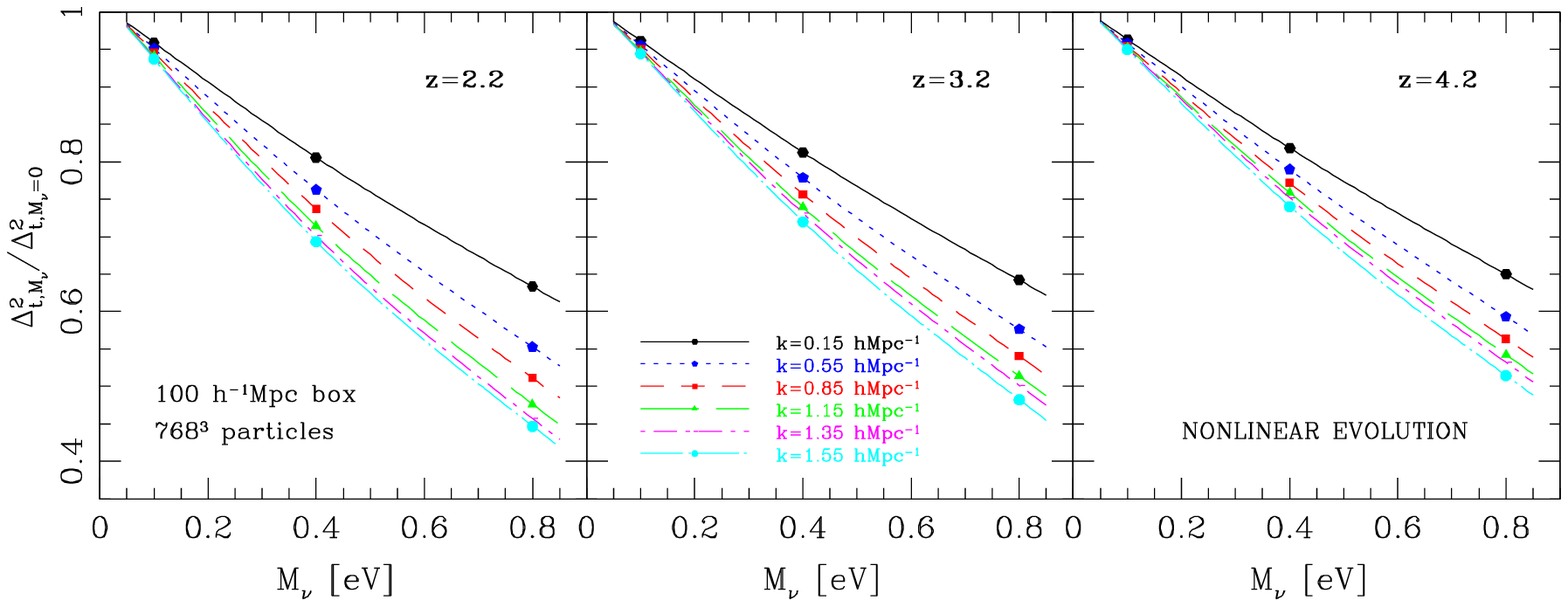}
\includegraphics[angle=0,width=0.95\textwidth]{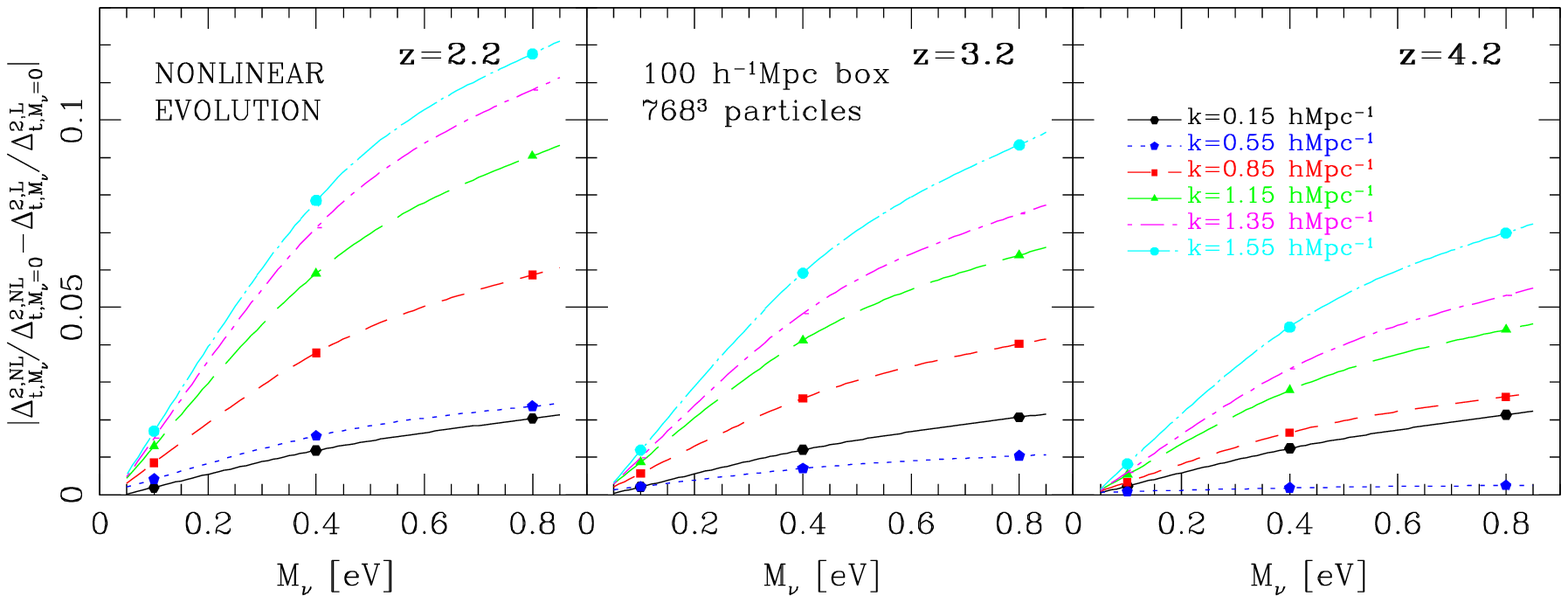}
\caption{[Top panels]
 Total nonlinear matter power spectra in simulations with massive neutrinos, normalized by the reference run with only a massless neutrino component as a function
of neutrino mass and  for different values of $k$  in the range
relevant for the BOSS Ly$\alpha$ forest data -- as specified in the panels. 
[Bottom panels] Evolution of
$f = |\Delta_{\rm t, M_{\nu}}^{\rm 2, NL}  /  \Delta_{\rm t, M_{\nu}=0}^{\rm 2, NL} -    \Delta_{\rm t, M_{\nu}}^{\rm 2, L}  /  \Delta_{\rm t, M_{\nu}=0}^{\rm 2, L}|$
for the same $k$- and $z$-intervals as considered in the top panels.
See the main text for more details.}
\label{fig_3d_matter_ps_comparisons_C}
\end{figure*}

\begin{figure*}
\centering
\includegraphics[angle=0,width=0.85\textwidth]{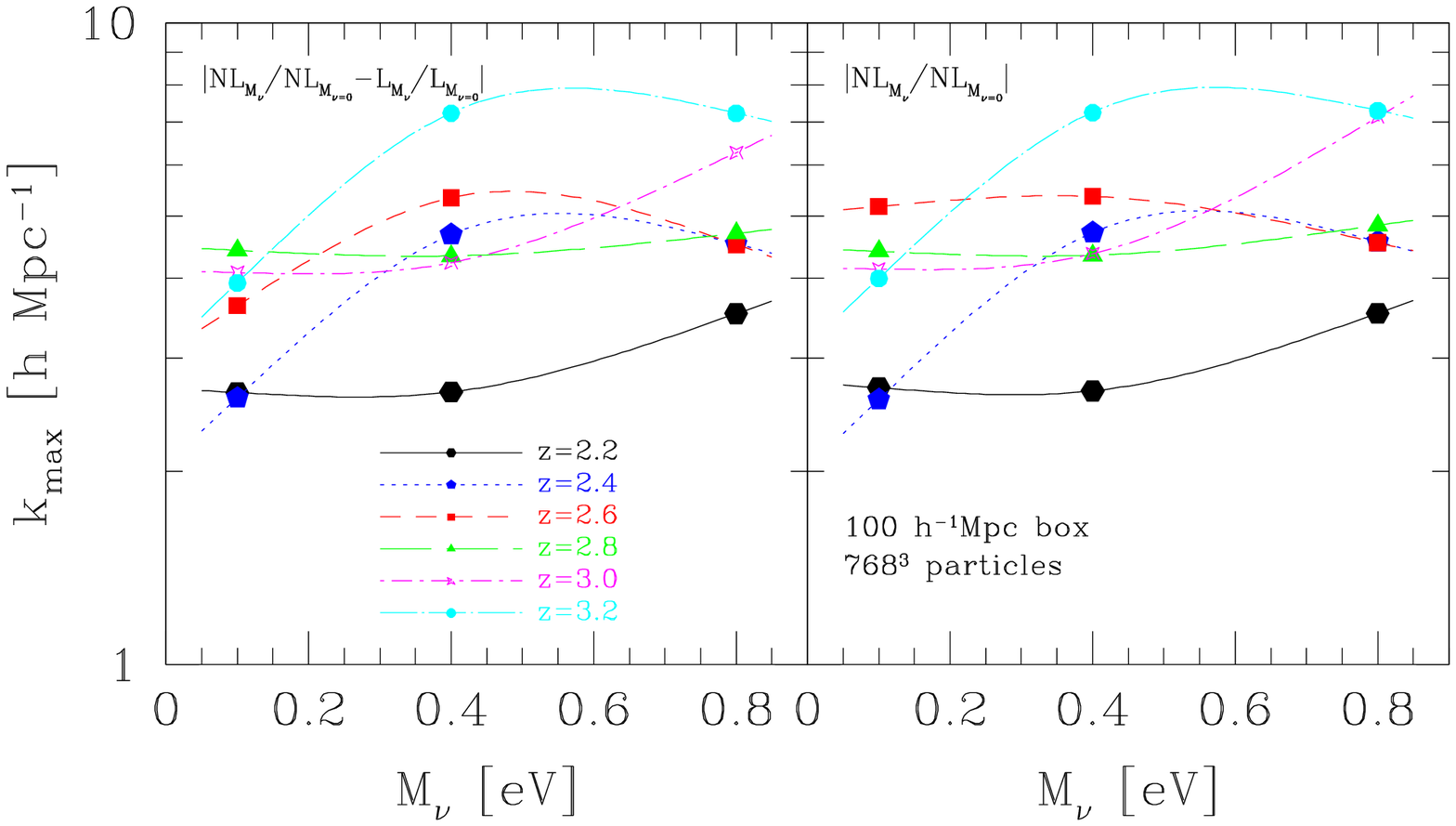}
\caption{[Left] Values of $k$
for which the difference between linear and nonlinear evolution in terms of ratios ($f$)
is maximized -- without restricting the wavenumber to the  one-dimensional BOSS Ly$\alpha$ range.
[Right] Same as the left panel, but when only the nonlinear evolution is considered. 
This plot is particularly useful because it allows determining which scales are more sensitive to the neutrino mass.
See again the main text for more details.}
\label{fig_3d_matter_ps_comparisons_D}
\end{figure*}


Figure  \ref{fig_3d_matter_ps_comparisons_B}  
shows all these effects more clearly: as a function of $k$, we plot
the total three-dimensional matter power spectra in presence of massive neutrinos,
normalized by their corresponding  power spectra from neutrino massless simulations at the same resolution.
The top panel shows $M_{\rm \nu}=0.1~{\rm eV}$, the middle panel 
$M_{\rm \nu}=0.4~{\rm eV}$, and the bottom panel  $M_{\rm \nu}=0.8~{\rm eV}$.
In the panels, the almost
straight lines are linear theory expectations:
the plateau of nearly $k$-independent suppression
predicted by linear theory is approximately described by $\Delta P/P \sim -8 f_{\rm \nu}$
and depends only very weakly on redshift. 
Clearly, 
the inclusion of nonlinear effects produces a characteristic $k$-dependent suppression
(i.e., the dips in the figure) on the three-dimensional matter power spectrum, 
 which varies as a function of mass; the higher the value of $M_{\rm \nu}$, the higher
 the $k$-mode where the dip occurs.
 Similarly, for a given neutrino mass, 
 at increasing redshifts the position
 of the maximum suppression deep
 is altered
 in a nontrivial way -- but typically  toward smaller scales.
 The trend we find here appears to be consistent with analogous results
in  Bird et al. (2012).
Note also that there is an upturn in the suppression, which 
was first reported and briefly discussed in
Brandbyge et al. (2010),  and was investigated in depth in Viel et al. (2010). In particular, 
according to Viel et al. (2012), it
appears to be related to the nonlinear collapse of halos decoupling from the large-scale
modes slightly differently in simulations with massive neutrinos than in simulations with only massless neutrinos, 
and has been shown by the same authors not to depend on the number of neutrino particles -- ruling out shot noise as a 
plausible cause. This {\rm finding} suggests that virialization of halos is slightly modified by the smoothly distributed neutrino component,
in a similar way as by dark energy where this is a well-known effect (Alimi et al. 2010).

In the top panels of Figure  \ref{fig_3d_matter_ps_comparisons_C} we study these effects in depth 
by displaying 
 the total nonlinear matter power spectra in simulations with massive neutrinos, normalized by the case with only a massless neutrino component, but now as a function
of neutrino mass ($M_{\rm \nu}=0.1,0.4,0.8~{\rm eV}$) and  for different values of $k$  in the range
relevant for the one-dimensional BOSS Ly$\alpha$ forest data. 
Specifically, we assumed
$k=0.15, 0.55, 0.85, 1.15, 1.35$, and $1.55~h{\rm Mpc^{-1}}$
 with different line styles, 
for three different redshift slices 
(from left to right, $z=2.2, 3.2$, and $4.2$).
For a given redshift interval, the detected trend at increasing $k$ is essentially linear, as 
expected from Figure \ref{fig_3d_matter_ps_comparisons_B}, with departures from
the best-guess simulations, which are more significant at lower redshift and for a larger neutrino mass.

Is of more interest to consider the evolution of the quantity $f$ defined by
\begin{equation}
f = \Big |   \Delta_{\rm t, M_{\nu}}^{\rm 2, NL}  /  \Delta_{\rm t, M_{\nu}=0}^{\rm 2, NL} -    \Delta_{\rm t, M_{\nu}}^{\rm 2, L}  /  \Delta_{\rm t, M_{\nu}=0}^{\rm 2, L}  \Big |,
\end{equation}
namely the difference between nonlinear and linear 3D power spectrum
predictions, expressed in terms of $\Delta^2_{\rm t}$ ratios (as defined before).
This is shown in the bottom panels of Figure  \ref{fig_3d_matter_ps_comparisons_C}
for the same redshift intervals and $k$-values as considered in the top panels.
Spline fits are used to connect points with the same $k$-value.
Clearly, at increasing redshift departures from linear theory are less significant, 
particularly for smaller neutrino masses and lower values of $k$.
Within the Ly$\alpha$ range of interest, it is clear that $f$ is maximized
at lower $z$ and higher values of $M_{\rm \nu}$.

Finally, in the left panel of Figure  \ref{fig_3d_matter_ps_comparisons_D} we find the value of $k$
for which the quantity $f$ (i.e., the difference between linear and nonlinear evolution in terms of ratios)
is maximized -- without restricting the wavenumber to the  BOSS Ly$\alpha$ range. 
Spline fits are again used to connect points with the same $k$ and different values of the
neutrino mass. 
This plot is particularly useful because at a given redshift it 
provides a quick way to determine at which $k$ there is more sensitivity to the neutrino mass, meaning that
it shows where the effects due to neutrino free-streaming are more pronounced.  
 The right panel of the same figure shows analogous information, but now determined by considering
the nonlinear evolution alone. Since the power 
suppression caused by neutrinos is essentially constant at scales $k>0.1$ (Lesgourgues \& Pastor 2006), 
using either the normalized differences between linear and nonlinear evolution (i.e., the quantity $f$ previously defined), or just 
the one given by the nonlinear evolution of the neutrino component in terms of the massless neutrino case
should not make a significant difference; this is in fact confirmed in the right panel of Figure  \ref{fig_3d_matter_ps_comparisons_D}.
The nonlinear power spectrum  strongly
depends on redshift and the dependence of scale becomes steeper with decreasing redshift.
It is interesting to see how these effects propagate into the Ly$\alpha$ flux power spectrum: we
briefly discuss this in the next section and treat the one-dimensional statistics in depth  
in a forthcoming publication.

Before moving on, we
note that there are several other numerical effects that can potentially impact the power spectrum: the number of neutrino particles, the 
velocities in the initial conditions, the sampling of the initial conditions  with neutrino pairs to balance momentum, and the starting redhsift.
All these effects have been investigated in Viel et al. (2010) and are not further discussed here.


\subsection{One-dimensional analysis: flux statistics} \label{sec_1d_flux_stat}

\begin{figure}
\centering
\includegraphics[angle=0,width=0.48\textwidth]{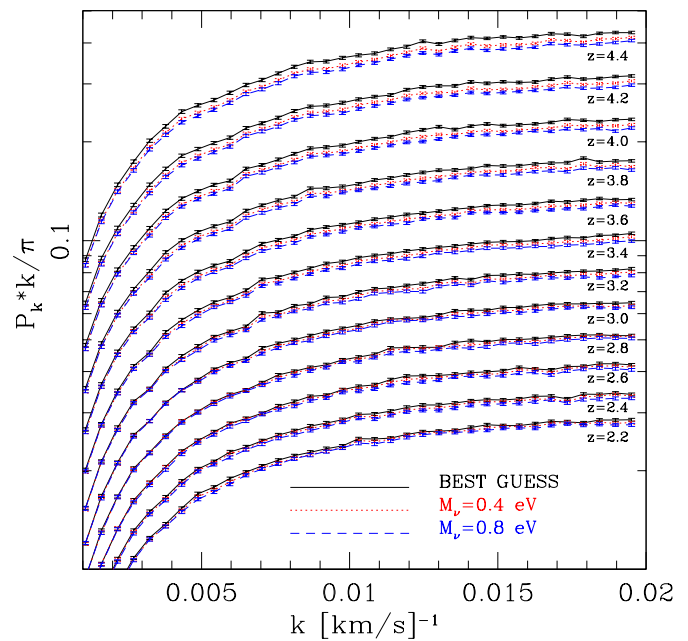}
\caption{One-dimensional flux power spectra, averaged over 10,000 lines of sight, without (solid) and with (dotted and dashed) massive neutrinos.
See the main text for more details.}
\label{fig_los_1d_flux_ps}
\end{figure}

The effect of neutrino free-streaming is a small scale-dependent suppression
of the total matter power, which is a function of redshift and mass of the neutrinos. 
In this part, we briefly address 
how this signal affects the statistical  properties of the transmitted flux fraction (the main observable along a number of quasar sightlines).
The Ly$\alpha$ transmitted flux $\cal{F}$, treated as a continuous field,  is defined as
\begin{equation}
\cal{F} = \exp (- \tau),
\end{equation}
where $\tau$ is the optical depth; the corresponding flux fraction power spectrum is
\begin{equation}
P_{\rm \cal{F}} (k) = | \tilde{\delta}_{\rm \cal{F}}(k)  |^2,
\end{equation}
where $\delta_{\rm F} = \cal{F} /\bar{\cal{F}}$  - 1.
Here $\cal{\bar{F}}$ is the mean flux and the tilde symbol denotes a Fourier-transformed quantity. 
The calibration of the mean flux level is the main systematic error, along with uncertainties in the thermal history of the IGM, 
and  the different scaling given by different simulations. 
The mean flux, a measure of the average density of neutral hydrogen, has a stong impact on the
amplitude of the flux power spectrum (Viel et al. 2010).

\begin{figure*}
\centering
\includegraphics[angle=0,width=0.85\textwidth]{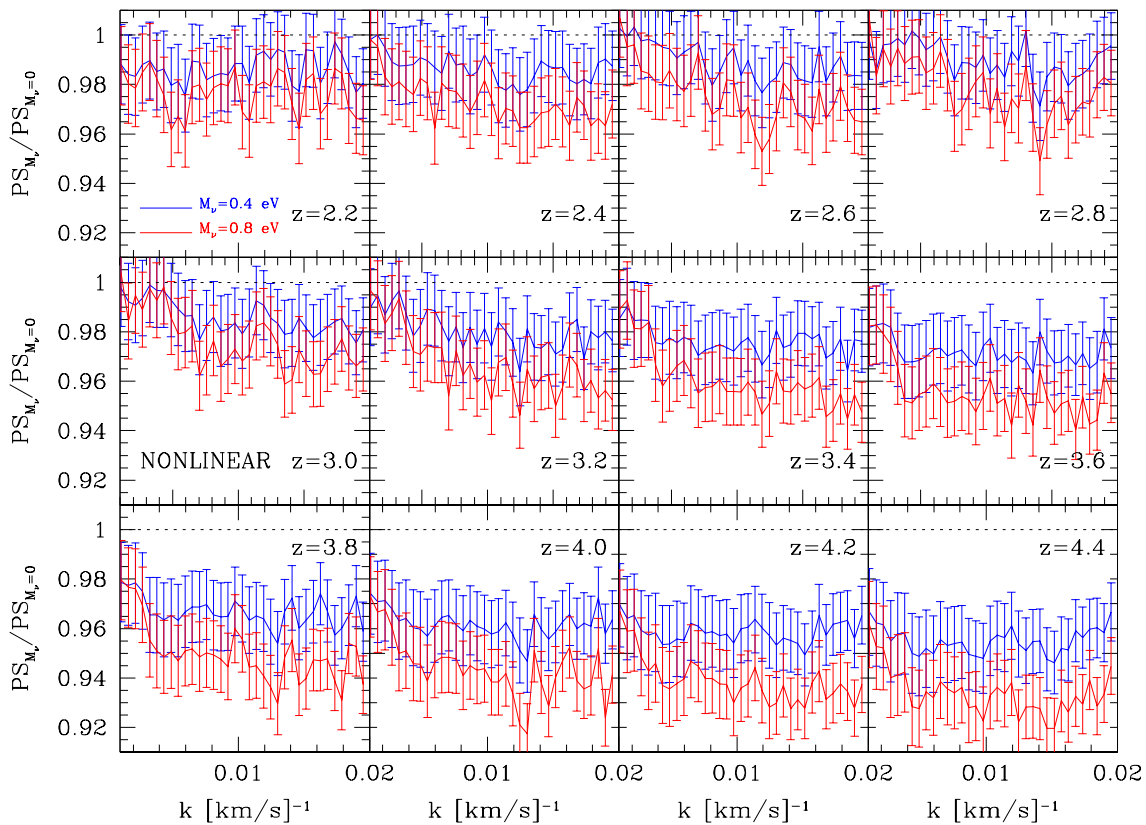}
\caption{Ratio of the averaged one-dimensional power spectra with and without massive neutrinos for two values of the neutrino mass -- as indicated in the panels.
Error bars are $1\sigma$ estimates derived from 10,000 LOS.}
\label{fig_los_1d_flux_ps_ratios}
\end{figure*}

Below, we mostly focus on the flux power spectrum, although one can explore the
flux statistics with a variety of tools such as the 
flux PDF and the flux bispectrum
(see for example Mandelbaum et al. 2003; Viel et al. 2004; Fang \& White 2004; Lidz et al. 2006; 
Bolton, Oh \& Furlanetto 2009; McQuinn et al. 2009; Viel et al. 2009).

The flux power spectrum of the Ly$\alpha$ forest  is sensitive to a wide range of cosmological and astrophysical parameters and instrumental effects
and  has been extensively used in the literature as a probe of the
primordial matter power spectrum on scales of $0.5 - 40~h^{-1}$Mpc at $2 \le z \le 4$;
it does not have a 
simple algebraic relationship to the matter power spectrum because of nonlinearities in the flux-density relation.
Note in fact that by $z \sim 3$ the most important absorbing structures are weakly nonlinear. 
The Ly$\alpha$ flux distribution depends on the spatial distribution, the peculiar velocity field, 
and the thermal properties of the gas.
Going from the observed flux distribution to the power spectrum of matter in LSS requires knowledge of the bias 
of gas to matter, which in turns demands the temperature-density relation
of the gas and its evolution over cosmic history, as well as the nature of  the ionizing background radiation.
Hence, the only way 
to compute it is to rely on hydrodynamical simulations.  
The flux power spectrum can also be used to constrain cosmological parameters and the nature of dark matter through its shape and redshift dependence
(Croft et al. 2002). In addition, the Ly$\alpha$ forest power spectrum at small scales allows much improved constraints on the inflationary spectral index $n$
the running of that index with scale, and neutrino masses.

The relation between the three- and one-dimensional power spectra
is given by
\begin{equation}
P_{\rm 1D}(k_{\parallel}) = \int_{0}^{\infty} {k_{\perp} \over 2 \pi} P_{\rm 3D}(k_{\parallel},k_{\perp}) {\rm d}k_{\perp},
\end{equation}
and in linear theory one has $P_{\rm 3D}(k_{\parallel}, k_{\perp}) = b^2_{\rm \delta} P(k)(1 + \beta k^2_{\parallel}/k^2)^2$
with $k^2 = k^2_{\parallel} + k^2_{\perp}$, $b_{\rm \delta}$ the density bias and $\beta$ the redshift distortion parameter.
As anticipated, one can rely on accurate high-resolution and large box-size hydrodynamical simulations
to model the bias function $b(k)$, which relates the flux to the linear
dark matter power spectrum: $P_{\rm F}(k)= b^2(k) P(k)$.

Figure \ref{fig_los_1d_flux_ps} shows an example of the
one-dimensional flux power spectra computed from our simulation sets (BG, NU04, NU08), without (solid) and with (dotted and dashed) massive neutrinos -- after application of the splicing technique.
In particular, we considered two neutrino masses, namely $M_{\rm \nu}=0.4$ and  $0.8$ eV. 
Note that here the wave vector $k=2 \pi / \Delta v$ is measured in (km/s)$^{-1}$.
To perform our analysis, we extracted 10,000 mock quasar absorption spectra from the simulation sets 
at various redshift intervals -- from $z=2.2-4.4$ with $\Delta z=0.2$. 
The optical depth was rescaled in the standard way to match the observed effective optical depth at $z=3$, as given by Schaye et al. (2003), 
that is, $\tau_{\rm eff} = 0.363$, 
and to reproduce the same mean flux level;
this procedure is justified because the HI photoionization rate adopted in 
the simulations is inversely proportional to the Ly$\alpha$ optical depth in 
our mock spectra  -- see Viel et al. (2010) for more details.

Figure \ref{fig_los_1d_flux_ps_ratios} displays similar quantities as
the previous figure, but now the flux power spectra are normalized by the
corresponding measurements obtained from simulations with massless neutrinos. 
Error bars are $1\sigma$ estimates derived from 10,000 LOS.
As in Viel et al. (2010), we also compared simulations with massive neutrinos against those with only a massless neutrino component
and a reduced overall amplitude of the matter power spectrum. 
This allows distinguishing the effect of the neutrino free-streaming on the shape of the flux power spectrum
and its evolution from the overall suppression of power due to the free-streaming.
The latter is responsible for the well-known degeneracy between neutrino mass and $\sigma_8$.
In essence, the differences in the matter power spectra translate into a 
difference in the flux power spectrum 
for neutrino masses with $\sum m_{\rm \nu}=0.4 -0.8$ eV, which varies with redshift
and is more pronounced at $z=4$  -- if simulations are normalized to have the same $\sigma_8$
in the initial conditions.
This very weak effect is difficult to detect from present Ly$\alpha$ data and,
according to Viel et al. (2010), 
 nearly perfectly degenerates with 
the overall amplitude of the matter power spectrum $\sigma_8$.
As in Viel et al. (2010), we found that the overall suppression of power
induced by massive neutrinos on the flux power spectrum 
becomes stronger with larger neutrino mass and at higher redshift values, 
while there is an upturn and a bump at smaller scales.

In closing, we note that 
all our comparisons between simulations with massive neutrinos and with only a massless neutrino component 
were made assuming the same initial random seed for both simulations, so that the  
contribution from cosmic variance is effectively removed.  
The reason behind our choice, following Viel et al. (2010), is that 
we aim at distinguishing the effect of  a varying neutrino mass from the additional complication
introduced by cosmic variance. 
Namely, in this work we are more concerned about quantifying the impact of changing neutrino masses and
how this translates both into the total matter power spectrum and into the flux power spectrum -- separating this latter effect from the contribution
caused by cosmic variance. However, when simulations are used to compare with data, it is important to quantify the effect of cosmic variance --
as done for example in Borde et al. (2014) to contrast 
simulation results with BOSS data (see their Section 6.1, and their Table 5). 
To this end, we ran
 simulations with two different initial random seeds
to show the order of magnitude of the cosmic variance effect and where the difference mostly resides: as expected,
the  derived power spectra for the two seeds agree excellently well
at small scales, while at larger scales they can differ up to 2 to 3 $\sigma$ at all redshifts because of
cosmic variance.  Therefore, at large scales
cosmic variance has an impact on the power spectrum 
that exceeds  the simulation statistical uncertainty and needs to be
included as a systematic uncertainty when applying our model to data.



\section{CONCLUSIONS} \label{sec:summary} \label{sec_conclusions}


The determination of the neutrino mass and the nature of the neutrino mass hierarchy
are key issues in particle physics today -- directly connected with the origin of mass.
To this end, cosmology offers the best sensitivity to the neutrino mass,
and by combining cosmological and particle physics results from solar, atmospheric, reactor, and accelerator observations of neutrino oscillations
the absolute mass scale of  
neutrinos  can probably be determined in the very near future.


Massive neutrinos impact the CMB power spectrum and affect the LSS -- depending on the epoch at which they have become non-relativistic.
Because of their free-streaming, they also alter the low-$z$ power spectrum and lead to a modified
redshift-distance relation. In essence,  
neutrinos suppress power in DM clustering on small scales, 
which erases their own fluctuations on scales below the free streaming length. 
In turn, this slows down the growth of CDM structure on the same scale, leaving an
imprint on the matter power spectrum.
The overall result is a model of the Universe different from the standard $\Lambda$CDM scenario, with
important consequences on the structure formation mechanism.
 

Typically, limits on neutrino masses from cosmology are directly obtained from the 
analysis of the CMB radiation via the CMB power spectrum, the ISW effect on polarization maps,  
or through gravitational lensing of the CMB by LSS.
Other popular methods for quantifying  the impact of massive neutrinos are based on galaxy clustering and exploit 
high-redshift surveys.
On the other hand,
fewer studies involve the Ly$\alpha$ forest, which is now 
emerging as a unique
window into the high-redshift Universe, because it is located at a redshift range inaccessible to other LSS probes and spans a wide interval in redshift. 
The Ly$\alpha$ forest is a powerful tool for constraining neutrino masses, since 
massive neutrinos 
impact the one-dimensional flux power spectrum, 
because they suppress the growth of cosmological structures on scales smaller than the neutrino free-streaming distance. 
For neutrino masses below $1$ eV, the full extent of the suppression
occurs on megaparsecs scales.
In addition, 
combined with CMB observations and other tracers sensitive to large scales, the power spectrum of the Ly$\alpha$ forest can provide
stringent constraints on the shape and amplitude of the primordial power spectrum,  and hence 
directly test models
of inflation
(Viel et al. 2005; Seljak et al. 2005; Viel \& Haehnelt 2006).  


Therefore, a detailed modeling of the line-of-sight power spectrum of the transmitted flux in the Ly$\alpha$
forest with massive neutrinos is required. 
The main goal of our study was indeed to provide a novel suite of state-of-the-art 
hydrodynamical simulations with cold dark matter, baryons, and massive neutrinos,  
targeted at modeling the low-density regions of the IGM as probed by the Ly$\alpha$ forest at high redshift.
Our simulations spanned volumes ranging from $(25~h^{-1} {\rm Mpc})^3$ to $(100~h^{-1} {\rm Mpc} )^3$, 
and were made using either $3 \times 192^3 \simeq 21$ million or $3 \times 768^3 \simeq 1.4$ billion particles --
with chosen cosmological parameters compatible with the latest Planck (2013) results. 

As explained in Section \ref{sec_implementing_neutrinos}, neutrinos were implemented as a new type of particle
in the $N$-body setup (on top of gas and DM), and we  considered three degenerate 
species with masses $\sum m_{\rm \nu} =0.1, 0.2, 0.3, 0.4$, and $0.8$ eV.
This more
direct and computationally intensive approach
is primarily driven by our goal to
accurately reproduce all the main features of the Ly$\alpha$ forest at the quality level of BOSS or  future 
deep Ly$\alpha$ surveys. 
Figure \ref{fig_neutrino_linear_theory_A} shows that
the
one-dimensional Ly$\alpha$ forest data provided by BOSS lies in a $k$-range where
nonlinear evolution of cosmological neutrinos cannot be neglected, and  
hence any attempt
to speed-up calculations by using approximated linear solutions for the neutrino component -- instead of
a full hydrodynamical treatment --
would
compromise our ability to reproduce accurately all the features of the forest.

Technical aspects of the new suite of hydrodynamical simulations, such as details on the code used for the  runs, initial conditions, 
optimization strategies and performance, along with various improvements and a description 
of the pipeline developed to extract the synthetic Ly$\alpha$ transmitted flux, were presented in Section \ref{sec_simulations_description} -- 
building upon the theoretical background 
of Sections \ref{sec_modeling_lya} and \ref{sec_implementing_neutrinos} (see also Tables \ref{tab_suite_sims_A} and \ref{tab_suite_sims_B}).
Since we are planning to make the simulations available to the scientific community upon request, 
this part may be regarded as a
guide for a direct use of the simulations.

We improved on previous studies in several directions, in particular 
with updated routines for IGM radiative cooling and heating processes, 
and initial conditions based on 2LPT instead of on Zel'dovich approximation.
Figures \ref{fig_sims_visualization_A} and \ref{fig_sims_visualization_B}
are visual
examples of a few snapshots at $z=2.2$ and $z=0$ for the
gas, dark matter, and neutrino components -- when present -- in a simulation
with $192^3$ particles per type and a box size of $25~h^{-1}$Mpc.


Using the splicing technique introduced by McDonald (2003), 
the resolution of our runs can be further enhanced 
to reach the equivalent of $3 \times 3072^3 \simeq 87$ billion particles in a $(100~h^{-1} {\rm Mpc} )^3$ box size.
This means that our simulations, specifically designed to meet the requirements of the BOSS survey
(which has already identified $\sim 150,000$ QSO over $10,000$ square degrees 
within $z=2.15-4.5$), are
also useful for upcoming or future experiments -- such as eBOSS and DESI.
In particular, the comoving volume of eBOSS will be nearly ten times that probed by the BOSS galaxy survey,
while DESI will exceed BOSS and eBOSS both in volume and in quasar density, increasing the total number of 
Ly$\alpha$ quasars by about a factor of 5.


In addition to providing technical details, in Section \ref{sec_first_results}
we  also performed a first analysis of our simulations; in particular, we characterized the  
nonlinear three- and one-dimensional matter and flux power spectra 
and the statistics of the transmitted flux in the Ly$\alpha$ forest with massive neutrinos. 
Massive neutrinos induce changes in the LSS clustering of DM and
thermal state of the gas (as evident from Figure \ref{fig_visualization_internal_energy}), affecting the $T_0-\gamma$ relation 
(equation \ref{eq_T_rho}).
In Section  \ref{sec_3d_matter_ps}, we investigated in more depth
the effect of massive neutrinos on the three-dimensional matter power spectrum, 
where linear and nonlinear evolutions
at different redshifts and for various neutrino masses are studied (Figures \ref{fig_3d_matter_ps_comparisons_A}-\ref{fig_3d_matter_ps_comparisons_D}).
The characteristic redshift- and mass-dependent suppression of the matter power spectrum caused by the massive
neutrino component is clearly seen in Figure \ref{fig_3d_matter_ps_comparisons_B}, and the values of $k$
most sensitive to the neutrino mass (i.e. the most relevant scales for detecting the power spectrum suppression due to neutrinos)
are  shown in Figure \ref{fig_3d_matter_ps_comparisons_D}. 
Finally, we briefly discussed how this feature propagates in the one-dimensional flux power spectrum
(Section \ref{sec_1d_flux_stat}, Figures \ref{fig_los_1d_flux_ps} and \ref{fig_los_1d_flux_ps_ratios}) and affects 
the statistical  properties of the transmitted flux fraction. 


This work represents the first of a series of papers dedicated 
to quantify the effects of massive neutrinos in the Ly$\alpha$ forest across different redshift slices and at nonlinear scales.
In particular, our 
primary next goal is to combine 
the Ly$\alpha$ one-dimensional power spectra at different redshifts obtained 
 from these simulations with analogous measurements derived from the BOSS Ly$\alpha$ forest data to
constrain  the sum of the masses of the three neutrino 
flavors and  the main cosmological parameters with improved sensitivity.
This is possible via a multidimensional likelihood analysis, a method
 pioneered by Croft et al. (1998, 2002) and used by  Viel \& Haehnelt (2006) on SDSS data,
 or more recently by Palanque-Delabrouille et al. (2013) on  SDSS-III/BOSS DR9 quasar spectra.
Clearly, at a later stage we will combine our Ly$\alpha$ measurements with Planck data and other available datasets (galaxy PS and BAO from BOSS, lensing PS) 
to derive tighter joint constraints on cosmological and astrophysical parameters, and on the neutrino mass. 


In addition, our simulations can be useful for a broader variety of cosmological and astrophysical applications, 
ranging from the three-dimensional modeling 
of the Ly$\alpha$ forest to cross-correlations between different probes, 
the study of dark energy and expansion history of the Universe 
in presence of massive neutrinos, and particle-physics-related topics.  
Examples include cross-correlation studies along the lines of Font-Ribera et al. (2013),
synergies between ground and space missions in constraining the neutrino mass (we note that DESI, DES, LSST, Euclid,
and CMB-stage 4 experiments will unambiguously detect the neutrino mass under both hierarchy scenarios), 
comparison of our results with different hydrodynamical codes and 
neutrino implementations, and studies of systematics affecting the Ly$\alpha$ forest as a tracer.
To this end, 
UV fluctuations at $z>4$, galactic winds, metal enrichment, re-ionization history, and the
thermal history of IGM  are all still major uncertainties in any analysis of the Ly$\alpha$ forest flux statistics,
along with instrument performance and survey design, and they deserve  a closer scrutiny.

The full suite of simulations presented in this paper will be made available to the scientific community upon request.



\section*{ACKNOWLEDGMENTS}

We acknowledge PRACE for awarding us access to resource Curie-thin nodes based in France at TGCC, for our project 2012071264.
This work was granted access to the HPC resources of CCRT under the allocation 2013-t2013047004 made by
GENCI (Grand Equipement National de Calcul Intensif).
A.B., N.P.-D., G.R. and Ch.Y.  acknowledge  support from grant ANR-11-JS04-011-01 of Agence Nationale de la Recherche.
The work of G.R. is also supported by the faculty research fund of Sejong University in 2014.
M.V. is supported by ERC-StG "CosmoIGM".
JSB acknowledges the support of a Royal Society University Research Fellowship.
We thank Volker Springel for making Gadget-3 available.



\appendix
\section{A SANITY CHECK}  

\begin{figure*}
\centering
\includegraphics[angle=0,width=0.45\textwidth]{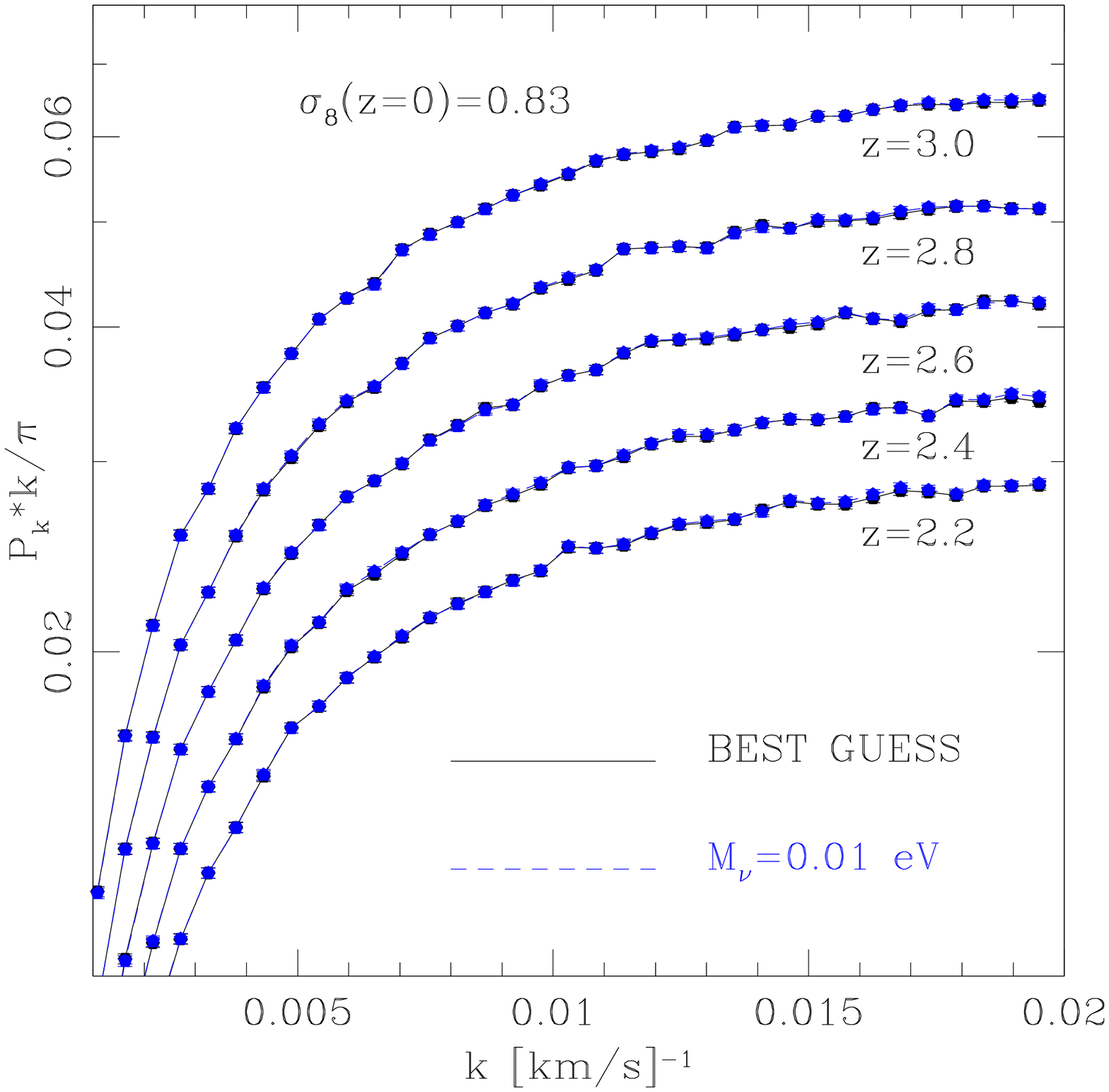}
\includegraphics[angle=0,width=0.45\textwidth]{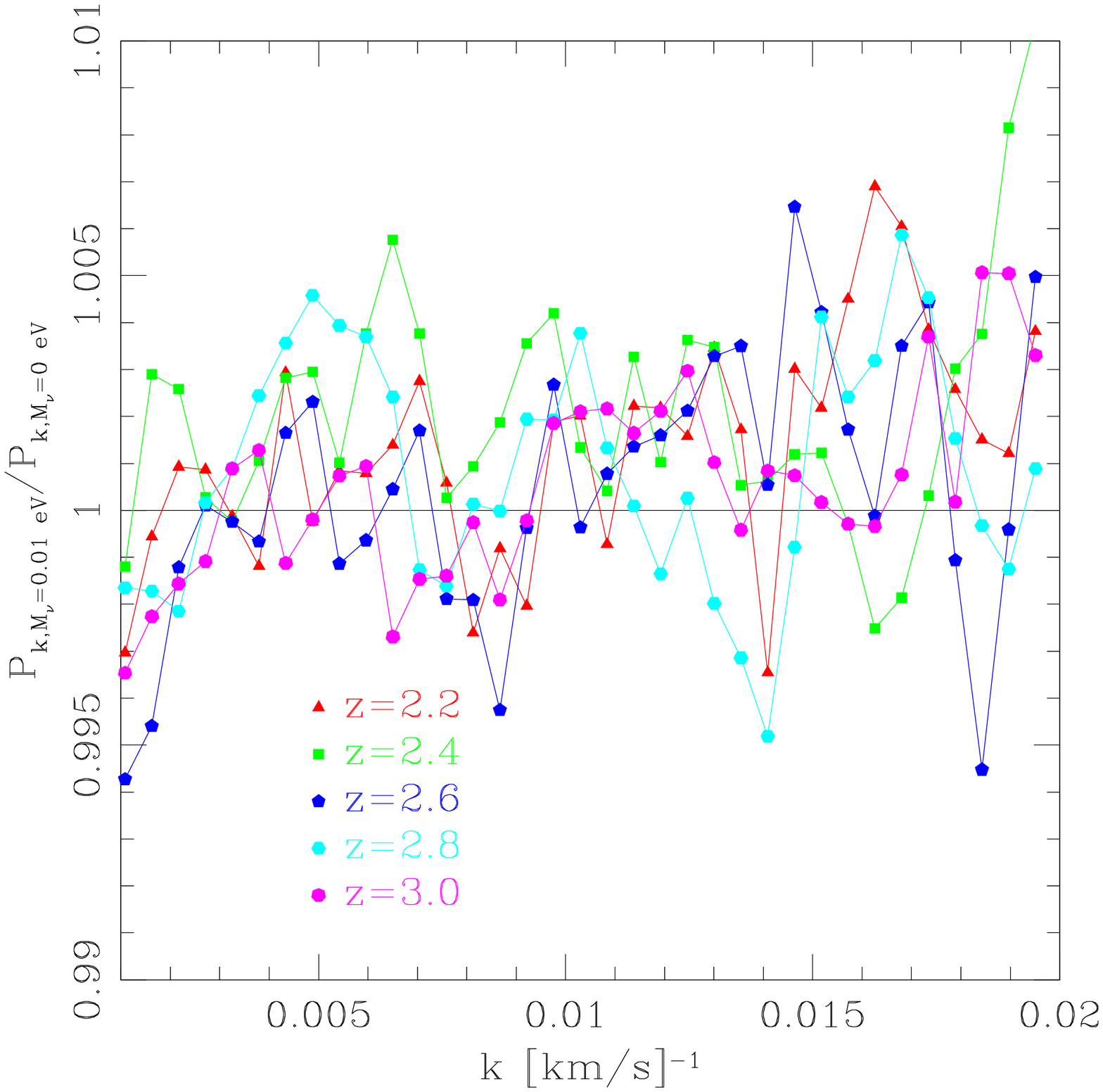}
\caption{[Left] Flux power spectra of the best-guess simulations and of the simulations with $M_{\rm \nu}=0.01$ eV for 6 values in redshift --
from $z=2.2$ till $z=3.0$, where $\Delta z=0.2$. [Right] Same as in the left panel, but now in terms of ratios between one-dimensional flux power spectra. In the range of interest, 
convergence is safely achieved.}
\label{fig_consistency_check}
\end{figure*}

As discussed in Section \ref{sec_convergence_tests}, 
achieving numerical convergence in the modeling of the Ly$\alpha$ flux power spectrum is nontrivial.
In addition, when we include massive neutrinos in hydrodynamical simulations,
the $N$-body setup is quite different from the case of gas and DM alone, since we
are dealing with an additional type of particle.
Clearly, for a very low value of the neutrino mass, we expect results to be consistent with
the case of massless neutrinos. 
To check that we indeed correctly recover the limit of massless neutrinos,
we ran a set of simulations with a very small neutrino mass, $M_{\rm \nu}=0.01$ eV (i.e., simulation set NUBG a,b,c, in Table \ref{tab_suite_sims_A}).
We then extracted the line-of-sight flux power spectra at different redshifts, as done in Section \ref{sec_1d_flux_stat},
and computed their average values across 10,000 random lines. These measurements were compared 
with analogous measurements obtained from the best-guess run, which did not include massive neutrinos.
Results are shown in Figure \ref{fig_consistency_check}, where it can be seen that
convergence is safely achieved in the range of interest.  




\label{lastpage}
\end{document}